	\documentclass[traditabstract]{aa} 
\usepackage{appendix}
\usepackage{latexsym}
\usepackage{graphicx}
\usepackage{natbib}
\usepackage{lscape}
\usepackage{longtable}
\usepackage{afterpage}
\usepackage[varg]{txfonts}

\begin{document}

\title{VLT/VIMOS integral field  spectroscopy of luminous and ultraluminous infrared galaxies: 2D kinematic properties}
   \author{Enrica Bellocchi\inst{1}, Santiago Arribas\inst{1}, Luis Colina\inst{1}, Daniel Miralles-Caballero\inst{2}  }
   \institute{$^1$ Centro de Astrobiolog\'ia, Departamento de Astrof\'isica, CSIC-INTA, Cra. de Ajalvir Km. 4, 28850 - Torrej\'on de Ardoz, Madrid, Spain \\  \email{bellochie@cab.inta-csic.es}\\ $^2$ Instituto de F\'isica Te\'orica, Universidad Aut\'onoma de Madrid, 28049, Madrid, Spain               }
                  
 \date{Received 29 December 2012 / Accepted 21 June 2013}
 
\abstract{We present and discuss the 2D kinematic properties of the ionized gas (H$\alpha$) in a sample of 38 local (ultra) luminous infrared galaxies [(U)LIRGs] (31 LIRGs and 7 ULIRGs, 51 individual galaxies) observed with VIMOS at the Very Large Telescope using optical integral field spectroscopy (IFS). This sample covers well the less studied LIRG luminosity range and includes the morphological types corresponding to the different  phases along the merging process (i.e., isolated disks, interacting systems, and mergers). The majority of the galaxies have two main kinematically distinct components. One component (i.e., {}{\it narrow} or {}{\it systemic}) extends over the whole line-emitting region and is characterized by small to intermediate velocity dispersions (i.e., $\sigma$ from 30 to 160 km s$^{-1}$). The second component ({\it broad}) has in general a larger velocity dispersion (up to 320 km s$^{-1}$); it is mainly found in the inner regions and is generally blueshifted with respect to the systemic component. The largest extensions and extreme kinematic properties of the broad component are observed in interacting and merging systems, and they are likely associated with nuclear outflows. The systemic component traces the overall velocity field showing a large variety of kinematic 2D structures, from very regular velocity patterns typical of pure rotating disks (29\%) to kinematically perturbed disks (47\%) and highly disrupted and complex velocity fields (24\%). Thus, most of the objects (76\%) are dominated by rotation. We find that rotation is more relevant in LIRGs than in ULIRGs. There is a clear correlation between the different phases of the merging process and the mean kinematic properties inferred from the velocity maps. In particular, isolated disks, interacting galaxies, and merging systems define a sequence of increasing mean velocity dispersion, and decreasing velocity field amplitude, characterized by average dynamical ratios (v$_{shear}^*/ \sigma_{mean}$) of 4.7, 3.0 and 1.8, respectively. We also find that the ratio between the nuclear ($\sigma_c$) and the mean velocity dispersions ($\sigma_{mean}$) vs. $\sigma_{mean}$ is an excellent discriminating plane between disks and interacting/merging systems: disks show a mean ratio a factor of 2 larger than those characterizing the other two classes. The LIRGs classified as isolated disks have similar velocity amplitudes but larger mean velocity dispersions (44 vs. 24 km s$^{-1}$) than local spirals, implying a larger turbulence and thicker disks. Interacting systems and mergers have values closer to those of low velocity dispersion ellipticals/lenticular galaxies (E/SOs). The subclass of (U)LIRGs classified as mergers have kinematic properties similar to those shown by the Lyman break analogs (LBAs), although the dynamical mass of LBAs is five times lower on average. Therefore, despite the difference in mass and dust content, the kinematics of these two local populations appears to have significant noncircular motions. These motions may be induced by the tidal forces, producing dynamically hot systems. The dynamical masses range from $\sim$ 0.04 m$_\star$ to 1.4 m$_\star$ (i.e., m$_\star$ = 1.4 $\times$ 10$^{11}$ M$_\odot$), with ULIRGs (M$_{dyn}$ $\sim$ 0.5 $\pm$ 0.2 m$_\star$) being more massive than LIRGs by, on average, a factor of about 2. The mass ratio of individual pre-coalescence galaxies is $<$ 2.5 for most of the systems, confirming that most (U)LIRG mergers involve sub-m$_\star$ galaxies of similar mass. }


\titlerunning{2D kinematics of (U)LIRGs}
\authorrunning{Bellocchi et al.}

\maketitle

%
\section{Introduction}

Luminous and ultraluminous infrared galaxies (LIRGs, L$_{IR}$ = [8-1000 $\mu$m] =10$^{11}$ - 10$^{12}$ L$_\odot$, and ULIRGs, L$_{IR} >$ 10$^{12}$ L$_\odot$, respectively) are important populations to study galaxy evolution.  They are systems of intense star formation (SF), whose spectral energy distributions (SEDs) are dominated by dust thermal emission arising from the reprocessing of ultraviolet photons produced by young massive stars and/or active galactic nucleus (AGN) heating  (e.g., \citealt{vivian12}, and references therein).  
It has been proposed that ULIRGs are systems that transform gas-rich disk galaxies into moderate-mass ellipticals through merger events (i.e., \citealt{genzel01, T02}), with some, but not all, evolving to quasars (QSOs; \citealt{Co01, sanders03}). These objects exhibit a large variety of morphologies, which suggest different dynamical phases: from mostly isolated disks for low-luminosity LIRGs to a majority of merger remnants for ULIRGs (e.g., \citealt{V02, A04}). Although (U)LIRGs are rare in the local universe (e.g., \citealt{lagache05}), they are much more numerous at high-z and are relevant contributors to the whole past star formation beyond z $\sim$ 1 (e.g., \citealt{LF05, PG05, PPG08}). Therefore, low-z (U)LIRGs offer a unique opportunity to study, at high linear resolution and signal-to-noise ratio (S/N), extreme SF events and compare them with those observed at high-z. 

The 2D kinematic characterization of galaxies is a key element to study the physical processes that govern their formation and evolution. For instance, it provides a powerful diagnostic i) to infer the main source of dynamical support (e.g., \citealt{puech07, Epi09}), ii) to distinguish between relaxed virialized systems and merger events (e.g., \citealt{Flores06, S08, YO2012}), iii) to detect and characterize radial motions associated with feedback mechanisms, like outflows (e.g., \citealt{S09}, \citealt{rupke13}), and iv) to infer fundamental galaxy quantities like dynamical masses (e.g., \citealt{Co05}) among other topics.  

Despite their relevance, the number of spatially resolved kinematic studies of low-z (U)LIRGs is rather limited. Most optical and near-IR integral field spectroscopy (IFS) analysis has been focused on individual objects (e.g., \citealt{Co99}, \citealt{tecza00}, \citealt{A01, GM06, Bed09, Piq12}) or small samples (e.g., \citealt{MI06}), generally at the highest luminosity (i.e., ULIRGs). \citealt{Co05} discussed IFS-based H$\alpha$ velocity maps for 11 ULIRGs, finding that in general their kinematics do not correspond to that of an ordered rotating system but are dominated by motions associated with tidal forces. Recently \cite{West12} have expanded this census by presenting optical (H$\alpha$) IFS of 18 ULIRGs observed with VIMOS at the Very Large Telescope (VLT), finding a larger fraction of objects dominated by rotation.

In this paper we significantly expand previous samples (in number and characteristics) by obtaining spatially resolved kinematics of 38 local (U)LIRG systems (51 individual galaxies) observed with the VIMOS/VLT integral field unit (IFU). The sample contains a large fraction (i.e., 31/38) of sources in the less studied LIRG luminosity range. This is relevant for two main reasons. On the one hand, it complements previous studies focused on ULIRGs, filling the gap between the extreme cases and the general population of local star-forming galaxies (SFGs). On the other hand, several authors have suggested that high-z LIRGs are scaled-up versions of low-z LIRGs (e.g., \citealt{pope06, papov07, elbaz10, nordon10, TAK10, elbaz11, nordon12}).

The main goals of the present paper are i) to present the atlas of ionized gas velocity fields and velocity dispersion maps for this sample of (U)LIRGs, ii) to characterize their main 2D kinematic properties, and iii) to compare such properties with those of other relevant low-z samples. More focused studies, including a detailed comparison with high-z samples, will be presented in future papers.   

This is part of a series of papers belonging to a large project aiming at characterizing the properties of (U)LIRGs on the basis of optical and infrared integral field spectroscopy. In particular, the first results of the VIMOS survey were presented in \cite{A08} (hereafter, Paper I), along with a detailed kinematic study of two galaxies representative, respectively, of interacting pairs (i.e., IRAS F06076-2139) and morphologically regular spirals (i.e., IRAS F12115-4656). \cite{MI10} (hereafter, Paper II) studied the nature and origin of the ionization mechanisms operating in the extra-nuclear regions of LIRGs as a function of the interaction phase and infrared luminosity. The morphologies of the stellar continuum and the ionized gas (H$\alpha$) emissions have been analyzed in \cite{RZ11} (hereafter, Paper III). In \cite{A12}, the analogy between 
local LIRGs and ULIRGs, observed with INTEGRAL/WHT and VIMOS/VLT and high-z massive SFGs is studied by comparing their basic H$\alpha$ structural characteristics.

The paper is structured as follows. In Sect. 2, we present the sample as well as some details about the observations, data reductions, line fitting and map construction. Sect. 3 is focused both on the 2D kinematic properties derived from the velocity maps and the global properties for the main kinematically distinct  components found. Sect. 4 discusses how the kinematic properties of (U)LIRGs depend on infrared luminosity and on the different dynamical phases along the merging process. This section is also devoted to deriving and discussing the dynamical masses of (U)LIRGs, which, together with the main kinematic properties, are compared with other local samples. Finally, the main results and conclusions are summarized in Sect. 5. In the Appendix, the kinematic maps, comments on the individual objects, and details on the method followed to derive the effective radii are presented. Throughout the paper we will consider H$_0$ = 70 km s$^{-1}$ Mpc$^{-1}$, $\Omega_M$ = 0.3 and $\Omega_\Lambda$ = 0.7.

\section{Observations, data reduction, data analysis}

\subsection{The sample}
\label{morph_class}

The sample analyzed here contains a total of 38 (U)LIRGs systems (51 individual galaxies) of the southern hemisphere drawn from the Revised Bright Galaxy Sample (RBGS, \citealt{sanders03}). Of these systems 31 are LIRGs (i.e., $<$L$_{IR}>$ = 2.9 $\times$10$^{11}$ L$_\odot$) with a mean redshift of 0.024 (corresponding to D $\sim$ 100 Mpc), and the remaining seven are ULIRGs (i.e., $<$L$_{IR}>$ =1.6 $\times$ 10$^{12}$ L$_\odot$) with a mean redshift of 0.069 (D $\sim$ 300 Mpc; see Table \ref{table_sample} and Paper I for details). Therefore, this sample includes a good representation of the relatively less studied LIRG luminosity range. It encompasses a wide variety of morphological types, suggesting different dynamical phases (isolated spirals, interacting galaxies, and ongoing- and post-mergers), and nuclear excitations (HII, Seyfert, and LINER). Some objects have evidence in their optical spectra of hosting an AGN, showing high  [NII]/H$\alpha$ values and/or broad H$\alpha$ emission lines (e.g., IRAS F05189-2524, IRAS F21453-3511; \citealt{MI10}, \citealt{A12}). The sample is complete neither in luminosity nor in distance. However, it covers well the relevant luminosity range and is representative of the different morphologies within the (U)LIRG phenomenon.

The morphology class was derived using ground-based images (i.e., Digital Sky Survey (DSS) and additional archival Hubble Space Telescope (HST) images when available). The sources have been morphologically classified following a simplified version of the scheme proposed by \cite{V02}, with three main classes instead of five (see \citealt{RZ11} for further details). Briefly, the three morphological classes are defined as follows:

\vspace{2mm}

\begin{itemize}
\item Class 0: objects that appear to be single isolated galaxies, with a relatively symmetric disk morphology and without evidence for strong past or ongoing interaction (hereafter $disk$).
\item Class 1: objects in a pre-coalescence phase with two well-differentiated nuclei separated a projected distance $>$ 1.5 kpc. For these objects, it is still possible to identify the individual merging galaxies and their corresponding tidal structures due to the interaction (hereafter $interacting$). 
\item Class 2: objects with two nuclei separated a projected distance $\leq$ 1.5 kpc or a single nucleus with a relatively asymmetric morphology suggesting a post-coalescence merging phase (hereafter $merger$).
\end{itemize}

In Table \ref{table_sample} we present the main properties of the sample. In some cases the properties of individual galaxies in multiple systems could be inferred separately and were therefore treated individually. As part of the analysis, additional galaxies with available IFS data have been included to increase the number of systems in the ULIRG luminosity range. These data (\citealt{Co05, GM07, GM09}) correspond to our own observations taken with INTEGRAL/WHT.

\begin{table*}
\centering
\caption{General properties of the (U)LIRGs sample.}
\label{table_sample}
\begin{scriptsize}
\begin{tabular}{c c c c c c c c c  c c} \\
\hline\hline\noalign{\smallskip}  
ID1{} &  ID2{} &   $z$ &  D&  scale{} & log L$_{IR}$  {}& Class  &  Notes   \\
IRAS & Other &  &  (Mpc) &  (pc/$^{\prime\prime}$)     & (L$_{\odot}$)    &     &   \\ 
(1) & (2) & (3) & (4) & (5) & (6) & (7) & (8)    \\
\hline\noalign{\smallskip} 	
F01159-4443  &   ESO 244-G012 		&  0.022903 &  99.8 	&  462  &11.48 (N) 		&  1  & a,d \\ 
F01341-3735  &  ESO 297-G011/G012 	&  0.017305 & 75.1  	&  352  & 11.18 (G11:10.99; G12:10.72) & 1  &  a,d \\
F04315-0840  &  NGC 1614 			& 0.015983  & 69.1  		& 325   & 11.69  		& 2 	&   \\
F05189-2524  &   					& 0.042563  & 188.2  		&  839  & 12.19  		& 2 	& \\
F06035-7102  &  						& 0.079465  &  360.7 		& 1501 & 12.26  		& 1 	&  \\
F06076-2139  &  						& 0.037446  & 165  		&  743  & 11.67 (N) 	& 1 	& a,d \\
F06206-6315  &  						&0.092441   & 423.3 		& 1720 &  12.27  		&1 	&\\
F06259-4780  & ESO 255-IG007 		& 0.038790  & 171.1  		& 769   &  11.91 (N)  	& 1 	&  b,d\\
F06295-1735  & ESO 557-G002 		& 0.021298  & 92.7  		&  431  &  11.27  		& 0  &\\
F06592-6313  &  						& 0.022956  & 100 		& 464  & 11.22 		& 0 	& \\
F07027-6011  &  AM 0702-601 		& 0.031322 & 137.4 		& 626  & 11.64 (S:11.51; N:11.04)& 0 & a,d \\
F07160-6215 & NGC 2369 			& 0.010807 & 46.7 		& 221  & 11.16 		& 0 	&\\
08355-4944   &  						& 0.025898 & 113.1 		& 521  & 11.60  		& 2 	& \\
08424-3130   & ESO 432-IG006 		& 0.016165 & 70.1 		&329   &11.04 (S) 	& 1 	& a,d \\
F08520-6850 & ESO 60-IG016 		& 0.046315 & 205.4 		& 909  & 11.83 		& 1 	& \\
09022-3615   &  						& 0.059641 & 267 		&1153 &12.32 		& 2 	& \\
F09437+0317& IC-563 / 564 			& 0.020467 & 89 			& 415  & 11.21(S:10.82; N:10.99)	& 1 (0) & a,c,d \\
F10015-0614 & NGC 3110 			& 0.016858 & 73.1 		&343   & 11.31 		&0 	&\\
F10038-3338 &  ESO 374-IG032		& 0.034100 & 149.9 		& 679  & 11.77	 	&2 	&\\ 
F10257-4339 & NGC 3256			& 0.009354 & 40.4 		& 192  & 11.69 		& 2 	& \\
F10409-4556 & ESO 264-G036 		& 0.021011 & 91.4 		& 425  & 11.26 		& 0 	&\\
F10567-4310 & ESO 264-G057 		& 0.017199 &  74.6  		&  350  & 11.07  		&  0 	&     \\
F11255-4120 &  ESO 319-G022 		& 0.016351 & 70.9 		& 333  & 11.04  		&  0 	&      \\
F11506-3851 & ESO 320-G030		& 0.010781 & 46.6 		& 221  & 11.30 		& 0 	& \\
F12043-3140 & ESO 440-IG058 		& 0.023203 & 101.1 		& 468  & 11.37 (S) 	& 1	& a,d \\ 
F12115-4656 & ESO 267-G030 		& 0.018489 & 80.3 		& 375  &11.11		& 0	& \\ 
12116-5615   & 						&  0.027102 & 118.5 		& 545  &11.61 		& 2 (0) &\\
F12596-1529 & MCG-02-33-098 		& 0.015921 & 69.0 		& 324  &11.07 		&1 	&  \\
F13001-2339 & ESO 507-G070 		& 0.021702 & 94.5 		& 439  & 11.48 		& 2 (0/1) & \\
F13229-2934 & NGC 5135 			&  0.013693 & 59.3		&  280 & 11.29 		& 0 	&  \\
F14544-4255 & IC 4518 				&  0.015728 & 68.2 		&  320 & 11.11(E:10.80; W:10.80) &  1& a,d \\
F17138-1017 & 						&  0.017335 & 75.2 		& 352  & 11.41		& 2 (0) & \\
F18093-5744 &  IC 4687/4686/4689	&  0.017345 & 75.3 		& 353  & 11.57 (N:11.47; C:10.87)& 1 &  b,d	\\
F21130-4446 &  						&  0.092554 & 423.9 		& 1722 & 12.09 		& 2 	& \\
F21453-3511 &  NGC 7130 			& 0.016151  & 70.0 		& 329   & 11.41  		& 2 	&     \\
F22132-3705 &  IC 5179 				& 0.011415  & 49.3 		& 234   & 11.22  		& 0 	&    \\
F22491-1808 &  						& 0.077760  & 352.5 		& 1471 &   12.17 		& 1  &  \\
F23128-5919 &  AM 2312-591			& 0.044601  & 197.5 		& 878   & 12.06   		&  1	&    \\
\hline\hline
\end{tabular}
\end{scriptsize}
 \begin{minipage}{16.8cm}
   \vskip0.1cm\hskip0.0cm
\footnotesize
\tablefoot{
Col (1): Object designation in the Infrared Astronomical Satellite (IRAS) Faint Source Catalog (FSC). 
Col (2): Other identification. 
Col (3): Redshift from the NASA Extragalactic Database (NED). 
Col (4): Luminosity distance assuming a $\Lambda$DCM cosmology with H$_0$ = 70 km s$^{-1}$ Mpc$^{-1}$, $\Omega_M$ = 0.3, and $\Omega_\Lambda$ = 0.7, using the E. L. Wright Cosmology calculator, which is based on the prescription given by \cite{Wri06}. 
Col (5): Scale. 
Col (6): Infrared luminosity (L$_{IR}$= L(8-1000) $\mu$m) in units of solar bolometric luminosity, calculated using the fluxes in the four IRAS bands as given in \cite{sanders03} when available. Otherwise, the standard prescription given in \cite{SM96} with the values in the IRAS Point and Faint Source catalogs was used. 
Col (7): Morphological class defined as follows: 0 identifies {\it isolated} objects, 1 {\it pre-coalescence} systems, and 2 stands for {\it merger} objects. For those objects for which the morphological classification is uncertain, the various possible classes are shown in the table with the preferred morphological classification indicated in the first place and the alternative classification within brackets (see text for further details). 
Col (8): Notes with the following code: 
(a) System composed of two galaxies. 
(b) System composed of three galaxies. 
(c) There are two VIMOS pointings for the northern source. 
(d) Interacting system (i.e., see notes $a$ and $b$) for which the total infrared luminosity L$_{IR}$ could be approximately assigned among the members of the system according to the MIPS/Spitzer photometry*. (The estimations for the individual sources are shown in parentheses in Col. (6)). When the whole infrared luminosity contribution is assigned only to one of the sources of the system, either Spitzer data at 24 $\mu$m are not available or the limited angular resolution does not allow the different galaxies to be separated. Each source is identified according to its position in the VIMOS FoV according to N, S, C, E, and W which stand for the northern, southern, central, eastern, and western object, respectively.\\ {}*MIPS/Spitzer data are available at the NASA/IPAC Infrared Science Archive ({\it http://irsa.ipac.caltech.edu}).} 
\end{minipage}
\end{table*}

\subsection{Observations}

The observations have been described in detail in previous papers (e.g., \citealt{MI10}). In brief, they were carried out  using the IFU of VIMOS (\citealt{Lfevre03}) at the VLT, covering the spectral range $(5250-7400)$ \AA\ with the high-resolution  grating GG435 (`HR-orange' mode) and a mean spectral resolution of 3470 (dispersion of 0.62 \AA\ pix$^{-1}$). The field of view (FoV) in this configuration is 27$^{\prime\prime}$ $\times$ 27$^{\prime\prime}$, with a spaxel scale of 0.67$^{\prime\prime}$ per fiber (i.e., 1600 spectra are obtained simultaneously from a 40 $\times$ 40 fiber array). A square 4 pointing dithering pattern was used. With a relative offset of 2.7$^{\prime\prime}$ (i.e., 4 spaxels), this pattern provides an effective FoV of 29.5$^{\prime\prime}$ $\times$ 29.5$^{\prime\prime}$. The exposure time per pointing ranges from 720 to 850 seconds, so that, the total integration time per galaxy is between 2880 and 3400 seconds. 

\subsection{Data reduction}
\label{reduc}

The VIMOS data were reduced with a combination of the pipeline {\it Esorex} (versions 3.5.1 and 3.6.5) which is included in the pipeline provided by ESO, and different customized IDL and IRAF scripts. The basic data reduction (i.e., bias subtraction, flat field correction, spectra tracing and extraction, correction of fiber and pixel transmission, and relative flux calibrations) is performed using the {\it Esorex} pipeline. The four quadrants per pointing are reduced individually and then combined into a single data cube. After that, the four independent dithered pointing positions are combined to end up with the final `supercube', containing 44 $\times$ 44 spaxels for each object (i.e., 1936 spectra). 

To estimate the absolute wavelength calibration accuracy and the spectral resolution (i.e., instrumental profile), we first fitted the [O I]$\lambda$6300.3 \AA\ sky line to a single Gaussian for all spectra of each individual source, obtaining the mean central wavelength and FWHM for each object. Since we are going to concentrate on a wavelength range around the H$\alpha$ and the [N II]$\lambda\lambda$6548.1, 6583.4 \AA\ emission lines, the [O I]$\lambda$6300.3 \AA\ sky line is suitable due to its proximity to these lines. The mean values for the central wavelength and FWHM for the whole sample were (6300.29 $\pm$ 0.07) \AA\ and (1.80 $\pm$ 0.07) \AA\ , respectively, which also gives a good estimate of the uncertainties due to calibration. The wavelength calibration, the instrumental profile, and fiber-to-fiber transmission correction are checked for all the galaxies using the aforementioned sky line. A more detailed description of the data reduction process is given in Paper I and Paper III.


\subsection{Line fitting and map generation}

The observed H$\alpha$ and [NII]$\lambda\lambda$6548, 6583 \AA\ emission lines of the individual spectra are fitted to Gaussian profiles using an IDL routine (i.e., MPFITEXPR, implemented by C. B. Markwardt). This algorithm derives the best set of lines that match the available data. In case of adjusting multiple lines, the line flux ratios and wavelengths of the different lines are fixed according to the atomic physics. The widths are constrained to be equal for all the lines and greater than the instrumental contribution ($\sigma_{INS}$).

We select spaxels where H$\alpha$ lines are detected with S/N~$>$ 4, fitting automatically all the lines to single Gaussian profiles (i.e., 1 component (1c) fit), which gave good results for the majority of the spectra. However, for certain regions of the galaxies, a multicomponent fit (e.g., 2 components (2c) per line) was required in order to adequately fit the data. In these cases, we can identify systemic (or {\it narrow}) and secondary (or {\it broad}) components, according to their widths. These components usually have different central velocities, confirming that they are two kinematically distinct components. The lines associated with different components were fitted simultaneously, maintaining for each component the assumptions mentioned before. The narrow component usually extends over the whole galaxy line-emitting region, while the broad component covers a smaller area corresponding to the nuclear regions of the galaxies.

\begin{figure*}
\includegraphics[width=0.5\textwidth]{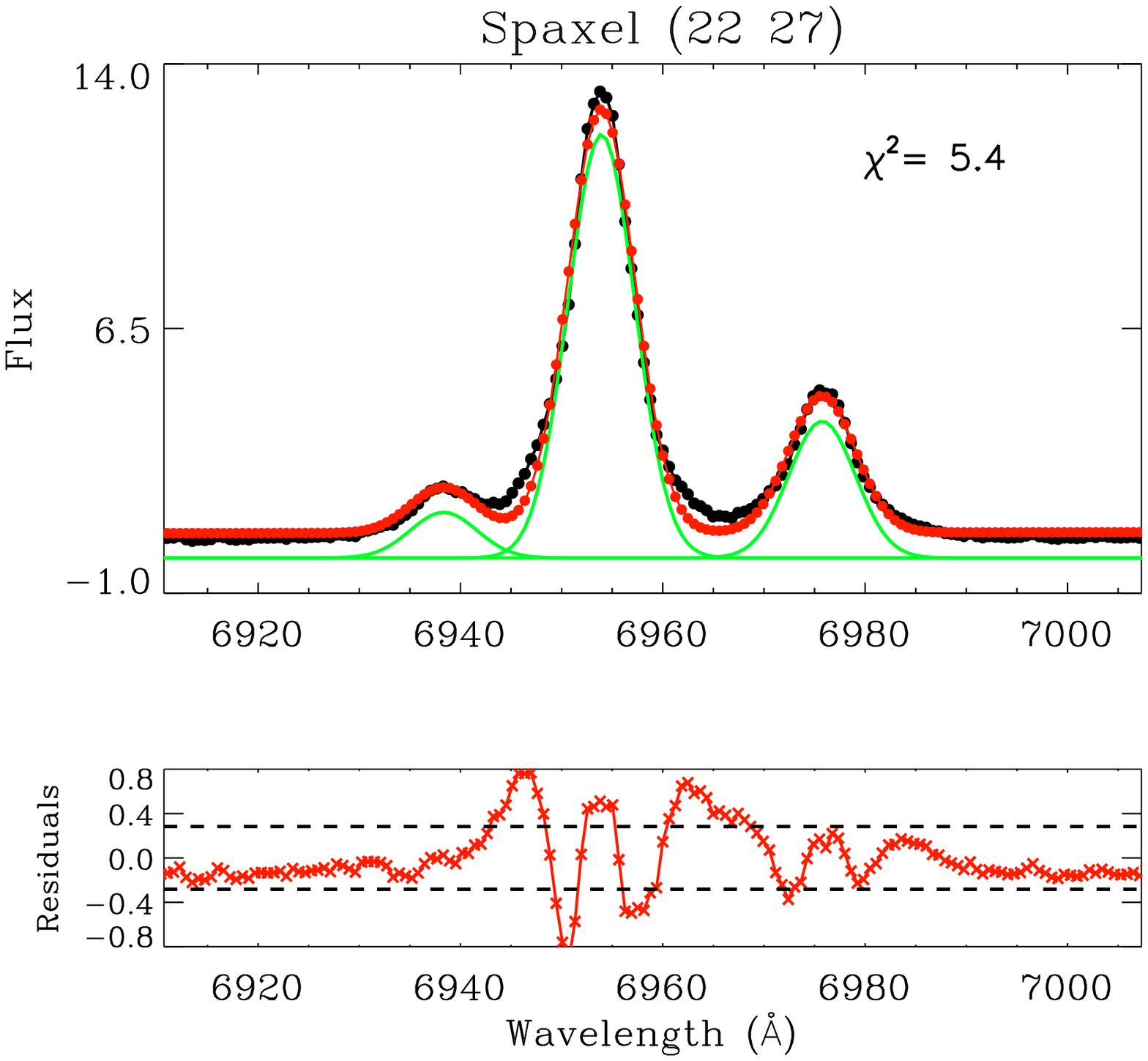}
\includegraphics[width=0.5\textwidth]{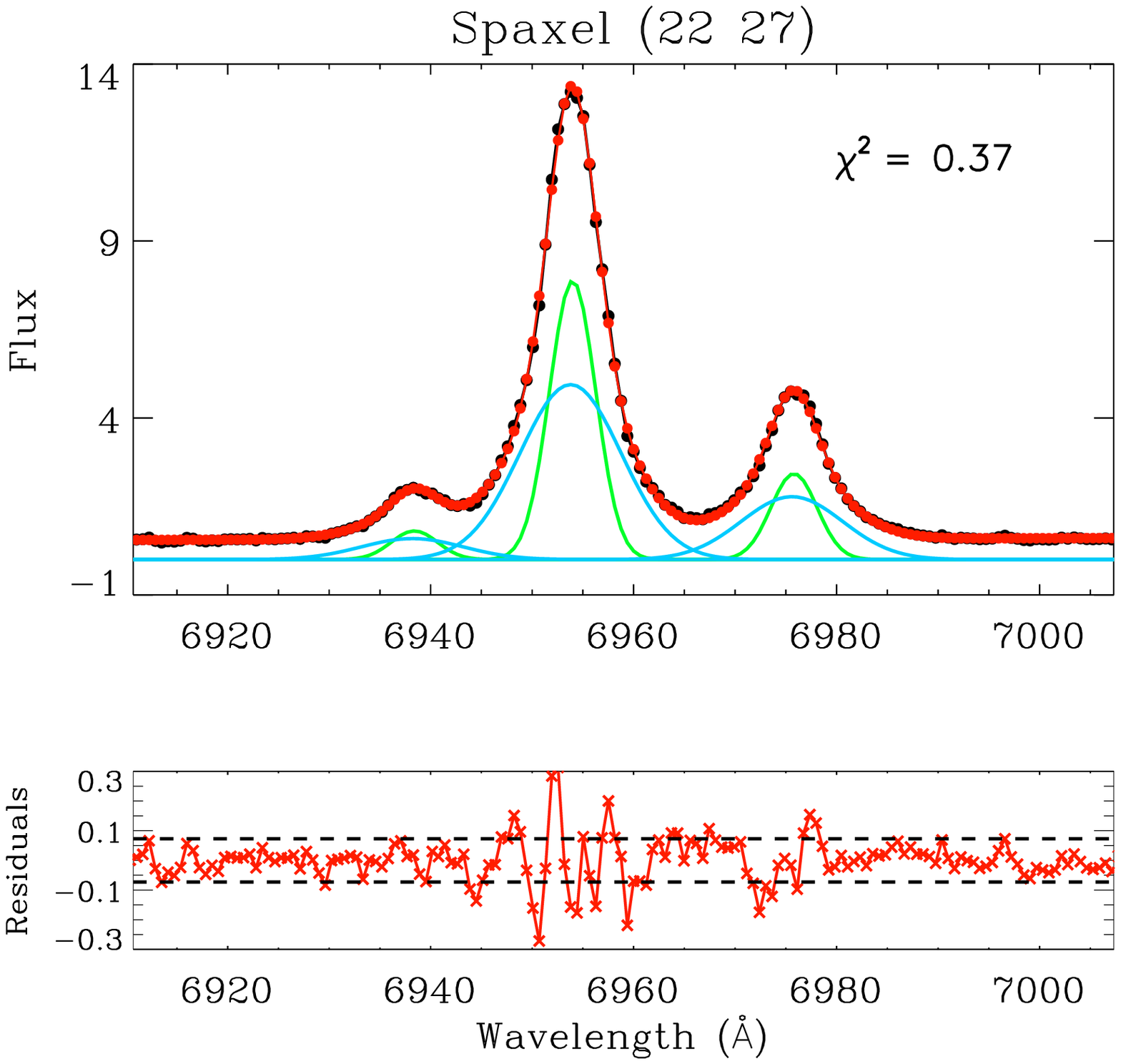}
\includegraphics[width=0.5\textwidth]{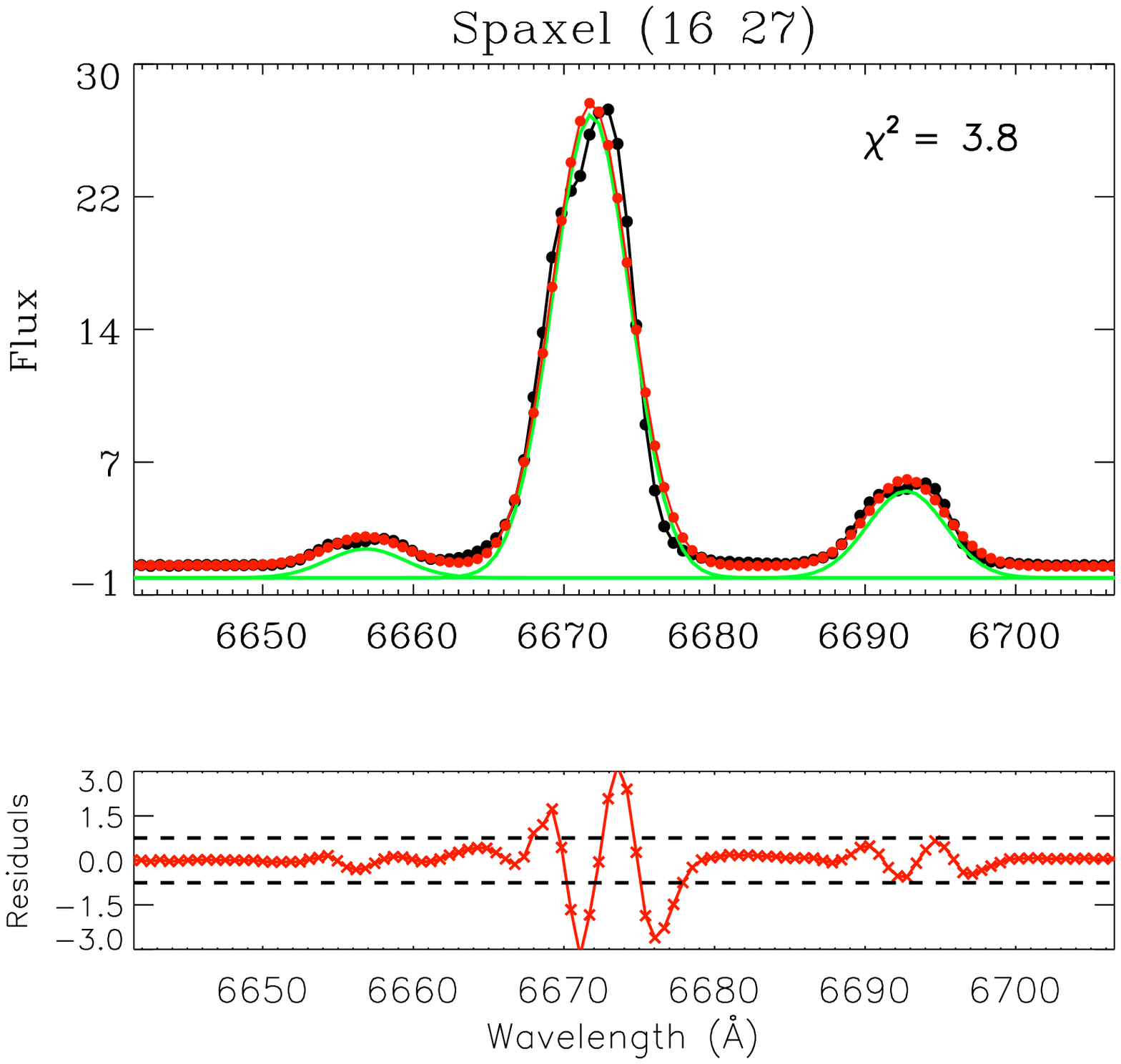}
\includegraphics[width=0.5\textwidth]{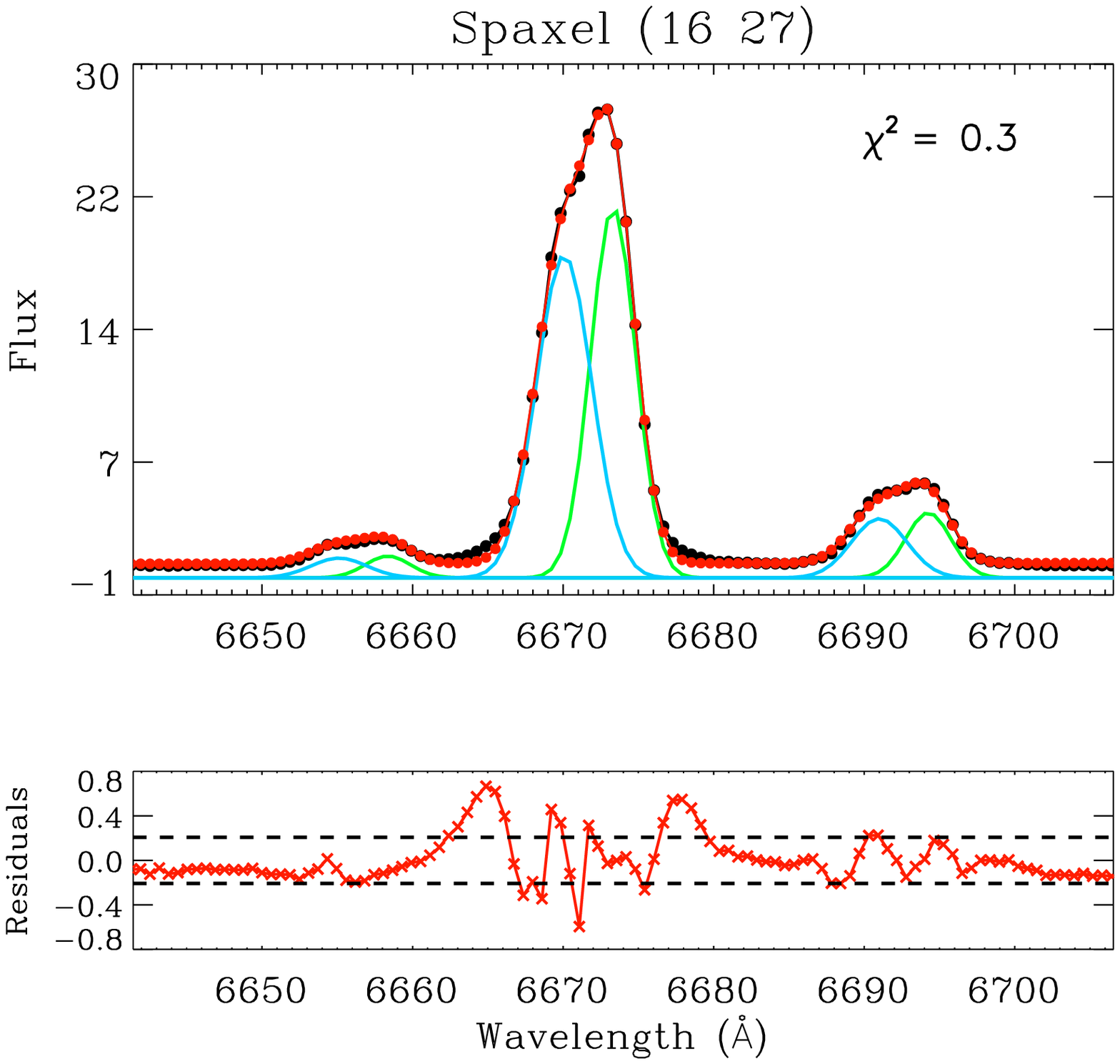}
\caption{\small Examples of one (1c, on the left) and two (2c, on the right) component Gaussian fits to the H$\alpha$ + [NII] profiles in two different spaxels in IRAS F09022-3615 (top) and IRAS F18093-5744 C (bottom). The $\chi^2$ for all the Gaussian fittings is shown for a direct comparison. The systemic (narrow) component is shown in green, while the broad one is shown in light blue. The red curve shows the total contribution coming from the H$\alpha$ + [NII] Gaussian fit. Below, the residuals (i.e., data - model) are presented: they clearly show when two components are needed to properly fit asymmetric line profiles. The standard deviation value of the residuals (i.e., $\pm$ 1$\sigma$) is shown using dashed lines. }
\label{1c_2c}
\end{figure*}

We check all the spectra by visual inspection. Looking at the H$\alpha$ line profile and the residuals of the fit, we decided to apply 1c or 2c fitting. Where the presence of one or two components was dubious the percentage of spectra was generally small (i.e., typically a  few percent) since it was clear that one or two components were preferred for a large majority. There are a few cases where decomposition introduces some uncertainties, and these are commented on Appendix \ref{App_maps}. Two components are considered when the 2c fits give a significantly lower $\chi^2$ value than that derived when using 1c fits (i.e., $\chi^2_{2c}$ / $\chi^2_{1c}<$ 0.5). A couple of examples showing two different kinds of fittings, 1c and 2c, are shown in figure \ref{1c_2c}. Only a few systems do not show clear signs of broad components (e.g., IRAS F06259-4780 S, IRAS F09437+0317 S, IRAS F12043-3140 N, and IRAS F22132-3705).  

For each emission line/component, we end up with the following information: line flux, central wavelength ($\lambda_c$), and intrinsic width ($\sigma_{line} = \sqrt{\sigma_{obs}^2 - \sigma_{INS}^2}$)\footnote{An average value of the $ \sigma_{INS}$ of (37 $\pm$ 2) km s$^{-1}$ is used (see Sect. 2.3).}, together with their corresponding fitting errors. However, a better estimate of the actual uncertainties associated with the kinematic parameters is obtained by combining in quadrature fitting and calibration errors (see Sect. 2.3). For high S/N spectra fitting errors are small (typically, $\Delta\lambda$ $<$ 0.1 \AA; $\Delta\sigma$ $<$ 0.1 \AA) and calibration errors can be considered the main source of uncertainty (and vice versa for spectra with S/N $<$ 5). 

Flux intensity, velocity field, and velocity dispersion maps for the narrow and broad components (see Appendix A) are obtained using IDL procedures (i.e, {\bf jmaplot}, \citealt{MA04}). As reference, we also include the continuum image generated from the VIMOS-IFU data cube. When HST imaging is available, it is also shown.

We note that due to the different resolving power of the original spectra, authors use a different number of line components to fit the observed profiles. For instance, \cite{Co05} consider only one component for most of their sample due to the relatively low spectral resolution of their INTEGRAL/WHT data (i.e., R = 1500), while \cite{West12} use two or three components (see Sect. \ref{compar_ifs} for details) thanks to the higher spectral resolution of the VIMOS/VLT data (i.e., R = 3100). When considering only one component, v and $\sigma$ are in general similar to those of the narrow component in a 2-Gaussian fit. We have used our sample to quantify the effects on the kinematic quantities (i.e., v$_{shear}$, $\sigma_{mean}$ (see Sect. 3.1.1) and $\sigma_{c}$ (see Sect. 4.2)). We find that the central velocity dispersion, $\sigma_c$, increases on average by $\sim$ 20\% if one (instead of two) component fitting is applied to about dozen cases (e.g., IRAS F09022-3615, IRAS F01341-3735 G11, IRAS F06035-7102, IRAS F07160-6215, IRAS F08355-4944, IRAS F14544-4255 W, IRAS F22491-1808) that are potentially affected by the selected approach. On the other hand, the mean velocity dispersion, $\sigma_{mean}$, only increases by a few percent (up to 10\% in the case of IRAS F08355-4944). The velocity shear, v$_{shear}$, is virtually unaltered by the fitting approach. Therefore, we note that the $\sigma_{mean}$ and v$_{shear}$ values, on which most of the analysis is based on, are not significantly affected by the fit approach considered. Although there is evidence in a few cases for the presence of more than two components in our spectra, these were faint and/or with a very complex structure.

\section{Results}

\subsection{Spatially resolved kinematics in (U)LIRGs: components properties}

Figures in Appendix A show the H$\alpha$ flux, velocity field, and velocity dispersion maps of the different kinematic components for each galaxy of the present sample. The characteristics of the velocity fields and velocity dispersion maps are discussed individually in that Appendix.

\subsubsection{Narrow (systemic) component}

The narrow component covers the whole H$\alpha$ line-emitting region of these galaxies and generally contains most of the H$\alpha$ flux. Therefore, it is morphologically similar to the total flux (i.e., narrow + broad) H$\alpha$ maps presented by \cite{RZ11}. The spatial extension of this component (10 - 350 kpc$^2$; see Table \ref{NARROW}) is similar to that of the continuum emission, though in some cases there are significant structural differences (e.g., F01341-3735 (S, N); see also \citealt{RZ11}). Its spatial distribution and kinematic properties represent those of the entire galaxy, and it is therefore identified as the systemic component.
 
This narrow (systemic) component shows a large variety of 2D kinematic properties. The kinematic maps of some sources show very regular patterns, with a spider-like velocity field and a centrally peaked velocity dispersion distribution, as expected for pure rotating disk (e.g., IRAS F11506-3851, IRAS F12115-4656). However, most of the objects have some kind of irregularities and distortions in their velocity maps, despite a large rotation component in many cases. These departures from a regular rotation pattern include: i) kinematic axes that are poorly defined (e.g., IRAS F13229-2934), nonperpendicular each to another (e.g., IRAS F11255-4120), and/or shifted with respect to the photometric axes (e.g., IRAS F10038-3338), ii) asymmetry between the approaching and receding regions (e.g., IRAS F13229-2934), and iii) off-nuclear kinematic center (e.g., IRAS F12596-1529). In addition, the velocity dispersion maps generally show an asymmetric structure. Although the highest velocity dispersion is usually coincident with the optical nucleus and therefore seems to trace the mass, there are cases where the absolute peak is associated with (tidal) structures located at large galactocentric distances (e.g., in IRAS F04315-0840 at 3.6 kpc;  in IRAS F10015-0614 at 4 kpc; in F12043-3140 S at 3.4 kpc). Interestingly, a significant number of  systems have ring-like structures of high velocity dispersion (i.e., $\sigma \sim$ 150 - 200 km s$^{-1}$) in their outer regions (e.g., IRAS F07027-6011S, IRAS 09022-3615, IRAS F12116-5615). These are regions of relatively low H$\alpha$ emission, which are likely ionized by shocks (\citealt{MI10} and references therein).

\begin{figure*}
\includegraphics[width=0.99\textwidth]{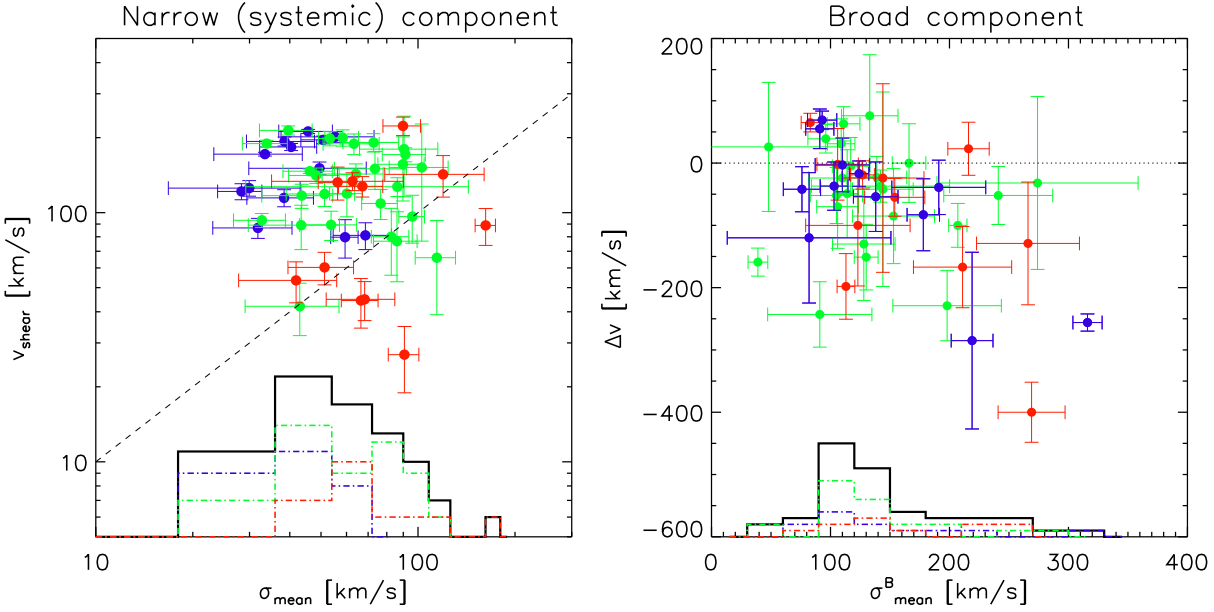}
\caption{\small {\bf Left panel:} Velocity shear, v$_{shear}$, versus the mean velocity dispersion, $\sigma_{mean}$, for the narrow (systemic) component. Blue, green, and red dots distinguish, respectively, morphological classes 0 (isolated disks), 1 (interacting), and 2 (mergers). The distribution of $\sigma_{mean}$ values is also shown, with the contribution of the different morphological classes. The 1:1 dot line is shown as reference to highlight the fact that most of the sources have v$_{shear}>\sigma_{mean}$. {\bf Right panel:} Velocity shift, $\Delta$v, between the broad and narrow components versus the mean velocity dispersion of the broad component, $\sigma^B_{mean}$. The horizontal dot line indicates the zero velocity shift. Most of the objects have a blueshifted broad component.}
\label{vs_comb}
\end{figure*}

We have used our IFS-based velocity maps to infer some global kinematic properties associated with this component (see Table \ref{NARROW}). Fig. \ref{vs_comb} shows the observed (i.e., uncorrected for inclination) velocity shear, v$_{shear}$\footnote{v$_{shear}$ is defined as half of the difference between the median value of the 5 percentile at each end of the velocity distribution (i.e., v$_{shear}$ = $\frac{1}{2}$ (v$^{5\%}_{max}$ - v$^{5\%}_{min}$)), following previous works (e.g., \citealt{Gon10}).}, and the spatially resolved uniformly weighted mean velocity dispersion, $\sigma_{mean}$, of the narrow (systemic) component over the whole line-emitting region. These two  properties have a relatively large range of values (30 $< v_{shear}<$ 220 km s$^{-1}$; 30 $<\sigma_{mean}<$ 160 km s$^{-1}$). However, the scatter in the figure is dominated not by errors associated with the observations and/or the analysis but by errors associated with different intrinsic properties among the objects. This reflects that the selected (U)LIRGs represent a kinematically diverse sample. In fact, the figure shows a clear dependence on the mean v$_{shear}$ and $\sigma_{mean}$ values with the morphological class, with isolated disks having relatively lower values than systems in a more evolved dynamical phase (interacting and merging systems; see Table \ref{NARROW}). In Sect. 4, we discuss in more detail the mean kinematic properties of (U)LIRGs as a function of their morphology/dynamical status.

\subsubsection{Broad component}

The broad component is generally restricted to the central regions, though in some cases it is found at very large galactocentric distances (up to $\sim$ 7 kpc in IRAS F23128-5919). Although in most of the cases, the (projected) area covered by the broad component is smaller than 3 kpc$^2$, there is a significant fraction (12/46) of objects with an extension larger than 5 kpc$^2$ (see Table \ref{BROAD}). In a few cases, the broad component can be dominant in terms of flux in the innermost spaxels (e.g., IRAS F05189-2524, IRAS F08355- 4944). In some cases the broad emission is offset with respect to the photometric nucleus, thus defining an asymmetric structure (e.g., IRAS F12043-3140 S, IRAS F18093-5744 N). At some locations the broad component can be due to the line-of-sight overlapping of two physically distinct structures (e.g., IRAS F06076-2139, where the emission from the northern galaxy overlaps with the ring of the southern source; see also Paper I).

Due to sampling limitations, compactness, and low-surface brightness constraints, the detailed study of the 2D kinematic structure of this component is sometimes difficult. However, in several sources it is clearly resolved, and the kinematic axes can be well defined. For instance, in IRAS F04315-0840 (NGC 1614) the kinematic axes associated with the broad component are nearly perpendicular to those of the narrow component, which is a clear feature of outflows perpendicular to the disk (\citealt{heck90, YO2012}). The velocity dispersion maps of several objects do not have a radial symmetry and/or have an off-nuclear kinematic center of symmetry (e.g., IRAS F04315-0840, IRAS F06035-7102, IRAS F07160-6215, IRAS F10038-3338).

The range of the global kinematic properties of the broad component (right panel of Fig. \ref{vs_comb}) in velocity dispersions is significantly larger than that of the narrow component, with values of up to $\sim$ 320 km s$^{-1}$ (i.e., FWHM $\sim$ 750 km s$^{-1}$ for IRAS F07027-6011N). However, a large fraction of sources (i.e., 26/46) are in the range 90 $<$ $\sigma^B_{mean}$ $<$ 150 km s$^{-1}$ (see Table \ref{BROAD}). The mean central velocity of the broad component is blueshifted with respect to that of the narrow component for a majority of sources, reaching values up to $\Delta$v = -400 km s$^{-1}$ (i.e., IRAS F05189-2524). There is however, a significant fraction of objects (9/46) with redshifted velocities, which seem to cluster around velocity dispersion values of $\sim$ 100 km s$^{-1}$.
 
The presence of AGNs may affect the kinematic properties of this component (e.g., \citealt{Lipari03, West12}, \citealt{rupke13}, Arribas et al. in prep.). One of the two class 0 objects with large FWHM and $\Delta$v offset (i.e., IRAS F07027-6011 N) has evidence of hosting an AGN (see \citealt{RZ11, A12}). Excluding this object, most of the systems with $\Delta$v $>$ 100 km s$^{-1}$ and / or $\sigma^B_{mean}$ $>$ 200 km s$^{-1}$ are class 1 and 2. Among these sources in an advance merging phase, several (i.e., IRAS F05189-2524, IRAS F06206-6315, IRAS F13229-2934, IRAS F14544-4255 W, IRAS F18093-5744 C, IRAS F21453-3511, and IRAS F23128-5919) also have evidence of hosting an AGN according to the optical spectroscopic classification (see Paper III) and/or the evidence for a broad H$\alpha$ line (see \citealt{A12}). 

As for the spatial extension of the broad component, there is a clear trend with the morphological class. Specifically, the median areas covered by the broad component for classes 0, 1, and 2 are 1.1 kpc$^2$, 1.5 kpc$^2$, and 5.2 kpc$^2$, respectively.

In summary, the present data show that an extended broad component is observed in most of LIRGs and ULIRGs. The mean velocity of most of the objects is blueshifted with respect to the narrow component, with values up to -400 km s$^{-1}$ and FWHM $\sim$ 750 km s$^{-1}$. The largest extensions and extreme kinematic properties ($\Delta$v, FWHM) are observed in interacting and merging systems (classes 1 and 2). The detailed study of the kinematic properties of the broad component will be done elsewhere (Arribas et al in prep.).

\subsubsection {Comparison with previous IFS studies}        
\label{compar_ifs}

As mentioned above, the detailed comparison of the component properties with previous IFS studies is not straightforward. On the one hand, most of the previous studies are focused on the ULIRG luminosity range (e.g., \citealt{Co05} and \citealt{West12}), while a large fraction of the present sources are LIRGs. On the other hand, there is some heterogeneity in the way in which the line fitting and component map construction are performed among the different works. In \cite{Co05} only one single component was in general adjusted to the observed profiles since the relatively low spectral resolution of the INTEGRAL/WHT data does not allow the presence of multiple components in the lines to be reliably distinguished on a spaxel by spaxel basis. However, in general the derived kinematical values of v and $\sigma$ in a 1-Gaussian fit are similar to those of the narrow component when applying a 2-Gaussian fit. In \cite{West12} multiple components were fitted for most of the sample thanks to the higher spectral resolution of the VIMOS/VLT data. The different components were assigned to the lines mainly using a line intensity criterion (rather than a line width criterion), regardless of the relative velocities of the components. Thus, they do not necessarily represent kinematically distinct components, as in our case. The morphological features seen in our kinematic maps are similar to those reported in these works, which also find asymmetries and departures from rotation for most of their objects. The range of $\sigma$ found by \cite{Co05} (i.e., 50 $<\sigma<$ 150 km s$^{-1}$) is similar to the one found here for the narrow component of the present sample. \cite{West12} also found it necessary to adjust multiple components over relatively extended regions. In addiction, they discovered motions associated with outflows for a relatively large fraction of objects in their sample (11/18). However, as we discuss in the next section, we find a larger fraction of rotating systems than in their samples.

\section{Discussion}

\subsection {Kinematical classification: the role of rotation in (U)LIRGs and its dependence on L$_{IR}$} 
\label{kine_class}

The present sample is formed by a group of morphologically diverse objects, classified on the basis of optical imaging into three groups according to their suggested dynamical status (i.e., isolated relaxed rotating disks: class 0, interacting systems: class 1, and mergers: class 2; see Sect. 2 for more detailed definitions). 
In this section we perform a simple kinematic classification of the (U)LIRG sample based on the velocity fields and the velocity dispersion maps of the systemic component. This classification is similar to the one proposed by \cite{Flores06} and distinguishes three main groups:  

\begin{enumerate}
\item Rotating disks (RD): both the velocity field and the velocity dispersion maps show the expected pattern for rotation. The major kinematic  axis in the velocity field follows the major photometric axis of the continuum map,  and the $\sigma$-map shows a peak near the galaxy nucleus. Regions of high velocity dispersions associated with the outer faint emission, such as those studied by \cite{MI10}, are not considered a significant deviation from the global rotation pattern. 
\item Perturbed disks (PD): the velocity field shows a general rotation pattern with well-defined approaching and receding regions, but the kinematic axes show some distortions and/or the $\sigma$-map has a peak shifted off the center;
\item Complex kinematics (CK): systems with both velocity field and velocity dispersion maps discrepant from the expected normal rotating disks showing the evidence of radial motions and/or disturbed velocity dispersion distribution.
\end{enumerate}

From a total of 51 galaxies in the sample, 49 could be classified according to this scheme\footnote{No kinematic classification has been possible for IRAS F08424-3130 N since it is located on the edge of the VIMOS FoV. For IRAS F12596-1529 the kinematic classification was not possible due to the limited linear resolution.}. Of these, 14 are undoubtably RD (29\%), while for 23 additional ones rotation is likely to be dominant, despite the presence of clear asymmetries (PD, 47\%). The remaining 12 show CK (24\%). Therefore, for 76\% of the galaxies (i.e., 37/49) rotation plays an important role.

As for the global properties, objects classified as RD, PD and CK have increasing values of $\sigma_{mean}$  (i.e., 50 $\pm$ 5, 58 $\pm$ 4, and 91 $\pm$ 8 km s$^{-1}$, respectively) and decreasing v$_{shear}$ (159 $\pm$ 12, 123 $\pm$ 10, and 112 $\pm$ 15 km s$^{-1}$, respectively), as shown in Table \ref{kin_prop_gen}.

The fraction of rotating systems is considerably larger than those reported previously in ULIRGs by \cite{Co05} and \cite{West12}. In particular, \cite{Co05} found while studying a sample of 11 ULIRGs that the velocity field of the ionized gas is dominated by tidally induced flows and generally does not correspond to rotationally supported systems. Only one system (i.e., IRAS 17208-0014, \citealt{A03}) shows a dominant rotation pattern when fitting the emission lines to a single Gaussian component, and two more (i.e., Arp 220, see \citealt{A01, Co04} and IRAS 08572+3915, see \citealt{A00}) could be considered as candidates for rotating systems after a kinematic decomposition similar to the one performed in the present paper. \cite{West12} find a larger fraction of rotating systems (i.e., 8/18) in their sample of ULIRGs. These fractions are significantly smaller than our finding of 37 out of 49 in the present (U)LIRG sample. This may be explained by the fact that the present work includes a large fraction of systems with lower luminosity (i.e., 31/38 LIRGs) than in the mentioned works, which only cover the ULIRG luminosity range.

In fact, we find that the kinematic class clearly correlates with the infrared luminosity L$_{IR}$. A large fraction (i.e., 6/7) of ULIRGs were classified as CK, while most of the objects with luminosity lower than 10$^{11.4}$ L$_\odot$ are classified as PD or RD (i.e., PD = 11/21 and RD = 7/21, respectively). Therefore, large deviations from rotation are generally associated with ULIRGs, while the lower end of the luminosity range (i.e., LIRGs) is generally populated by rotation-dominated systems. However, exceptions to this rule exist: there is a ULIRG (i.e., IRAS F06206-6315) with a regular rotation pattern and there are a few LIRGs (e.g., IRAS F13001-2339, IRAS F13229-2934) showing complex kinematics.

It is also worth mentioning that a quantitative analysis of the relative importance of rotation with respect to the kinematic asymmetries, based on the application of kinemetry methodology (\citealt{K06}), has been presented in \cite{YO2012} for four sources. This methodology will be applied to the whole sample in a forthcoming work (Bellocchi et al. 2013). However, it is important to note that the kinemetry methodology by itself is not able to kinematically classify objects in terms of disk and interacting and merging systems. It requires some references to interpret the kinematic asymmetries associated with the different dynamical phases. \cite{S08} and \cite{YO2012} propose different frontiers (i.e., total kinematic asymmetry) to distinguish disks from mergers, indicating that this topic deserves further investigation.

\subsection{Global kinematic properties along the merging process: $\sigma_c$/$\sigma_{mean}$ vs. $\sigma_{mean}$  a good discriminator between disks and interacting/merging systems}

The kinematic classification described above shows that rotation is more frequent in systems classified as isolated (class 0) than in those classified as interacting (class 1) and mergers (class 2). The fraction of RD decreases from 46\% to 32\% for isolated and interacting systems, while none of the mergers are classified as RD. In contrast, only one isolated object presents CK, while the fraction of CK increases to 20\% and 55\% for interacting systems and mergers, respectively (see Table \ref{kin_prop_gen}).

\begin{figure}
\includegraphics[width=0.5\textwidth]{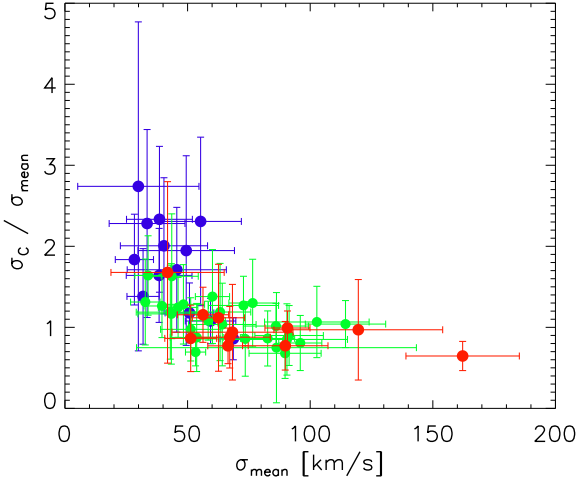}
\caption{\small Relationship between the central velocity dispersion, $\sigma_c$, of the systemic component and its mean velocity dispersion over the observed area, $\sigma_{mean}$, for the three different morphological classes. The color code is the same as the one used in figure \ref{vs_comb}.}
\label{ratio_S}
\end{figure}

When the central (nuclear) velocity dispersion\footnote{The H$\alpha$ central velocity dispersion is derived as the mean value of the 4 $\times$ 4 spaxels around the VIMOS continuum flux peak.}, $\sigma_{c}$, of the systemic component is compared with its spatially resolved mean value, $\sigma_{mean}$, a clear segregation between isolated disks and interacting/merging systems is found (Fig. \ref{ratio_S}). Specifically, the derived mean ${\sigma_c}$/$\sigma_{mean}$ ratios for disks, as well as for interacting and merging systems, are, respectively, (1.8 $\pm$ 0.15), (1.1 $\pm$ 0.05), and (0.97 $\pm$ 0.09). On the one hand, this trend can be explained as a consequence of the tidal effects on the gas velocity dispersion. Isolated relaxed disks (class 0) have relatively low velocity dispersions outside the nuclear region, with a $\sigma_{mean}\approx$ 44 km s$^{-1}$ (i.e., a factor of $\sim$ 2 larger than that derived for local spiral galaxies, see Tables \ref{kin_prop_gen} and \ref{kin_prop_lowz}) and, generally, centrally peaked velocity dispersion maps, making the mentioned ratio relatively large. On the other hand, as in principle the beam smearing of the central velocity gradient in a disk might be significant to enhance the $\sigma_c$/$\sigma_ {mean}$ ratio (see \citealt{davies11}), the comparison of samples in this diagram should be done ideally at similar linear resolution. For our particular case, the effects on this ratio are expected to be small (i.e., $\sim$ 10\% in the most extreme case), as further discussed in Sect. 4.3.1. In the pre-coalescence (class 1) objects, the velocity dispersion outside the nucleus increases because the main body of the system is perturbed by tidal forces associated with interactions, reducing significantly such a ratio. The largest $\sigma_{mean}$ values are reached for the ongoing, postcoalescence mergers (class 2). However, though $\sigma_c$ and, to a lesser degree, $\sigma_{mean}$ may (slightly) depend on the fitting approach (i.e., to one or two components), the general trend observed in the plot does not depend on the fitting approach\footnote{We note that the narrow component of a 2-Gaussian fit has, in general, similar properties compared to those of a single Gaussian fitted line.}.

We note that this could be a good kinematic diagnostic diagram for classifying spatially resolved high-z sources, though similar linear resolutions to the ones considered here should be used for a direct comparison. This is the case of several AO-assisted IFS works of high-z SFG samples (e.g., \citealt{Law07a,FS09, swin12}), which can reach a sub-kpc scale of the order of those of low-z (U)LIRGs observed under seeing-limited conditions.

In the following section we discuss the dynamical ratio v/$\sigma$ and its dependence on the kinematical and morphological classes.

\subsection{Dynamical support in (U)LIRGs}
\label{dyn_support}

The dynamical ratio between the velocity amplitude and the local velocity dispersion v/$\sigma$ of a system is a useful parameter to kinematically characterize a system. It allows the distinction between rotation-dominated v/$\sigma$ $>$ 1 and random-motion-dominated systems v/$\sigma$ $<$ 1 (e.g., \citealt{Epi12} and references there in). We note that for systems whose kinematic is dominated by tidal forces and/or radial motions (e.g., outflows), v$_{shear}$ may not be tracing rotation. Similarly, in those cases $\sigma_{mean}$ does not necessary give a measure of velocity dispersion, but it can be the result of integrating several kinematically distinct components. However, since only a small fraction of systems show complex kinematics (see Sect. 4.1) and a kinematic decomposition has been applied (Sects. 3.1.1 and 3.1.2.), v$_{shear}$ and $\sigma_{mean}$ should trace, respectively, rotation and dispersion in most of the cases.

\subsubsection{From rotating disks to dispersion-dominated systems}      
\label{rot_disp_systs}

\begin{figure}[h]
\includegraphics[width=0.5\textwidth]{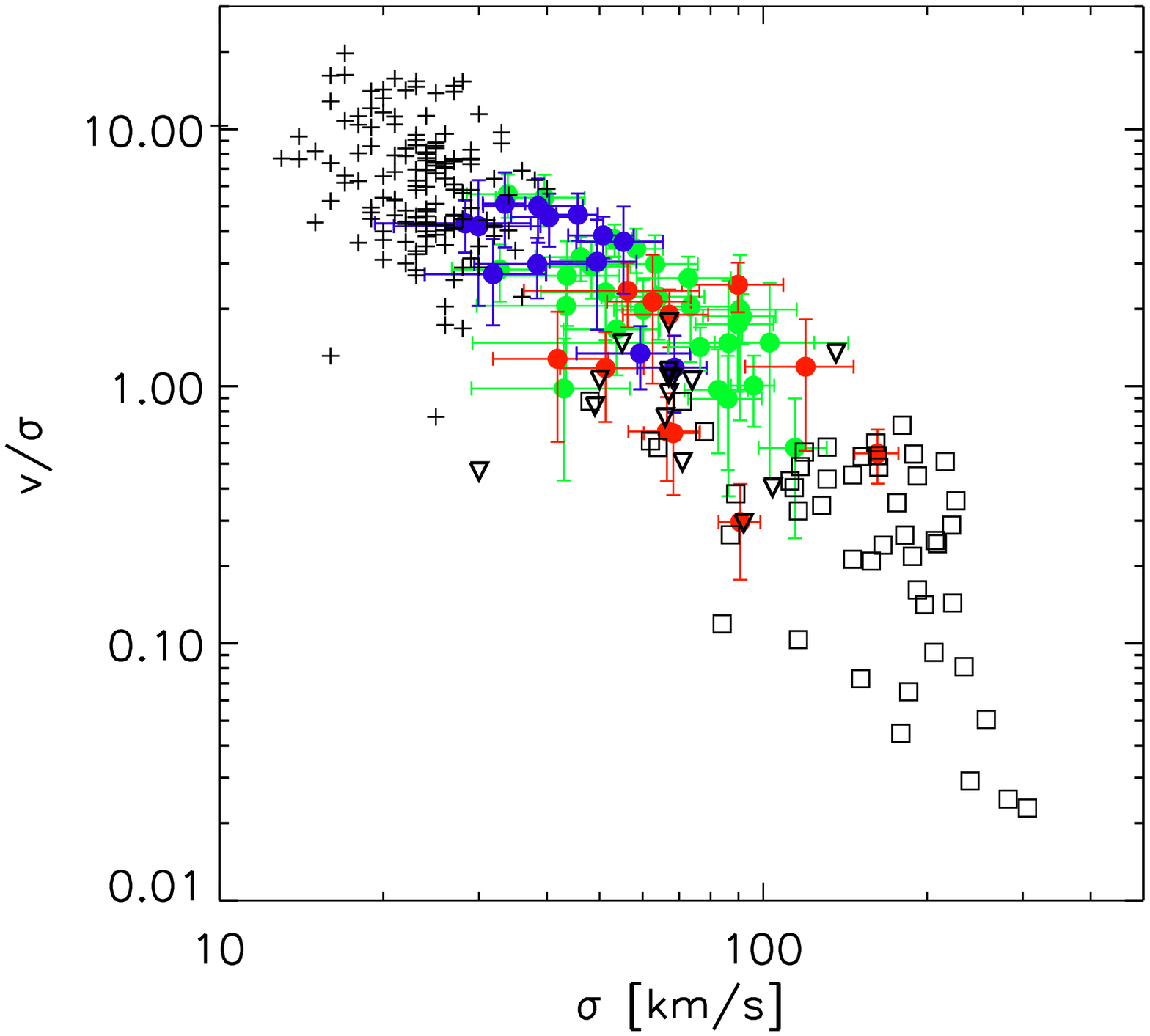}
\caption{\small Relationship between the observed dynamical ratio v/$\sigma$, taking here the v$_{shear}$/$\sigma_{mean}$ (see text) and the mean velocity dispersion. Color code is the same as the one adopted in previous figures. The plus signs represent spiral GHASP galaxies (i.e., \citealt{Epi10}), empty squares E/SO objects (i.e., \citealt{Cap07}), and top-down empty triangles LBAs (i.e., \citealt{Gon10}). For the spirals the $v$ parameter is defined as the maximum amplitude of the rotational curve within the extent of the velocity field along the major axis, and $\sigma$ is the average of the velocity dispersion map. 
For E/S0, v and $\sigma$ are luminosity-weighted square quantities derived, respectively, from the velocity field and velocity dispersion maps (see details in \citealt{Cap07}). For the LBAs, the velocity shear v$_{shear}$ has been defined as in this work, while the velocity dispersion is the flux-weighted mean value.}
\label{epinat_1}
\end{figure}

In figure \ref{epinat_1} the dynamical ratio v/$\sigma$ with respect to the mean velocity dispersion is presented. A clear trend is shown among the different morphological classes with $disk$ (class 0) galaxies having higher v/$\sigma$ than objects in a more advanced interaction/merging phase (classes 1, 2; see Table \ref{kin_prop_gen})\footnote{Although there are two class 0 objects with relatively low v/$\sigma$ (high $\sigma$), we note that one (IRAS F13229-2934, v/$\sigma$ $\sim$ 1.4) is known to be affected by the presence of an AGN (\citealt{A12}).}.
 
Kinematical data of samples of nearby galaxies with available IFS are also included for comparison: elliptical/lenticular galaxies drawn from the SAURON project, studying the stellar kinematics (e.g., \citealt{Cap07}), local spiral galaxies selected from the Gassendi H$\alpha$ survey of SPirals (GHASP, \citealt{Epi10}), and LBAs at redshift $\sim$ 0.2 observed in H$\alpha$ with OSIRIS/Keck (i.e., \citealt{Gon10})\footnote{While the LBAs have star formation rates (SFRs) similar to our local LIRGs, the spiral and E/S0 samples have considerably smaller rates (see Table \ref{kin_prop_lowz}).}. We note that these works use slightly different definitions for $v$ and $\sigma$, but these have no significant impact on the interpretation of the v/$\sigma$ over $\sigma$ relationship (see caption of Fig. \ref{epinat_1}).

The (U)LIRGs fill the gap between rotation-dominated spirals and dispersion-dominated ellipticals in the v/$\sigma$-$\sigma$ plane. In particular, there is a clear transition among (U)LIRGs, with isolated (class 0), interacting (class 1), and merging (class 2) sources tending to increase their mean velocity dispersion while decreasing the velocity amplitude, therefore moving from high-$\sigma$ and low-v/$\sigma$ spirals to low-$\sigma$ and high-v/$\sigma$ ellipticals. Moreover, systems classified as class 0 (LIRGs in all cases) have mean velocities v$_{shear}$ = (150 $\pm$ 14) km s$^{-1}$ similar to those of spirals (mean velocity amplitude of $\sim$ 160 km s$^{-1}$) but with much higher velocity dispersion $\sigma_{mean}$ (i.e., 44 $\pm$ 4 km s$^{-1}$, while 24 km s$^{-1}$ for spirals, see Table 3). This suggests that class~0 LIRGs are rotating disks, more turbulent and thicker than those of normal spirals. Therefore, the main difference between the global kinematic properties of these two samples, rather than the rotational amplitude, is the mean velocity dispersion, which is higher for LIRG disks. On the other hand, interacting (class 1) and merging systems (class 2) have lower dynamical ratios, with kinematical values closer to those obtained for the ellipticals. The trend observed in this plot is consistent with the idea that the process of merging in (U)LIRGs possibly transforms spiral galaxies into intermediate-mass ellipticals (e.g., \citealt{genzel01}; \citealt{T02}; \citealt{dasyra06}). These class 1 and 2 systems even share similar kinematical properties with the LBAs (see Tables 2 and 3), discussed in detail in Sect. \ref{LBAcomp}.

When the number of resolution elements that characterizes a galaxy is small (i.e., galaxy size $\sim$ PSF), the determination of the $\sigma_{mean}$ values may be affected by the smearing of the velocity field and may lead to artificial peaks in the $\sigma$-map (see \citealt{davies11}). Due to the linear resolution limitations, this effect becomes common at higher redshift, as discussed in previous works (e.g., \citealt{FS09, Epi10}). However, for local samples, which are observed with much higher linear resolution, this effect is generally neglected. Indeed, this correction was not applied for the local comparison samples used here (i.e., spiral and elliptical/lenticular galaxies). Our LIRGs have spatial scales somewhat larger than those characterizing those systems\footnote{The typical linear resolutions (seeing FWHM $\sim$ 1$^{\prime\prime}$) of GHASP spirals and E/S0 galaxies are, respectively, 0.27 kpc and 0.09 kpc, while for our LIRGs and ULIRGs they are 0.48 kpc  and 1.32 kpc, respectively.}, but the smearing effects are still in general very small\footnote{To estimate the smearing effects in our data, we have calculated the velocity gradients along the central regions of our systems and derived the typical range in velocities across a spaxel ($\delta$v). When quadratically subtracting $\delta$v from $\sigma$, it was confirmed that the effects of smearing are negligible (a few percent) for most of the systems. In the worst case (i.e., IRAS F06259-4780 S) the effect was $\sim$ 15\% (see Tab \ref{NARROW}).} and corrections for the individual galaxies have not been applied. 

On the other hand, there could be some bias when directly comparing our results with those of LBAs at z $\sim$ 0.2 (i.e., \citealt{Gon10}). They actually compute the mean velocity dispersion using a flux-weighted quantity. As stated in \cite{davies11}, a flux-weighted mean velocity dispersion is biased towards bright regions, typically closer to the center, where the intrinsic rotation curve has a steeper gradient: the smearing is potentially more important and the mean velocity dispersion derived in \cite{Gon10} might be therefore considered as an upper limit\footnote{The values of the flux-weighted mean velocity dispersion can be a factor of 1.5 - 3 larger than those derived for the uniformly weighted mean velocity dispersion (see \citealt{davies11}).}.

  
\begin{table*}
\centering\tabcolsep=1mm
\vspace{1cm}
\caption{Mean (and median) kinematic properties for the different (U)LIRG samples.} 
\label{kin_prop_gen}
\begin{scriptsize}
    \begin{tabular}{l ccccccccc}   
\hline\hline\noalign{\smallskip}
Sample          		&          $<log L_{IR}>$     &     SFR			& 		v$_{shear}$          	&       v$_{shear}^*$   	&   $\sigma_{mean}$  	& v$_{shear}^*/ \sigma_{mean}$    & R$_{eff}^{IR}$	& M$_{dyn}^{IR}$   & 	K class* \\ 
{\smallskip}
          			&     	(L$_\odot$)  		&    (M$_\odot$ yr$^{-1}$)	&	(km s$^{-1}$)  		& (km s$^{-1}$)    		& (km s$^{-1}$)     		&    & (kpc)	& (10$^{10}$ M$_\odot$)      		&	(RD; PD; CK)/T \\
				(1) & (2) &(3) &(4) &(5) &(6) & (7)  & (8) & (9) & (10)\\
\hline\noalign{\smallskip}                                             
         ALL     		& 11.4 $\pm$ 0.4 &   88 $\pm$ 15 (45)		&  130 $\pm$ 7 (128)    &    171 $\pm$ 9 (177)  & 	63 $\pm$ 4  (58)    	&     3.2 $\pm$ 0.3 (2.6)      & 2.5 $\pm$ 0.2 (2.4)	&  4.9 $\pm$ 0.6 (3.5) 	 	&	(14; 23; 12)/49       \\                        
\hline\noalign{\smallskip}
           LIRG 		&  11.3 $\pm$ 0.3 &  44 $\pm$ 6 (34)		& 130 $\pm$ 8 (128)     &  170 $\pm$ 9 (167)  &	59 $\pm$ 3 (54)  	&     3.4 $\pm$ 0.5 (3.3)      & 2.5 $\pm$ 0.2 (2.4)	&  4.5 $\pm$ 0.5 (3.5) 		&	(14; 22; 6)/42           \\                        
         ULIRG		&  12.2 $\pm$ 0.1 &  278 $\pm$ 20 (272)   	&  130 $\pm$ 26 (150)  &   178 $\pm$ 24 (190) &	95 $\pm$ 12 (90)      &     2.0 $\pm$ 0.4 (1.9)        & 2.8 $\pm$ 0.6 (2.4)	& 7.1 $\pm$ 2.6 (4.8)		&	(0; 1; 6)/7           \\
\hline\noalign{\smallskip}    
(U)LIRG class 0    	&  11.3 $\pm$ 0.3 &   48 $\pm$ 18 (32) 		&  150 $\pm$ 14 (171) &      195 $\pm$ 17 (182)  &  	44 $\pm$ 4 (46)       	&   4.7 $\pm$ 0.5 (4.3) 	&  2.7 $\pm$ 0.3 (3.0)	&  5.1 $\pm$ 1.0 (4.9)		&	(6; 6; 1)/13         \\
(U)LIRG class 1    	&  11.4 $\pm$ 0.5  &  91 $\pm$ 25 (41)  	&  135 $\pm$ 7 (141) &      174 $\pm$ 8 (190)  &		67 $\pm$ 3 (63)       	&   3.0 $\pm$ 0.3 (2.5) 	& 2.6 $\pm$ 0.2 (2.4)	&  4.8 $\pm$ 0.5 (3.5)		&	(8; 12; 5)/25       \\
(U)LIRG class 2    	&  11.7 $\pm$ 0.3 &  129 $\pm$ 32 (86)		&  98 $\pm$ 18 (89)     &   134 $\pm$ 21 (157)	 &	80 $\pm$ 11 (67)      	&  1.8 $\pm$ 0.3 (1.5)	& 2.6 $\pm$ 0.5 (2.0)	&  4.7 $\pm$ 1.9 (1.7)  		&	(0; 5; 6)/11         \\
\hline\noalign{\smallskip}   	                  
(U)LIRG RD		& 11.4 $\pm$ 0.3 &  65 $\pm$ 25 (35)		&   159 $\pm$ 12 (180)  &      207 $\pm$ 16 (207) &	50 $\pm$ 5 (46)       	&     4.7 $\pm$ 0.5 (5.4)      &  2.7 $\pm$ 0.4 (2.5)	&  5.8 $\pm$ 1.0 (5.1) 	  	& 	(14; 0; 0)/49        \\                        
(U)LIRG PD       	& 11.3 $\pm$ 0.4 &  56 $\pm$ 15 (36)		&   123 $\pm$ 10 (119) &     160 $\pm$ 13 (163)   &	58 $\pm$ 4 (53)      	&   3.1 $\pm$ 0.3 (3.1)      &  2.4 $\pm$ 0.2 (2.3)	&  4.1 $\pm$ 0.6 (3.5) 	  	& 	(0; 23; 0)/49      \\
(U)LIRG CK       	& 11.7 $\pm$ 0.5&   161$\pm$ 36 (202)		&   112 $\pm$ 15 (127) &     155 $\pm$ 17 (157)  &	91 $\pm$ 8 (90)      	&   1.7 $\pm$ 0.2 (1.7)      &  2.7 $\pm$ 0.5 (2.3)	&  5.9 $\pm$ 1.8 (3.4)		& 	(0; 0; 12)/49     \\
\hline\hline\noalign{\smallskip}                      
\end{tabular} 
\vskip0.2cm\hskip0.0cm
\end{scriptsize}
\begin{minipage}[h]{18cm}
\tablefoot{
Col (1): Sample: class 0 identifies isolated disks, class 1 denotes pre-coalescence interacting systems,  and class 2 stands for mergers (see Sect. \ref{morph_class}). RD stands for rotation dominated, PD perturbed disks and CK complex kinematic (see Sect. \ref{kine_class}). 
Col (2): Infrared luminosity.
Col (3): Star formation rate derived from the infrared luminosity listed in Col (2) applying the \cite{Ken98} relation.
Col (4): Observed velocity shear as defined in the text. 
Col (5): Intrinsic velocity shear (i.e., inclination corrected). 
Col (6): Mean velocity dispersion. 
Col (7): Dynamical ratio between the intrinsic velocity shear and the mean velocity dispersion. 
Col (8): Infrared effective radius, derived as explained in Appendix B.
Col (9): Dynamical mass derived as explained in the text. 
Col (10): Number of galaxies classified as RD, PD, CK with respect to the total number of galaxies in the sub-sample. \\
{\tiny(*)} The total number of galaxies in the present sample is 51, but for the galaxy IRAS F08424-3130 N and IRAS F12596-1529 the kinematical classification was not possible as explained in the text.  So, the number of galaxies with kinematical classification is 49.} 
\end{minipage}
\end{table*}

\begin{table*}
\centering
\vspace{1cm}
\caption{Mean (and median) observed kinematic properties for low-z comparison samples from the literature.}
\label{kin_prop_lowz}
\begin{scriptsize}
    \begin{tabular}{l cccccc } 
\hline\hline\noalign{\smallskip}
Sample          	&         SFR 						& 	          v            	&        $\sigma$ 		& 	v/ $\sigma$         & 	 R$_{eff}$	&	M$_{dyn}$   \\ 
{\smallskip}
         			&      [M$_\odot$ yr$^{-1}$]	&	(km s$^{-1}$)  	& (km s$^{-1}$)     	&           				& 		(kpc)		&   (10$^{10}$ M$_\odot$) 	 \\
			(1) 	&		 (2) 							&		(3) 			&		(4) 			&		(5) 			&		(6)    		& 	(7)			\\
\hline\noalign{\smallskip}                                             	
         Spiral       	&    	 1.8 $\pm$ 0.4 (0.7)  	 	& 162 $\pm$ 7 (143)	&	24 $\pm$ 0.5 (24)	& 7.0 $\pm$ 0.3 (6.3)		& 5.5 $\pm$ 0.3 (4.8)	&11.5 $\pm$ 1.5 (4.9)		\\                        
\hline\noalign{\smallskip}      
        E/S0   		&     0.12 $\pm$ 0.04 (0.08)  	& 47 $\pm$ 4 (44)	&	162 $\pm$ 9 (163)	& 0.34 $\pm$ 0.03 (0.34)	& 2.9 $\pm$ 0.3 (2.4)	&13.8 $\pm$ 3.2 (6.9)		\\                        
\hline\noalign{\smallskip}                                             
         LBA       	&      26.7 $\pm$ 7.1 (17.0)	& 67 $\pm$ 11 (63)	&	71 $\pm$ 6 (67)		& 0.95 $\pm$ 0.11(1.1)	& 1.4 $\pm$ 0.1 (1.5)	&1.0 $\pm$ 0.2 (1.0)		\\                                  
  \hline\hline\noalign{\smallskip}                      
\end{tabular} 
\vskip0.2cm\hskip0.0cm
\end{scriptsize}
\begin{minipage}[h]{18cm}
\tablefoot{Col (1): Low-z sample: Spiral sample is drawn from \cite{Epi10}, E/S0 from \cite{Cap07} and LBAs from \cite{Gon10}. 
Col (2): Star formation rate (SFR). 
For spirals, it has been derived using only 50 objects, for which the SFRs have been computed from H$\alpha$ measurements by \cite{james04} using Salpeter IMF (\citealt{Sal55}) and applying the \cite{Ken94} relation. 
For E/S0 galaxies we considered the SFR computed in \cite{shapiro10} for a subsample of 13 sources using the Spitzer/IRAC data at 8.0 $\mu$m.
They have calibrated this relation from the \cite{Yun01} relation between SFR and radio continuum. If the \cite{Ken98} SFR-H$\alpha$ conversion is applied to calibrate the 8.0 $\mu$m SFR estimator, the SFR would lower by $\sim$ 13\% (see \citealt{Wu05}). 
 For the LBAs, the star formation rates are measured by \cite{Gon10} from combined H$\alpha$ and MIPS 24 $\mu$m data using \cite{Kroupa08} IMF. The resulting SFRs are lower by a factor of $\sim$ 1.5 compared to a \cite{Sal55} IMF.
Col (3): Observed velocity amplitude obtained as follows: for spirals, it is the observed maximum rotational velocity; for the E/S0 sample it is the luminosity-weighted squared velocity (i.e., v = $\sqrt{<v^2>}$) within 1 R$_{eff}$; for LBAs the velocity amplitude has been defined as the v$_{shear}$ in this work.
Col (4): Velocity dispersion derived as follows: for spirals, it is the (uniformly weighted) mean velocity dispersion; for E/S0s it is the luminosity-weighted squared velocity dispersion(i.e., $\sigma$ = $\sqrt{<\sigma^2>}$) within 1 R$_{eff}$; for LBAs the velocity dispersion is the average velocity dispersion of each spaxel, weighted by flux.
Col (5): Dynamical ratio determined as the ratio between the values shown in column (3) and (4).
Col (6): Effective radius for the different samples. For spirals it has been derived as half of the optical radius (R$_{opt}$) associated with the isophotal level at 25 mag arcsec$^{-2}$ in the B-band, where R$_{opt}$ = 1.9 $\times$ R$_{eff}$ (see \citealt{Epi09}); for E/S0 the half-light radius is measured in the I-band from WFPC2/HST images; for LBAs the effective radius was derived by selecting N spaxels with S/N $>$ 6 such that R$_{eff}$ = $\sqrt{(N/\pi)}$ (see \citealt{Law07a}). 
Col (7): Dynamical mass determined as follows: for spirals we compute the total mass using the formula employed in their paper for rotation-dominated objects; we compute the dynamical masses for E/S0 using the formula $K \sigma^2$ R$_{eff}$ /G as explained in \cite{Cap06}; finally, the dynamical masses of LBAs have been derived in \cite{Gon10} using the formula 5 $\sigma^2$ R$_{eff}$/G.} 
\end{minipage}
\end{table*}

\subsection{(U)LIRGs dynamical masses}

\subsubsection{Derivation of dynamical mass} 

Dynamical mass determinations, which include the stellar, gas, and dark matter contributions, are relevant in the context of (U)LIRGs for several reasons. First, these objects have large amounts of gas and, therefore, the stellar mass gives only a partial estimate to their total mass (i.e., \citealt{Dow98, Com11}). Second, in order to infer the stellar masses, the SED modeling is difficult due to the presence of several stellar populations (\citealt{RZ10}), large amounts of dust (e.g., \citealt{cunha10}), and the complex structure of the extinction (\citealt{GM09, Piq13}). Therefore, stellar mass measurements not only have their own uncertainties, but intrinsically represent a lower limit to the mass of these systems. 

\citealt{Co05} have found that the warm gas kinematics in ULIRGs is in general coupled to that of the stellar component; therefore, it can be used as tracer of the gravitational potential. The kinematic properties of the ionized gas obtained from our IFS velocity maps are used in the following to estimate the dynamical mass, M$_{dyn}$.

As shown in previous sections, our sample contains systems with a wide range of kinematic properties, from regular rotating disk to merging systems. Hence, for deriving dynamical mass estimates we have used different approximations, distinguishing systems dominated by random motions from those with a large rotation component.      
   
For spherical, bound, and dynamically relaxed systems with random motion, the dynamical mass can be estimated via the virial theorem as
 
\begin{equation}
\hspace{1cm} M_{dyn, vir} = \frac{K  \hspace{1mm}R_{hm}\hspace{1mm}\sigma^2\hspace{1mm}}{G} ,
\label{MASSA1}
\end{equation} 
\vspace{0.5cm}
  
where $R_{hm}$ is the half-mass radius, $\sigma$ is the mean velocity dispersion, the factor $K$ takes into account the effects of structure on dynamics, and G is the gravitational constant (i.e., 4.3$\times$10$^{-3}$ pc M$_\odot^{-1}$ (km s$^{-1})^{2}$). $R_{hm}$ cannot be obtained directly and is estimated by measuring the half-light radius, $R_{eff}$, which encloses half luminosity (see Sect. \ref{sec_radii}). \cite{Taylor10} have recently studied the differential effects of nonhomology in the dynamical mass determinations based on this formula. In particular, they analyzed the dependence of the $K$ factor on the galaxy structure, parametrized by the S\'{e}rsic index $n$, assuming a spherical mass distribution that is dynamically isotropic and non-rotating, and which, in projection, follows a S\'{e}rsic (1963, 1968) surface density profile. According to their expression, the $K$ factor ranges between 7.4 (n=1, exponential profile) and 4.7 (n=4, de Vaucouleurs) and gives an idea of the differential effects due to nonhomology. We have considered a mean value of $K$ = 6, which is also close to the one found by other authors\footnote{We note that the value of $K$ = 6 corresponds to m=1.4 in Eq. (1) given in \cite{Co05}, who used m=1.75, and therefore, the masses derived here are $\sim$ 20\% lower than those derived with their formula.} (e.g., \citealt{Cap06}).  Since this formula is only valid for nonrotating systems, we have applied it to galaxies with kinematics compatible with slow rotation (i.e., v$^*_{shear}$/$\sigma<$ 2, where the v$^*_{shear}$ corresponds to the v$_{shear}$ value corrected by inclination $i$) and complex kinematic (CK) similar to those found by \cite{Co05} in ULIRGs. 

For the rest of the systems showing a significant rotation component, we have derived the dynamical mass from the circular velocity, obtaining the mass enclosed within a given radius. In particular,  we consider the total M$_{dyn}$ = 2 M$_{1/2}$, where M$_{1/2}$ is the dynamical mass within $r_{1/2}$, the radius that contains the three-dimensional half-light volume\footnote{This radius is related to the effective radius by the relation $r_{1/2} \approx$ 1.33 $\times$ R$_{eff}$, for a wide range of radial profiles (e.g., \citealt{wolf10}).}. Thus, for an oblate spheroid (\citealt{Williams10}),

\begin{equation}
\hspace{1cm}M_{1/2} \approx \frac{v_c^2 \hspace{1mm} R_{eff} }{G}\hspace{1mm}, 
\label{MASSA2}
\end{equation} 
\vspace{0.5cm}

where v$_c$ is the circular velocity at r$_{1/2}$. To infer the circular velocity we have considered both the rotation and dispersion motions. In particular, we included the ``asymmetric drift'' term, which represents an extra component due to the dispersion of the gas around the disk of the galaxy, following the prescriptions by \cite{Epi09} and \cite{Williams10}. Assuming a disk scale length, h = $\frac{r_{1/2}}{\sqrt{2 ln2}}$ (\citealt{Epi09}): 

\begin{equation}
\hspace{1cm}M_{dyn} \approx  \frac{2 \hspace{1mm} R_{eff} \hspace{1mm} (v_{rot}^2+1.35\hspace{1mm} \sigma^2)}{G} \hspace{1mm}, 
\label{MASSA3}
\end{equation} 
\vspace{0.5cm}

where $v_{rot}$ and $\sigma$ are the rotation and dispersion velocities and are taken here as the inclination-corrected v$^*_{shear}$, and $\sigma_{mean}$, respectively.

Mass determination can be affected by large uncertainties. In addition to those related to the implicit assumptions of the above expressions, the observed parameters R$_{eff}$, $v_{shear}$, and $\sigma_{mean}$ have their own uncertainties. The use of different tracers (e.g., IR-continuum, H$\alpha$) may lead to discrepancies (see Sect. \ref{sec_radii} for further details). Rotation velocities for individual objects can be significantly affected by the inclination correction, which therefore has an important impact on the M$_{dyn}$ determination due to its quadratic dependence on v$^*_{shear}$ in Eq. (\ref{MASSA3}). For a relatively small number of objects (i.e., 8/51) we applied an average inclination correction following \citealt{Law09}. Those have a relatively large uncertainty in $i$, but we apply Eq. (\ref{MASSA3}) instead of Eq. (\ref{MASSA1}) only for four. As for $\sigma_{mean}$, the fact that we are considering the values corresponding to the narrow component (Sect. 3.1) minimizes possible effects due to radial motions like outflows, which are generally associated with the broad component. We therefore estimate that the present masses should be corrected on average within a factor of $\sim$ 3.

\subsubsection{Effective radii of (U)LIRGs} 
\label{sec_radii}

We derived the effective radii using near-IR continuum images (i.e., H- and K-bands), which trace the bulk of the galaxy stellar component (see Appendix \ref{App_rad}). This spectral range also has the advantage of reducing the effects of the reddening, which can be significant in (U)LIRGs (i.e., \citealt{GM09}, \citealt{Piq13}). Specifically, we based our determinations on the 2MASS and  HST imaging, which cover a wide FoV and have high angular resolution, respectively (see details in Appendix \ref{App_rad}). The mean R$_{eff}$ for the (U)LIRGs sample is (2.6 $\pm$ 1.3) kpc. For the ULIRG subsample, we obtain a mean R$_{eff}$ of (3.3 $\pm$ 0.5) kpc, which is in excellent agreement with the value of (3.48 $\pm$ 1.39) kpc obtained by \cite{V02} using ground-based K-band imaging for most of the 118 ULIRGs of the IRAS 1 Jy sample. 

\cite{V02} also obtained R-band-based effective radii, which are on average 38 percent larger than in the K-band (i.e., R$_{eff}^R$= 4.80 $\pm$ 1.37 kpc). This difference can be attributed to the difference in extinction between the two bands, since after inspecting the original images (i.e., \citealt{Kim02}) it is clear that the K-band images are considerably shallower than the R ones. Therefore, one should expect systematic differences lower than 40 percent when comparing visible and near IR continuum-based radii for these objects. 

\begin{figure*}
\includegraphics[width=1\textwidth]{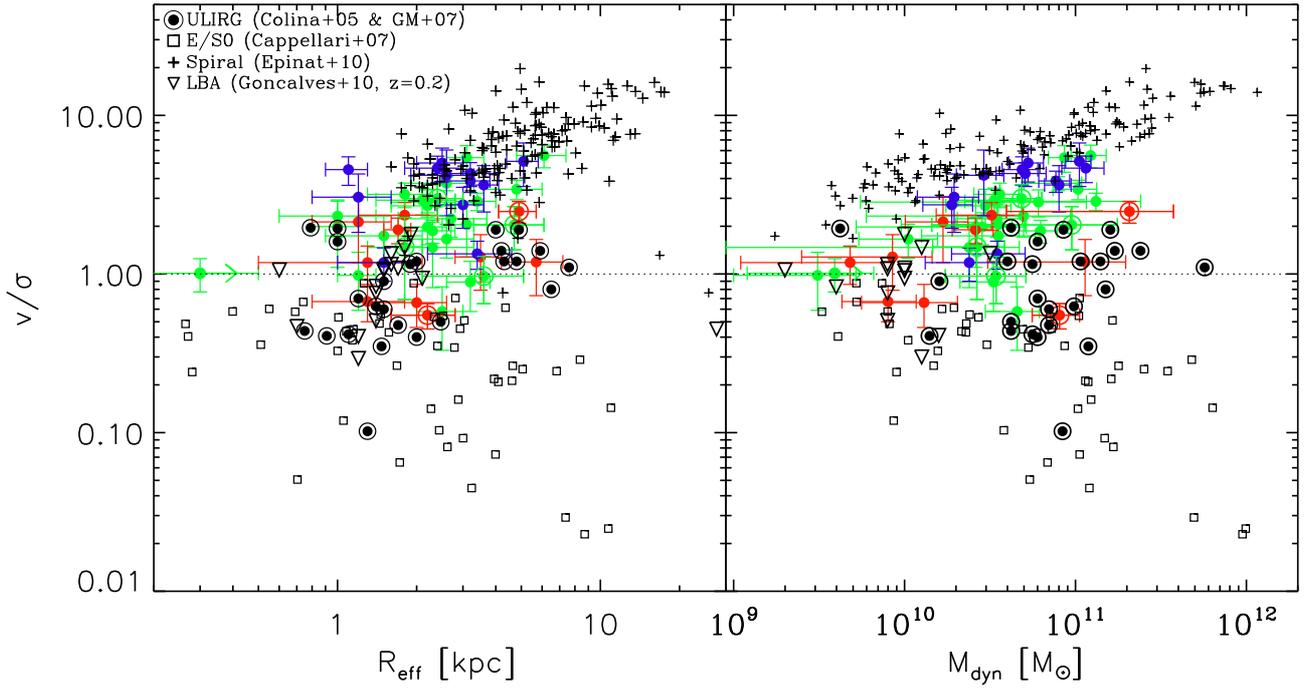}
\caption{\tiny Observed (i.e., not inclination corrected) v/$\sigma$ ratio as a function of the effective radius R$_{eff}$ (left panel) and dynamical mass M$_{dyn}$ (right panel) is shown (see text). The colors used are the same as in the previous figures (i.e., blue for isolated disks, green for interacting galaxies and red for mergers). We identify ULIRGs with a second circle around the dots.
The other samples are represented as follows: the nearby isolated spiral galaxies from the GHASP survey (see \citealt{Epi10}) are represented with cross symbols. Local ellipticals and lenticular objects, drawn from \cite{Cap07}, are shown with empty squares. LBAs, drawn from \citealt{Gon10}, are represented with empty downward triangles, with values derived using the Pa-$\alpha$ line. For these three samples, the effective radii R$_{eff}$, dynamical ratios v/$\sigma$, and dynamical masses M$_{dyn}$ have been derived as explained in the caption of table \ref{kin_prop_lowz}. We finally add the ULIRG systems studied in \cite{Co05} and those studied in \cite{GM07} (i.e., empty circle surrounding the black solid dot). In general, \cite{Co05} derived their effective radius from the H-band measured using F160W Near Infrared Camera and Multi-Object Spectrometer (NICMOS/HST) images (only for two sources were the F814W WFPC2/HST images used instead). In \cite{GM07} the effective radius considered is the one derived from the I-band (i.e., F814W) images. For both of the samples, their v/$\sigma$ ratio is obtained considering the half peak-to-peak velocity difference, and the central velocity dispersion $\sigma_c$, applying one Gaussian component fitting to the H$\alpha$ emission line data. Their dynamical masses are derived using a constant of $K = 7.5$ (see footnote 9 for further details) and the central velocity dispersion $\sigma_c$ (\citealt{Co05}).}
\label{low_z_fig}
\end{figure*}

The sizes of the (U)LIRGs are compared with those of other local samples in Figure \ref{low_z_fig} (left panel). On average, (U)LIRGs have smaller radii than local spirals, though the intrinsic scatter is rather large in both samples. In the present sample of (U)LIRGs, objects with radii as large as 7-20 kpc, such as those in the GHASP survey (i.e., derived in the B-band), are not seen. The E/SOs cover a range in size (measured using the H-band) larger than the (U)LIRGs, with a significant number of sources with R$_{eff}$ smaller than 0.5 kpc and larger than 7 kpc. The ULIRG sample observed with INTEGRAL/WHT (i.e., \citealt{Co05, GM09}) covers similar sizes as those of the present sample, with somewhat larger scatter. The derived LBA radii using Pa$\alpha$ emission are on average smaller by a factor of 0.7 than (U)LIRG radii. We note that H$\alpha$ sizes are also somewhat smaller than IR-continuum-based determinations (i.e., \citealt{A12}).

\subsubsection{(U)LIRGs dynamical masses, and mass ratios in interacting systems}

The dynamical mass estimates for the whole (U)LIRG sample range from $\sim$ 5 $\times$ 10$^9$ M$_\odot$ to 2 $\times$10$^{11}$ M$_\odot$, with a mean (median) value of (4.8 $\pm$ 0.6) $\times$ 10$^{10}$ M$_\odot$ (3.3 $\times$ 10$^{10}$ M$_\odot$), confirming that (U)LIRGs are intermediate mass systems like previously suggested (i.e., \citealt{T02, Co05}; though \citealt{Rothberg10} and \citealt{Rothberg13} find that (U)LIRGs masses are consistent with those of the most massive elliptical galaxies). In terms of m$_\star$ (i.e., m$_\star$ = 1.4 $\times$ 10$^{11}$ M$_\odot$; see \citealt{Cole01}), these values correspond to a mean value of (0.35 $\pm$ 0.04) m$_\star$ and median of 0.24 m$_\star$. The range of mass covered by the present (U)LIRGs overlaps with those of galaxies in the GHASP and SAURON, although these surveys also include galaxies with masses of $\sim$ 7 m$_\star$ (i.e., $\sim$ 10$^{12}$ M$_\odot$), which are not seen in our sample of (U)LIRGs.

The ULIRGs are more massive than LIRGs  by, on average, a factor of about 2 (see Table \ref{kin_prop_gen}). The present mean value for ULIRGs of (0.51 $\pm$ 0.19) m$_\star$ is between those obtained by \cite{Co05} (i.e., 0.4 m$_\star$) and  \cite{T02} (i.e., 0.86 m$_\star$). If the present sample of ULIRGs is combined with those of \cite{Co05} and \cite{GM07} a mean and median mass values of (0.5 $\pm$ 0.06) m$_\star$ and 0.3 m$_\star$ are, respectively, derived. These are also in good agreement with those found in \cite{dasyra06}, where ULIRGs show dynamical masses between sub- and $\sim$ m$_\star$.

The different morphological/kinematical classes do not show a clear trend with M$_{dyn}$ (Table \ref{kin_prop_gen}). The mean M$_{dyn}$ for classes 0, 1, and 2 define a weak decreasing sequence, with relatively low statistical significance. Under the assumption that class 2 objects are made by the merger of two class 0 (or individuals classified as 1), one should expect masses for the mergers that are larger (by a factor 2) than those of individual pre-coalescence galaxies, contrary to what is found. To analyze this topic further, we selected the individual galaxies of ten pre-coalescence multiple systems (i.e., 19 LIRGs), and derived a mean dynamical mass and the respective mass ratio between each merging galaxy pair\footnote{For triple systems we consider the ratio between the two more massive sources. The dynamical mass estimation for the northern galaxy of the system F07027-6011 was not possible due to the presence of an AGN.}. The mean dynamical mass derived for these sources is (0.33 $\pm$ 0.05) m$_\star$: it seems too high when compared with the mean mass for the post-coalescence mergers (i.e., M$_{dyn}$ = 0.34 $\pm$ 0.14 m$_\star$). However, we point out the limited statistical significance of this result, especially taking into account that two objects (i.e., IRAS F09437+0317 N and IRAS F14544-4255 E) deviate about 3 standard deviation from rest. If these are excluded, a typical mass of  (0.28 $\pm$ 0.04) m$_\star$ is obtained for pre-coalescence galaxies, a value that is still high if compared with that derived for class 2 objects. This can be explained as due to our L$_{IR}$ selection: a low-mass system may exceed a given infrared luminosity threshold only in the most intense starburst phase, while a higher mass system may pass this threshold even in a more quiescent phase. Since we select sources with L$_{IR} >$ 10$^{11}$ L$_\odot$, the fraction of large mass systems in the less active phases (e.g., type 0) is expected to be larger than in the most active ones (e.g., type 2).

A large fraction of the pre-coalescence systems (seven out of ten) have a mass ratio $\le$ 2.4, with a mean value of 2.0 $\pm$ 0.4, while the other three have ratios of 3, 4.3, and 20 (i.e., IRAS F06259-4780, IRAS F14544-4255, and IRAS F12043-3140\footnote{ Although the relatively high dynamical mass ratio derived for the system IRAS F12043-3140 would suggest that the northern source might be considered as a satellite-galaxy of the southern one, the relative stellar distribution (i.e., continuum map) suggests that the northern system is almost face-on and the mass ratio may be smaller.}, respectively). These results are in good agreement with the findings by \cite{Dasy_b_06}, who found that the majority of ULIRGs are triggered by almost equal-mass major mergers of 1.5:1 average ratio, which indicates that ULIRGs are mainly the products of almost equal mass mergers (i.e., sub-m$_\star$ galaxies; e.g., \citealt{T02, Co05, dasyra06})

\subsubsection{Low-z (U)LIRGs versus Lyman Break Analogs}
\label{LBAcomp}

While low-z (U)LIRGs represent extreme IR-luminous, dusty starbursts in our nearby universe, other type of galaxies, the UV-luminous (i.e., LBA, \citealt{heck05}), represent the class of (almost) extinction-free starbursts, showing SFRs similar to those of some local LIRGs (see Table \ref{kin_prop_lowz}). Both low-z IR and UV-bright galaxies are also considered local analogs of SFGs at cosmological distances, SMGs, and LBGs, respectively. So, it is important to establish a direct comparison of the overall kinematics and dynamical mass estimates of both samples based on IFS. 
Mean kinematic properties and dynamical masses for both samples are given in Tables 2 and 3, respectively. The corresponding values for local spirals and ellipticals based on IFS are also given for comparison.

On average, IR-luminous galaxies are much more massive than the UV-luminous galaxies with LIRGs and ULIRGs about five and ten times more massive than LBAs, respectively. Differences also clearly exist between (U)LIRGs and LBAs in their kinematics. The observed mean dynamical ratio (v/$\sigma$) for all (U)LIRGs is 2.5 larger than for LBAs, indicating that in general IR-luminous galaxies appear to be more rotation-supported than the UV-luminous galaxies. This is particularly true when only (U)LIRGs classified as isolated (class 0) are considered, while interacting (class 1) and merging (class 2) galaxies have, respectively, a ratio of 2.3 $\pm$ 0.2 and 1.3 $\pm$ 0.2, with the latter being much closer to that derived for LBAs (i.e., 0.95 $\pm$ 0.11). Detailed HST imaging of LBAs detected faint tidal features and companions around all LBAs imaged (i.e., \citealt{over08}), suggesting that UV-bright starbursts are the product of an interaction or merger. Looking independently at the kinematic tracers (i.e., velocity amplitude and velocity dispersion), there is a clear trend among (U)LIRGs such that the amplitude of the velocity field decreases from isolated to interacting and merging systems while the velocity dispersion increases. The LBAs have a mean velocity dispersion (i.e., 71 $\pm$ 6 km s$^{-1}$) much higher than that of isolated (U)LIRGs (i.e., 44 $\pm$ 4 km s$^{-1}$); however they are similar to that of interacting (i.e., 67 $\pm$ 3 km s$^{-1}$) and mergers (i.e., 80 $\pm$ 11 km s$^{-1}$). These velocity dispersions are factors of 3 to 4 larger than those of normal spirals (24 $\pm$ 0.5 km s$^{-1}$, see \citealt{Epi10}), and indicate that the starburst and the interaction/merging process play a dominant role in establishing the kinematic properties of the gas, likely due to combined effect of stellar winds and tidal forces.

So, independent of the overall mass of the system and of whether it is dusty (i.e., class 1 and 2 (U)LIRG) or not (i.e., LBA), the kinematics of low-z starburst galaxies is dominated by the star formation and the tidal forces, and produces dynamically hot systems characterized by an average $\sigma\sim$ 70-80 km s$^{-1}$ and v/$\sigma$ $\sim$ 1-2.

\section {Summary and conclusions}

Using the VIMOS IFS on the VLT, we have obtained spatially resolved kinematics (i.e., velocity fields and velocity dispersion maps) of 38 local (z $<$ 0.1) LIRGS and ULIRGs (51 individual sources). The sample includes sources with different morphological types (i.e., spirals, interacting systems, merger remnants) and is therefore well suited to study the kinematic properties of the (U)LIRG population showing different dynamical phases. A large fraction of the sample galaxies (31/38) covers the relatively less studied LIRG luminosity range, which fills the gap between the ULIRGs and the population of star-forming galaxies at large. This sample has been supplemented by a sample of low-z ULIRGs with their own available IFS (i.e., INTEGRAL/WHT data). 

The main results of the present study can be summarized as follows:

\begin{enumerate}
\item The H$\alpha$ emission line profile is well fitted with one or two Gaussians per emission line in most of the spectra, which have allowed us to identify two kinematically distinct components in the systems. One of these components (i.e., `systemic' or `narrow' component) is found over the whole line-emitting region of the system is characterized by spatially resolved mean velocity dispersions of $\sigma_{mean}\approx$ 30-160 km s$^{-1}$. The second component (`broad' component) is found in the inner (spatially resolved) regions of most of LIRGs and ULIRGs. It is characterized by relatively large line widths $\sigma_{mean}$ $\sim$ 320 km s$^{-1}$, and its central velocity is blueshifted with respect to the systemic component for most of the objects with values up to $\sim$ 400 km s$^{-1}$. The largest extensions and extreme kinematic properties of the broad component are observed in interacting and merging systems. This component likely traces nuclear outflows.
\item The systemic component traces the overall velocity field and shows a large variety of kinematic 2D structures, from very regular velocity patterns typical of pure RD (29\%) to kinematically PD (47\%), and highly disrupted and CK (24\%). Thus, most of the objects (76\%) are dominated by rotation. This fraction is larger than previously found in samples of ULIRGs. In fact, we found that the importance of rotation anti-correlates with the infrared luminosity, with a higher fraction of objects with complex kinematics among ULIRGs than in LIRGs (respectively, 6/7 vs. 6/42).
\item We find a clear correlation between the different phases of the merging process and the mean kinematic properties inferred from the velocity maps. In particular, isolated disks, interacting galaxies, and merging systems define a sequence of increasing mean velocity dispersion and decreasing velocity field amplitude, which is characterized by intrinsic average dynamical ratios (v*/$\sigma$) of 4.7, 3.0, and 1.8, respectively. We also find that the  $\sigma_c$/$\sigma_{mean}$ vs. $\sigma_{mean}$  plane is an excellent discriminator between disks and interacting/merging systems, with disks showing higher $\sigma_c/\sigma_{mean}$ and lower $\sigma_{mean}$ values by factors of 1.5 to 2, on average.
\item In terms of dynamical support (v/$\sigma$), the sample covers the gap between local spirals and E/SOs. The LIRGs classified as isolated disks partially overlap with the values of local spirals, with similar velocity amplitudes but larger mean velocity dispersions $\sigma_{mean}$ (44 vs. 24 km s$^{-1}$), implying a larger turbulence and thicker disks. Interacting systems and mergers have values closer to those of low-velocity dispersion E/SOs, though with higher velocity amplitudes.
\item The dynamical mass estimates for the present (U)LIRG sample range from $\sim$ 0.04 to 1.4 m$_\star$, with ULIRGs ($\sim$ 0.5 $\pm$ 0.2 m$_\star$) more massive than LIRGs by, on average, a factor of about 2. The mass ratio of individual pre-coalescence galaxies are $<$ 2.5 for most of the systems, confirming that most (U)LIRG mergers involve sub-m$_\star$ galaxies of similar mass.  
\item The subclass of (U)LIRGs classified as mergers and the LBAs share similar kinematic properties, although the dynamical mass of LBAs is a factor 5 smaller, on average. This is a clear indication that, independent of the mass of the system and of whether it is dusty or dust-free, starburst winds and tidally induced forces produce dynamically hot systems characterized by $\sigma\sim$ 70-80 km s$^{-1}$ and v/$\sigma$ $\sim$ 1-2.
\end{enumerate}
 

\begin{table*}
 \centering
\caption{Kinematic properties of the narrow component for the (U)LIRGs sample.}
\label{NARROW}
\begin{scriptsize}
   \begin{tabular}{c cccccccccccc}
\hline\hline\noalign{\smallskip}
 Galaxy ID             &    v$_{sys}$     & v$_{amp}$ & v$_{shear}$  &  $\sigma_{mean}$  &  $\sigma_ c$  &  v$_{shear}^*$                      &     i   &   $\delta$v   &K class     & M$_{dyn}$    &  {}Com\\
{\smallskip}
(IRAS code)          & (km s$^{-1}$)  &(km s$^{-1}$) &(km s$^{-1}$)     & (km s$^{-1}$)  & (km s$^{-1}$) & $\overline{\sigma_{mean}}$          &      degree  &  (km s$^{-1}$ spx$^{-1}$)    &        & (10$^{10}$ M$_{\odot}$)      &       \\
  (1) & (2) & (3) & (4) & (5) & (6) & (7) & (8) &(9)& (10) &(11) &(12) \\
\hline\noalign{\smallskip}
        F06295-1735          & 6373             & 107 $\pm$ 15        &   87 $\pm$ 8      &      32 $\pm$ 9  	& 44 $\pm$ 7          &   3.5 $\pm$ 1.1          &  52 $\pm$ 7       &  4.6      	& PD      &   1.9 $\pm$ 0.6      	& vii        \\
        F06592-6313          & 6942             & 160 $\pm$ 11        &   151$\pm$ 9     &      49 $\pm$ 20  	&  96 $\pm$ 20       &   3.6 $\pm$ 1.5         &  58 $\pm$ 8		&  12.7   	& PD      &   2.0 $\pm$ 0.8      	&  v  \\
        F07027-6011 S      & 9270             &  95 $\pm$ 34         &   81 $\pm$ 10    &      69 $\pm$ 14  	&  59 $\pm$ 5         &   2.4 $\pm$ 0.7         &  29 $\pm$ 4         & 12.7   		& RD      &   2.4 $\pm$ 1.1      	&  v \\
        F07027-6011 N      & 9582            &  124 $\pm$ 14        &   115 $\pm$ 9    &      38 $\pm$ 9  	& 63 $\pm$ 10        &   3.9 $\pm$ 1.0        	    &  50 $\pm$ 5         &  10.4    	& PD      &   -                   		&   iii, v   \\
        F07160-6215          & 3297             &  237 $\pm$ 28        &   202 $\pm$ 10 &       55 $\pm$ 18  	& 128 $\pm$ 17      &   3.8 $\pm$ 1.2         &  75 $\pm$ 3        &  6.2  		& PD (CK)&   8.0 $\pm$ 1.4        & iv,vii \\
        F10015-0614          & 5143             &  218 $\pm$ 29        &  196 $\pm$ 7     &      51 $\pm$ 7  	& 60 $\pm$ 10        &   4.3 $\pm$ 0.7         &  63 $\pm$ 3        &  5.3 		& PD      &   7.7 $\pm$ 1.6       &  iv,vii      \\
        F10409-4556          & 6241             &  214 $\pm$ 12        &   193 $\pm$ 8    &      39 $\pm$ 9 	& 90 $\pm$ 13        &   5.4 $\pm$ 1.3         &   68 $\pm$ 6      &   7.2        	& RD      &   5.3 $\pm$ 1.2       & iv,v   \\
        F10567-4310          & 5129             &  149 $\pm$ 11        &  122 $\pm$ 9     &    28 $\pm$ 4 		& 52 $\pm$ 8          &   6.4 $\pm$ 1.1         &  42 $\pm$ 2       &   5.9    		& RD      &   5.1 $\pm$ 1.7       & iv,v     \\
        F11255-4120          & 4958             &  149 $\pm$ 13        &   125 $\pm$ 9    &      30 $\pm$ 13  	& 82 $\pm$ 25        &   5.1 $\pm$ 2.3         &  56 $\pm$ 2       &   5.4    		& PD       &   2.9 $\pm$ 0.9       & v       \\
        F11506-3851          & 3114             &  202 $\pm$ 9         &  184 $\pm$ 6     &    40 $\pm$ 8  		& 81 $\pm$ 18        &   7.6 $\pm$ 1.6         &   37 $\pm$ 3      &   13.7  		& RD      &   4.9 $\pm$ 1.2        & v       \\
        F12115-4656          & 5282             &  221$\pm$ 29        &  212 $\pm$ 4     &    46 $\pm$ 8  		& 78 $\pm$ 20        &   6.9 $\pm$ 1.4         &  42 $\pm$ 3       &   12.8      	& RD      &   11.5 $\pm$ 3.2      & v     \\
        F13229-2934          & 4059             &  115 $\pm$ 24       &  80 $\pm$ 14     &    59 $\pm$ 6  		& 64 $\pm$ 10        &   2.2 $\pm$ 0.5         &  37 $\pm$ 3       &   5.6      	& CK      &   3.5 $\pm$ 1.6     	 & vii    \\
        F22132-3705          & 3500             &  186 $\pm$ 5         &  172 $\pm$ 3     &    33 $\pm$ 10           & 76 $\pm$ 15        &   6.2 $\pm$ 1.9      &   56 $\pm$ 4      &   7.7     		& RD      &   10.5 $\pm$ 1.8      &  iv,v  \\
        \hline\noalign{\smallskip}
      F01159-4443 S         & 6930       & 130 $\pm$ 17 &        117 $\pm$ 13    &       46 $\pm$ 11   &     71 $\pm$ 16        &       3.7 $\pm$ 1.1       	& 46 $\pm$ 8    & 13.6 		& PD    & 3.0 $\pm$ 1.2    &  ii,iv,vii  \\
      F01159-4443 N         & 6792       & 145 $\pm$ 25 &        119 $\pm$ 13    &       51 $\pm$ 12   & 	50 $\pm$ 8         &       6.3 $\pm$ 3.0      	& 22 $\pm$ 9    & 16.6 		& PD    & 5.0 $\pm$ 4.4   	&  ii,vii        \\
    F01341/ESO G12        & 5240      & 199 $\pm$ 59 &       156 $\pm$ 45      &       90 $\pm$ 15  &       61 $\pm$ 18      &       2.2 $\pm$ 0.8  	     	& 52 $\pm$ 10  & 6.4 			& PD    & 6.1 $\pm$ 5.5   	&  i,vii     \\
    F01341/ESO G11        & 5200      & 111 $\pm$ 11 &         93 $\pm$ 6        &       33 $\pm$ 6  	&       43 $\pm$ 9        &        7.3 $\pm$3.5       	& 23 $\pm$ 11  & 10.8 			& PD    & 3.5 $\pm$ 2.3   	&  vii  \\
        F06035-7102           & 23867     & 184 $\pm$ 27 &       150 $\pm$ 26     &       74 $\pm$ 19  &        63 $\pm$ 18     &       2.6 $\pm$ 1.4       	& 52 $\pm$ 26  & 12.9 		& CK    & 9.5 $\pm$ 8.1  	&   vi    \\
     F06076-2139 S          & 11810     & 118 $\pm$ 38 &        89 $\pm$ 18      &       43 $\pm$ 14   &       51 $\pm$ 11      &       3.4 $\pm$ 1.3       	& 37 $\pm$ 3    & 10.7		& PD    & 3.5 $\pm$ 1.5    &    ii,v \\
      F06076-2139 N         & 11199     & 83 $\pm$ 52 &         77 $\pm$ 24       &       86 $\pm$ 13   &     87 $\pm$ 11        &       1.2 $\pm$ 0.5       	& 46 $\pm$ 11  & 17.2 		& PD    & 3.3 $\pm$ 1.2  	&    i, ii, iv, v    \\
        F06206-6315           & 27669     & 203 $\pm$ 21 &      189 $\pm$ 18      &       64 $\pm$ 13   &       75 $\pm$ 23      &       3.1 $\pm$ 0.7      	& 76 $\pm$ 10  & 10.3  		& PD    & 4.8 $\pm$ 1.5 	&   v     \\
    F06259-4780 S           &11736      & 213 $\pm$ 19 &        198 $\pm$ 13    &       53 $\pm$ 4     &       37 $\pm$ 6        &       4.7 $\pm$ 0.7       	& 53 $\pm$ 9    & 22.5  		& RD   & 7.9 $\pm$ 2.9   	&   i,v    \\
    F06259-4780 C           & 11587     & 104 $\pm$ 22 &        89 $\pm$ 12      &       54 $\pm$ 11   &  	47 $\pm$ 13     &       1.9 $\pm$ 0.5        	& 62 $\pm$ 11  & 6.6  		& RD   & 1.1 $\pm$ 0.5	& ii, v    \\
    F06259-4780 N           & 11935     & 229 $\pm$ 41 &      180 $\pm$ 64      &       91 $\pm$  25  &       77 $\pm$ 19     &       2.3 $\pm$ 1.1  		& 58 $\pm$ 9    & 20.7  		& RD   & 2.6 $\pm$ 1.7   	&    ii, v    \\
     08424-3130 S            & 4914       & 133 $\pm$ 31 &      119 $\pm$ 23       &        60 $\pm$  7   &     83 $\pm$ 25       &        2.5 $\pm$ 1.2  		& 52 $\pm$ 26  & 11.0    		& PD   & 2.8 $\pm$ 2.3    	&   ii, iv,vi \\
       08424-3130 N          & 5029       & 67 $\pm$ 50 &        66 $\pm$ 27         &     114 $\pm$ 16   &    119 $\pm$ 17     &        0.7 $\pm$ 0.5  		& 52 $\pm$ 26  &  -      		&  -      & 4.6 $\pm$ 1.6   	&   ii,iv,vi,viii \\
   F08520-6850 E          & 13518     & 123 $\pm$ 10 &       109 $\pm$ 15     &       77 $\pm$ 10     &     100 $\pm$ 40      &       1.6 $\pm$ 0.3           	& 62 $\pm$ 6  	& 7.9   		& RD   	& 2.0 $\pm$ 0.7  &   ii, vii  \\
   F08520-6850 W         & 13736     & 196 $\pm$ 12 &       191 $\pm$ 16     &       73 $\pm$ 14     &     92 $\pm$ 20      &       2.7 $\pm$ 0.6           	& 77 $\pm$ 4  	& 12.6   		& PD (RD)   & 11.3 $\pm$ 2.8  &  ii, vii   \\
       F09437+0317 S       & 5966       & 236 $\pm$ 22 &        214 $\pm$ 9      &        40 $\pm$ 7     &        50 $\pm$ 9     &       6.1 $\pm$ 1.2  		& 63 $\pm$ 4    & 8.0   		& RD   & 8.6 $\pm$ 1.7      &    i,v     \\
       F09437+0317 N       & 5983       & 134 $\pm$ 11 &       189 $\pm$ 6       &        34 $\pm$ 5     &     56 $\pm$ 8        &       6.0 $\pm$ 1.0  		& 68 $\pm$ 5    & 4.1  		& RD   & 12.3 $\pm$ 2.8  	&    iv,v         \\
        12043-3140 S          & 7009       & 191 $\pm$ 35 &        171 $\pm$ 11    &     101 $\pm$ 15   &  82 $\pm$ 21        &       2.4 $\pm$ 0.4           	& 52 $\pm$ 2    & 12.7  		& PD    & 6.3 $\pm$ 1.6   	&   ii,v    \\
       12043-3140 N          & 7030       & 54 $\pm$ 10 &        42 $\pm$ 10         &    43 $\pm$ 14  	&     51 $\pm$ 8        &       1.2 $\pm$ 0.7  		& 52 $\pm$ 26  &  6.5  		& PD (CK) & 0.3 $\pm$ 0.2   	&   ii,vi    \\
        F12596-1529           & 4875       & 163 $\pm$ 22 &      143 $\pm$ 14      &    64 $\pm$ 14      & 66 $\pm$ 18      	&          2.3 $\pm$ 0.6        & 75 $\pm$ 3    &  5.9 		& -  		& 3.5 $\pm$ 1.4   	& iv,vii, viii \\
          F14544-4255 E     & 4714       & 220 $\pm$ 88 &      200 $\pm$ 15      &    60 $\pm$ 7         & 63 $\pm$ 6       	&      3.5 $\pm$ 0.5           		& 78 $\pm$ 4    &  7.6  		& PD    & 10.4 $\pm$ 3.0 	&  i,vii   \\
          F14544-4255 W    & 4800       & 146 $\pm$ 35 &      127 $\pm$ 11      &    86 $\pm$ 37  	&       82 $\pm$ 16    &      1.7 $\pm$ 0.8  		& 59 $\pm$ 7    & 8.7 	 		& CK (PD)   & 2.4 $\pm$ 2.3  	&  i,vii      \\
          F18093-5744 S     & 5328       & 155 $\pm$ 7 &        141 $\pm$ 10      &      48 $\pm$ 9  	&       62 $\pm$ 13    &      1.2 $\pm$ 0.3  		& 53 $\pm$ 6    & 10.3  		& RD    & 3.4 $\pm$ 1.3   	&   v      \\
          F18093-5744 C     & 4969       & 130 $\pm$ 25 &       96 $\pm$ 21      &   96 $\pm$ 9           & 77 $\pm$ 25    &      4.4 $\pm$ 0.8            		& 56 $\pm$ 4    & 7.3  		& CK (PD)    & $>$ 0.4 $\pm$ 0.3 &  i,vii  \\
          F18093-5744 N     & 5184       & 159 $\pm$ 9 &      147 $\pm$ 3          &    46 $\pm$ 8  	&       58 $\pm$ 7      &      3.7 $\pm$ 0.8  			& 47 $\pm$ 3    & 7.0   		& RD    & 3.6 $\pm$ 1.1     &   v    \\
       F22491-1808            & 23283     & 88 $\pm$ 44 &        80 $\pm$ 24        &    83 $\pm$ 11  	&       71 $\pm$ 19    &      1.4 $\pm$ 0.5  		& 44 $\pm$ 3    & 10.6  		& CK (PD)    & 3.4 $\pm$ 1.7    &  v      \\
       F23128-5919            & 13448     & 260 $\pm$ 62 &       152 $\pm$ 75    &    103 $\pm$ 21  	&     110 $\pm$ 23    &      1.9 $\pm$ 1.0  		& 53 $\pm$ 8    & 14.2     		&  CK    &2.7 $\pm$ 1.2	&  vii \\
      \hline\noalign{\smallskip}
        F04315-0840            & 4668         & 162 $\pm$ 36     &  134 $\pm$ 11    &      63 $\pm$ 28    &   70 $\pm$ 28    	&     2.5 $\pm$ 1.1      & 58 $\pm$ 4   &     5.6   		& CK     & 1.7 $\pm$ 0.7     &   vii   \\
        F05189-2524            & 12808       & 30 $\pm$ 19       &  27 $\pm$ 8         &      91 $\pm$ 10   &        90 $\pm$ 10  &     0.9 $\pm$ 0.5       & 18 $\pm$ 9  &     6.3   		& CK     & -                    	  &   iii,vii    \\
        08355-4944              & 7782         & 73 $\pm$ 14        &  60 $\pm$ 9        &      51 $\pm$ 12    &     44 $\pm$ 12    	&     1.5 $\pm$ 0.8      & 52 $\pm$ 26 &     5.0   		& PD     & 0.5 $\pm$ 0.4     &   vi   \\
        09022-3615              &17918        & 110 $\pm$ 36     &  89 $\pm$ 15      &     162 $\pm$ 12    & 105 $\pm$ 12      &    1.1 $\pm$ 0.2        &  29 $\pm$ 2  &    12.1   		& CK     & 8.1 $\pm$ 2.5     &   v     \\
        F10038-3338            &10287       & 54 $\pm$ 18        &  45 $\pm$ 8        &      68 $\pm$ 16     &    64 $\pm$ 16   	&     0.8 $\pm$ 0.4      & 52 $\pm$ 26 &     5.8   		& CK     & 1.3 $\pm$ 0.7     &   vi   \\
        F10257-4339            & 2813         & 170 $\pm$ 34     & 132 $\pm$ 20     &      56 $\pm$ 7       &         65 $\pm$ 7    &     3.3 $\pm$ 0.9      & 45 $\pm$ 10  &     6.5    		& PD     & 3.2 $\pm$ 1.7     &    iv,vii  \\
        12116-5615              & 8163         & 52 $\pm$ 19       &  44 $\pm$ 10      &     66 $\pm$ 9        &         51 $\pm$ 9    &     0.9 $\pm$ 0.2     & 48 $\pm$ 4   &     7.3  		& PD     & 0.8 $\pm$ 0.4     &    v   \\
        F13001-2339            & 6540         & 169 $\pm$ 110   &  142 $\pm$ 27    &    120 $\pm$ 25    &   116 $\pm$ 34     &     1.3 $\pm$ 0.5       & 65 $\pm$ 4    &   11.9   		& CK     & 11.4 $\pm$ 8.3   &    v     \\
        F17138-1017            & 5274         & 144 $\pm$ 21     &  128 $\pm$ 12    &    67 $\pm$ 11      &    59 $\pm$ 11       &     2.4 $\pm$ 0.5      & 52 $\pm$ 8    &   12.9    		& PD     & 2.6 $\pm$ 0.7     &   vii       \\
        F21130-4446            & 28028       & 228 $\pm$ 28     &  223 $\pm$ 19    &      90 $\pm$ 12    &        69 $\pm$ 12   &     3.2 $\pm$ 1.5      & 52 $\pm$ 26  &   22.1  	 	& CK     & 20.7 $\pm$ 16.8 &   vi     \\
        F21453-3511            & 4816         & 78 $\pm$ 29       &  53 $\pm$ 10      &      42 $\pm$14     &         70 $\pm$ 14   &     1.6 $\pm$ 0.6     & 51 $\pm$ 2    &    3.7    		& PD     & 0.9 $\pm$ 0.6      &  iv,vii   \\
\end{tabular}
\vskip0.2cm\hskip0.0cm
\end{scriptsize}
\begin{minipage}[h]{18cm}
\tablefoot{Derived H$\alpha$ kinematic quantities for the narrow (systemic) component. Horizontal lines distinguish class 0, 1, and 2 (from top to bottom, respectively).
Col (1): IRAS name.
Col (2): Systemic velocity calculated as the H$\alpha$ radial velocity. A mean error of 5-7 km s$^{-1}$ should be added when calibration and fitting errors are considered.
Col (3): H$\alpha$ velocity amplitude defined as the half of the observed `peak-to-peak' velocity (i.e., half the difference between the maximum and minimum values is considered without applying the inclination correction).
Col (4): H$\alpha$ velocity shear defined as the half of the difference between the median of the 5 percentile at each end of the velocity distribution (i.e., v$^{5\%}_{max}$ and v$^{5\%}_{min}$), as in \cite{Gon10} (not corrected for the inclination).
Col (5): H$\alpha$ mean velocity dispersion.
Col (6): H$\alpha$ central velocity dispersion, derived as the mean of 4 $\times$ 4 spaxels around the VIMOS continuum flux peak.
Col (7): Dynamical ratio between the velocity shear (corrected for the inclination) and the mean velocity dispersion values.
Col (8): Inclination of the galaxy, defined as $i$ = cos$^{-1}$(b/a) (i.e., where a and b are, respectively, the major and minor axes of the object). When possible, it has been inferred directly from the H$\alpha$ map or continuum images (i.e., DSS, HST and 2MASS). When the morphology of the object was too complex, we assumed a mean inclination of about 52$^{\circ}\pm$ 26$^\circ$, equivalent to $<sin\hspace{1mm} i>$ = 0.79 $\pm$ 0.35 (see appendix in \citealt{Law09}). 
In this case, the uncertainty of the inclination has been computed as the standard deviation of the distribution of the different measures derived when $i$ varies from 0$^\circ$ to 90$^\circ$, using step of 1$^\circ$. See details in column 12, notes (v, vi, vii).
Col (9): Velocity gradient (km s$^{-1}$) associated with one spaxel derived as explained in the text.
Col (10): Kinematical classification inferred from the kinematic maps. RD stands for rotating disk, PD perturbed disk and CK are systems with complex kinematics (see text for details). The galaxy IRAS 08424-3130 N is in the edge of the FoV, and its kinematic classification was not possible. For those objects for which the kinematical classification is controversial, the possible classifications are shown in the table.
Col (11): Dynamical masses (see text).
Col (12): Comments with the following code:
(i) After distributing the total IR luminosity between the individual objects forming the system, this galaxy does not qualify as LIRG as its luminosity is lower 10$^{11}$ L$_{\odot}$. The galaxies IRAS F01341-3735 N and IRAS F09437+0317 N are in the limit to be classified as LIRG (see column 6 in Table 1).
(ii) Systems for which a mask has been applied to the data-cube in order to analyze separately the sources (e.g., N and S stand for northern and southern galaxies).
(iii) No R$_{eff}^{IR}$ reliable due to AGN contamination (see \citealt{A12} and App. \ref{App_rad} for details).
(iv) For this galaxy two pointings have been combined to derive the kinematic values.
(v) The inclination has been derived using the H$\alpha$ map.
(vi) A mean inclination (i.e., 52$^{\circ}\pm$ 26$^\circ$) has been considered.
(vii) The inclination has been derived using continuum available images (i.e., HST or DSS or 2MASS). 
(viii) The kinematic classification was not possible due to the poor linear resolution for IRAS F12596-1529, while the emission of the IRAS F08424-3130 N is only partially included in the FoV.}
\end{minipage}
\end{table*}

\begin{table*}
\centering
\vspace{1cm}
\caption{Kinematic properties of the broad component for the (U)LIRGs sample.}
\label{BROAD}
\begin{scriptsize}
    \begin{tabular}{l ccccr } 
\hline\hline\noalign{\smallskip}
Galaxy ID          &   v$_{amp}$          &  $\sigma_{mean}$           & Area          & v$_{amp}/ \sigma_{mean}$         & $\Delta$v offset \\ 
{\smallskip}
{(IRAS code)}  &     (km s$^{-1}$)  & (km s$^{-1}$)     & (kpc$^2$)     &  & (km s$^{-1}$)\\
{(1)} & {(2)} &{(3)} &{(4)} &{(5)} &{(6)}\\
\hline\noalign{\smallskip}                                             
         F06295-1735              &    90 $\pm$ 35      &  110 $\pm$ 23     &     1.33      &      0.8 $\pm$ 0.4      &  -3\\                        
         F06592-6313              &   217 $\pm$ 133  &  219 $\pm$ 18     &     1.2      &      1.0 $\pm$ 0.6       & -285 \\
         F07027-6011S            &    59 $\pm$ 10     &  124 $\pm$ 8       &     6.0      &      0.5 $\pm$ 0.1      & -17 \\
         F07027-6011N            &    17 $\pm$ 5       &  316 $\pm$ 12     &     5.1      &      0.05 $\pm$ 0.02   &  -256 \\
         F07160-6215              &   152 $\pm$ 22    &   91 $\pm$ 12       &     0.3      &      1.7 $\pm$  0.3     &  55\\
         F10015-0614              &    66 $\pm$ 32     &  103 $\pm$ 12      &     1.1      &      0.6 $\pm$  0.3     &   -37\\
         F10409-4556              &   90 $\pm$ 97      &  82 $\pm$ 69        &      0.5     &      1.1 $\pm$  1.5     &  -120 \\
         F10567-4310              &    65 $\pm$ 47     &  138 $\pm$ 19      &     1.0      &      0.5 $\pm$  0.4     & -54 \\
         F11255-4120              &    52 $\pm$ 49     &  178 $\pm$ 14      &     0.5      &      0.3 $\pm$  0.3     &  -83\\
         F11506-3851              &    77 $\pm$ 30     &   76 $\pm$ 16       &     0.3      &      1.0 $\pm$  0.5     &  -42\\
         F12115-4656              &    21 $\pm$ 20     &   93 $\pm$ 12       &     1.0      &      0.2 $\pm$  0.1     &  69\\
         F13229-2934              &  143 $\pm$ 30     &  191 $\pm$ 40     &     2.5      &      0.7 $\pm$  0.2      &  -39 \\
         F22132-3705              &   -                 		& -                    		&      -           &     -             & - \\
         \hline\noalign{\smallskip}                                    
            F01159-4443 S           &    76 $\pm$ 10      &    39 $\pm$ 8          &     1.5          &    2.0 $\pm$ 0.5    &    -159  \\
            F01159-4443 N           &     84 $\pm$ 13     &    142 $\pm$ 58      &     3.5         &    0.6 $\pm$ 0.3     &  -37\\
          F01341/ESO G11          &     71 $\pm$ 20    &    110 $\pm$ 28      &     1.5          &    0.7 $\pm$ 0.3     &  -24\\
          F01341/ESO G12          &    142 $\pm$ 25   &    114 $\pm$ 10      &     1.1          &    1.3 $\pm$ 0.3     & -49\\
           F06035-7102               &     19 $\pm$ 13     &    139 $\pm$ 25      &    16.2         &    0.1 $\pm$ 0.1   & -28\\
           F06076-2139 S           &    38 $\pm$ 35      &    91 $\pm$ 44        &      2.5          &    0.4 $\pm$ 0.4    & -243  \\
            F06076-2139 N          &      7 $\pm$ 11      &    208 $\pm$ 8        &     1.7           &    0.03 $\pm$ 0.05 & -100\\
           F06206-6315              &    42 $\pm$ 38       &    198 $\pm$ 46     &      6.6          &    0.2 $\pm$ 0.2  &  -229   \\
           F06259-4780 S             &   -           &   -              &   -           &   -                  &   -   	\\
          F06259-4780 C     &     16 $\pm$ 15     &    111 $\pm$ 14      &     2.1           &    0.14 $\pm$ 0.14 	& 63 \\
         F06259-4780 N     &     87 $\pm$ 34    &   133 $\pm$ 24    &     3.5               &   0.7 $\pm$ 0.4    	& 76        \\
           08424-3130 S           &     22 $\pm$ 17     &    109 $\pm$ 12      &     0.5        &    0.7 $\pm$ 0.1    	&   31\\
           08424-3130 N           &     69 $\pm$ 64     &    128 $\pm$ 26      &     0.3        &    0.2 $\pm$ 0.6       	&-130\\
         F08520-6850 E             &      -             &   -                 &      -              &    -              &    -  			\\
         F08520-6850 W             &     52 $\pm$ 40    &    130 $\pm$ 10      &     3.7        &    0.5 $\pm$ 0.3      	&-151\\
         F09437+0317 S          &    -             &   -                 &      -              &    -              &    -  \\
           F09437+0317 N      &    77 $\pm$  53      &  106 $\pm$ 60      &     1.3        &    0.8 $\pm$ 0.6  		& -70 \\     
           12043-3140 S           &    203 $\pm$ 145     &    144 $\pm$ 26     &      4.5          &    1.4 $\pm$ 1.0 	&    -42  \\
           12043-3140 N            &    -                &   -                  &     -             &   -                  &    -   \\
         F12596-1529               &    145 $\pm$ 49      &    166 $\pm$ 14      &     0.9        &    0.9 $\pm$ 0.3 	& 0\\
            F14544-4255 E         &    189 $\pm$ 89      &    48 $\pm$ 62        &      0.6          &    3.9 $\pm$ 5.4  &    26  \\
            F14544-4255 W         &    175 $\pm$ 36    &    241 $\pm$ 46      &     1.5        &    0.7 $\pm$ 0.2  	& -52\\
            F18093-5744 C         &     47 $\pm$ 24     &    105 $\pm$ 22      &     1.5        &    0.5 $\pm$ 0.3 		&  -52 \\
            F18093-5744 N         &     34 $\pm$ 20     &     96 $\pm$ 15      &     0.8        &    0.4 $\pm$ 0.2  	& 39\\
            F18093-5744 S         &     68 $\pm$ 66     &    153 $\pm$ 34      &     0.6        &    0.4 $\pm$ 0.5  	& -85\\
           F22491-1808             &     22 $\pm$ 22      &    117 $\pm$ 15      &     9.7        &    0.2 $\pm$ 0.2  	&    -5\\
           F23128-5919             &    435 $\pm$ 63     &    274 $\pm$ 85     &    54.7        &    1.6 $\pm$ 0.5  	&  -32\\
       \hline\noalign{\smallskip}       
         F04315-0840             &   260 $\pm$ 54      &  211 $\pm$ 41   &    8.6             	&       1.2 $\pm$ 0.4 	&        -167\\
         F05189-2524             &    59 $\pm$ 40       &  269 $\pm$ 28   &   14.5             &       0.2 $\pm$ 0.2  	&    -400\\
         08355-4944                &    63 $\pm$ 14      &  127 $\pm$ 10   &    7.9             	&       0.5 $\pm$ 0.1 	&        -19\\
         09022-3615                &    35 $\pm$ 28      &  216 $\pm$ 17   &   11.9             &       0.16 $\pm$ 0.13 	&        23\\
         F10038-3338              &    82 $\pm$ 89      &  123 $\pm$ 44   &    2.5             &       0.7 $\pm$ 0.8 	&        -100 \\
         F10257-4339              &   144 $\pm$ 38     &  106 $\pm$ 17   &     5.2            &       1.4 $\pm$ 0.4 	&     -2\\
         12116-5615                &    50 $\pm$ 20       &  154 $\pm$ 25  &    2.4            	&       0.32 $\pm$ 0.14  	&     -55\\
         F13001-2339             &    137 $\pm$ 109  &  144 $\pm$ 38     &     1.6            &       1.0 $\pm$ 0.8 	&        -24  \\
         F17138-1017             &     4 $\pm$ 3           &   83 $\pm$ 8    &    0.3     		&       0.05 $\pm$ 0.04 		&  	  65    \\
         F21130-4446             &    30 $\pm$ 34       &  113 $\pm$ 8     &    8.0           	&       0.3 $\pm$ 0.3 		&    -198    \\
         F21453-3511             &   113 $\pm$ 89      &  266 $\pm$ 43   &    1.3             &       0.4 $\pm$ 0.3 	&      -129\\
\end{tabular} 
\vskip0.2cm\hskip0.0cm
\end{scriptsize}
\begin{minipage}[h]{18cm}
\tablefoot{Derived H$\alpha$ kinematic quantities for the broad component. Horizontal lines distinguish class 0, 1, and 2 (from top to bottom), respectively).
Column (1): IRAS name. 
Column (2): Velocity amplitude. 
Column (3): Mean velocity dispersion. 
Column (4): Area covered by the broad component. 
Column (5): Dynamical ratio between the velocity amplitude (Col. 2) and the mean velocity dispersion (Col. 3). 
Column (6): Velocity offset, computed as the difference between the mean velocity of the broad and the narrow components. Negative values represent blueshifted values while positive ones are redshifted. }
\end{minipage}
\end{table*}


\newpage
\begin{acknowledgements}

We thank the anonymous referee for useful comments and suggestions that helped to improve the quality and presentation of the paper. Thanks are also due to Javier Rodr\'iguez-Zaur\'in and Mark Westmoquette for the useful discussions. This work was funded in part by the Marie Curie Initial Training Network ELIXIR of the European Commission under contract PITN-GA-2008-214227. This work has been supported by the Spanish Ministry of Science and Innovation (MICINN) under grant ESP2007-65475-C02-01. It is based on observations carried out at the European Southern observatory, Paranal (Chile), Programs 076.B-0479(A), 078.B-0072(A), and 081.B-0108(A). This research made use of the NASA/IPAC Extragalactic Database, which is operated by the Jet Propulsion Laboratory, California Institute of Technology, under contract with the National Aeronautic and Space Administration.

\end{acknowledgements}

\bibliographystyle{aa} 
\bibliography{biblio}

\begin{thebibliography}{88}
\expandafter\ifx\csname natexlab\endcsname\relax\def\natexlab#1{#1}\fi

\bibitem[{{Arribas} {et~al.}(2004){Arribas}, {Bushouse}, {Lucas}, {Colina}, \&
  {Borne}}]{A04}
{Arribas}, S., {Bushouse}, H., {Lucas}, R.~A., {Colina}, L., \& {Borne}, K.~D.
  2004, \aj, 127, 2522

\bibitem[{{Arribas} \& {Colina}(2003)}]{A03}
{Arribas}, S. \& {Colina}, L. 2003, \apj, 591, 791

\bibitem[{{Arribas} {et~al.}(2012){Arribas}, {Colina}, {Alonso-Herrero},
  {Rosales-Ortega}, {Monreal-Ibero}, {Garc{\'{\i}}a-Mar{\'{\i}}n},
  {Garc{\'{\i}}a-Burillo}, \& {Rodr{\'{\i}}guez-Zaur{\'{\i}}n}}]{A12}
{Arribas}, S., {Colina}, L., {Alonso-Herrero}, A., {et~al.} 2012, \aap, 541,
  A20

\bibitem[{{Arribas} {et~al.}(2000){Arribas}, {Colina}, \& {Borne}}]{A00}
{Arribas}, S., {Colina}, L., \& {Borne}, K.~D. 2000, \apj, 545, 228

\bibitem[{{Arribas} {et~al.}(2001){Arribas}, {Colina}, \& {Clements}}]{A01}
{Arribas}, S., {Colina}, L., \& {Clements}, D. 2001, \apj, 560, 160

\bibitem[{{Arribas} {et~al.}(2008){Arribas}, {Colina}, {Monreal-Ibero},
  {Alfonso}, {Garc{\'{\i}}a-Mar{\'{\i}}n}, \& {Alonso-Herrero}}]{A08}
{Arribas}, S., {Colina}, L., {Monreal-Ibero}, A., {et~al.} 2008, \aap, 479, 687

\bibitem[{{Bedregal} {et~al.}(2009){Bedregal}, {Colina}, {Alonso-Herrero}, \&
  {Arribas}}]{Bed09}
{Bedregal}, A.~G., {Colina}, L., {Alonso-Herrero}, A., \& {Arribas}, S. 2009,
  \apj, 698, 1852

\bibitem[{{Bellocchi} {et~al.}(2012){Bellocchi}, {Arribas}, \&
  {Colina}}]{YO2012}
{Bellocchi}, E., {Arribas}, S., \& {Colina}, L. 2012, \aap, 542, A54

\bibitem[{{Cappellari} {et~al.}(2006){Cappellari}, {Bacon}, {Bureau}, {Damen},
  {Davies}, {de Zeeuw}, {Emsellem}, {Falc{\'o}n-Barroso}, {Krajnovi{\'c}},
  {Kuntschner}, {McDermid}, {Peletier}, {Sarzi}, {van den Bosch}, \& {van de
  Ven}}]{Cap06}
{Cappellari}, M., {Bacon}, R., {Bureau}, M., {et~al.} 2006, \mnras, 366, 1126

\bibitem[{{Cappellari} {et~al.}(2007){Cappellari}, {Emsellem}, {Bacon},
  {Bureau}, {Davies}, {de Zeeuw}, {Falc{\'o}n-Barroso}, {Krajnovi{\'c}},
  {Kuntschner}, {McDermid}, {Peletier}, {Sarzi}, {van den Bosch}, \& {van de
  Ven}}]{Cap07}
{Cappellari}, M., {Emsellem}, E., {Bacon}, R., {et~al.} 2007, \mnras, 379, 418

\bibitem[{{Cole} {et~al.}(2001){Cole}, {Norberg}, {Baugh}, {Frenk},
  {Bland-Hawthorn}, {Bridges}, {Cannon}, {Colless}, {Collins}, {Couch},
  {Cross}, {Dalton}, {De Propris}, {Driver}, {Efstathiou}, {Ellis},
  {Glazebrook}, {Jackson}, {Lahav}, {Lewis}, {Lumsden}, {Maddox}, {Madgwick},
  {Peacock}, {Peterson}, {Sutherland}, \& {Taylor}}]{Cole01}
{Cole}, S., {Norberg}, P., {Baugh}, C.~M., {et~al.} 2001, \mnras, 326, 255

\bibitem[{{Colina} {et~al.}(1999){Colina}, {Arribas}, \& {Borne}}]{Co99}
{Colina}, L., {Arribas}, S., \& {Borne}, K.~D. 1999, \apjl, 527, L13

\bibitem[{{Colina} {et~al.}(2004){Colina}, {Arribas}, \& {Clements}}]{Co04}
{Colina}, L., {Arribas}, S., \& {Clements}, D. 2004, \apj, 602, 181

\bibitem[{{Colina} {et~al.}(2005){Colina}, {Arribas}, \&
  {Monreal-Ibero}}]{Co05}
{Colina}, L., {Arribas}, S., \& {Monreal-Ibero}, A. 2005, \apj, 621, 725

\bibitem[{{Colina} {et~al.}(2001){Colina}, {Borne}, {Bushouse}, {Lucas},
  {Rowan-Robinson}, {Lawrence}, {Clements}, {Baker}, \& {Oliver}}]{Co01}
{Colina}, L., {Borne}, K., {Bushouse}, H., {et~al.} 2001, \apj, 563, 546

\bibitem[{{Combes} {et~al.}(2011){Combes}, {Garc{\'{\i}}a-Burillo}, {Braine},
  {Schinnerer}, {Walter}, \& {Colina}}]{Com11}
{Combes}, F., {Garc{\'{\i}}a-Burillo}, S., {Braine}, J., {et~al.} 2011, \aap,
  528, A124

\bibitem[{{da Cunha} {et~al.}(2010){da Cunha}, {Charmandaris},
  {D{\'{\i}}az-Santos}, {Armus}, {Marshall}, \& {Elbaz}}]{cunha10}
{da Cunha}, E., {Charmandaris}, V., {D{\'{\i}}az-Santos}, T., {et~al.} 2010,
  \aap, 523, A78

\bibitem[{{Dasyra} {et~al.}(2006{\natexlab{a}}){Dasyra}, {Tacconi}, {Davies},
  {Genzel}, {Lutz}, {Naab}, {Burkert}, {Veilleux}, \& {Sanders}}]{Dasy_b_06}
{Dasyra}, K.~M., {Tacconi}, L.~J., {Davies}, R.~I., {et~al.}
  2006{\natexlab{a}}, \apj, 638, 745

\bibitem[{{Dasyra} {et~al.}(2006{\natexlab{b}}){Dasyra}, {Tacconi}, {Davies},
  {Naab}, {Genzel}, {Lutz}, {Sturm}, {Baker}, {Veilleux}, {Sanders}, \&
  {Burkert}}]{dasyra06}
{Dasyra}, K.~M., {Tacconi}, L.~J., {Davies}, R.~I., {et~al.}
  2006{\natexlab{b}}, \apj, 651, 835

\bibitem[{{Davies} {et~al.}(2011){Davies}, {F{\"o}rster Schreiber}, {Cresci},
  {Genzel}, {Bouch{\'e}}, {Burkert}, {Buschkamp}, {Genel}, {Hicks}, {Kurk},
  {Lutz}, {Newman}, {Shapiro}, {Sternberg}, {Tacconi}, \& {Wuyts}}]{davies11}
{Davies}, R., {F{\"o}rster Schreiber}, N.~M., {Cresci}, G., {et~al.} 2011,
  \apj, 741, 69

\bibitem[{{Downes} \& {Solomon}(1998)}]{Dow98}
{Downes}, D. \& {Solomon}, P.~M. 1998, \apj, 507, 615

\bibitem[{{Elbaz} {et~al.}(2011){Elbaz}, {Dickinson}, {Hwang},
  {D{\'{\i}}az-Santos}, {Magdis}, {Magnelli}, {Le Borgne}, {Galliano},
  {Pannella}, {Chanial}, {Armus}, {Charmandaris}, {Daddi}, {Aussel}, {Popesso},
  {Kartaltepe}, {Altieri}, {Valtchanov}, {Coia}, {Dannerbauer}, {Dasyra},
  {Leiton}, {Mazzarella}, {Alexander}, {Buat}, {Burgarella}, {Chary}, {Gilli},
  {Ivison}, {Juneau}, {Le Floc'h}, {Lutz}, {Morrison}, {Mullaney}, {Murphy},
  {Pope}, {Scott}, {Brodwin}, {Calzetti}, {Cesarsky}, {Charlot}, {Dole},
  {Eisenhardt}, {Ferguson}, {F{\"o}rster Schreiber}, {Frayer}, {Giavalisco},
  {Huynh}, {Koekemoer}, {Papovich}, {Reddy}, {Surace}, {Teplitz}, {Yun}, \&
  {Wilson}}]{elbaz11}
{Elbaz}, D., {Dickinson}, M., {Hwang}, H.~S., {et~al.} 2011, \aap, 533, A119

\bibitem[{{Elbaz} {et~al.}(2010){Elbaz}, {Hwang}, {Magnelli}, {Daddi},
  {Aussel}, {Altieri}, {Amblard}, {Andreani}, {Arumugam}, {Auld}, {Babbedge},
  {Berta}, {Blain}, {Bock}, {Bongiovanni}, {Boselli}, {Buat}, {Burgarella},
  {Castro-Rodriguez}, {Cava}, {Cepa}, {Chanial}, {Chary}, {Cimatti},
  {Clements}, {Conley}, {Conversi}, {Cooray}, {Dickinson}, {Dominguez},
  {Dowell}, {Dunlop}, {Dwek}, {Eales}, {Farrah}, {F{\"o}rster Schreiber},
  {Fox}, {Franceschini}, {Gear}, {Genzel}, {Glenn}, {Griffin}, {Gruppioni},
  {Halpern}, {Hatziminaoglou}, {Ibar}, {Isaak}, {Ivison}, {Lagache}, {Le
  Borgne}, {Le Floc'h}, {Levenson}, {Lu}, {Lutz}, {Madden}, {Maffei}, {Magdis},
  {Mainetti}, {Maiolino}, {Marchetti}, {Mortier}, {Nguyen}, {Nordon},
  {O'Halloran}, {Okumura}, {Oliver}, {Omont}, {Page}, {Panuzzo},
  {Papageorgiou}, {Pearson}, {Perez Fournon}, {P{\'e}rez Garc{\'{\i}}a},
  {Poglitsch}, {Pohlen}, {Popesso}, {Pozzi}, {Rawlings}, {Rigopoulou},
  {Riguccini}, {Rizzo}, {Rodighiero}, {Roseboom}, {Rowan-Robinson},
  {Saintonge}, {Sanchez Portal}, {Santini}, {Sauvage}, {Schulz}, {Scott},
  {Seymour}, {Shao}, {Shupe}, {Smith}, {Stevens}, {Sturm}, {Symeonidis},
  {Tacconi}, {Trichas}, {Tugwell}, {Vaccari}, {Valtchanov}, {Vieira},
  {Vigroux}, {Wang}, {Ward}, {Wright}, {Xu}, \& {Zemcov}}]{elbaz10}
{Elbaz}, D., {Hwang}, H.~S., {Magnelli}, B., {et~al.} 2010, \aap, 518, L29

\bibitem[{{Epinat} {et~al.}(2010){Epinat}, {Amram}, {Balkowski}, \&
  {Marcelin}}]{Epi10}
{Epinat}, B., {Amram}, P., {Balkowski}, C., \& {Marcelin}, M. 2010, \mnras,
  401, 2113

\bibitem[{{Epinat} {et~al.}(2009){Epinat}, {Contini}, {Le F{\`e}vre},
  {Vergani}, {Garilli}, {Amram}, {Queyrel}, {Tasca}, \& {Tresse}}]{Epi09}
{Epinat}, B., {Contini}, T., {Le F{\`e}vre}, O., {et~al.} 2009, \aap, 504, 789

\bibitem[{{Epinat} {et~al.}(2012){Epinat}, {Tasca}, {Amram}, {Contini}, {Le
  F{\`e}vre}, {Queyrel}, {Vergani}, {Garilli}, {Kissler-Patig}, {Moultaka},
  {Paioro}, {Tresse}, {Bournaud}, {L{\'o}pez-Sanjuan}, \& {Perret}}]{Epi12}
{Epinat}, B., {Tasca}, L., {Amram}, P., {et~al.} 2012, \aap, 539, A92

\bibitem[{{Flores} {et~al.}(2006){Flores}, {Hammer}, {Puech}, {Amram}, \&
  {Balkowski}}]{Flores06}
{Flores}, H., {Hammer}, F., {Puech}, M., {Amram}, P., \& {Balkowski}, C. 2006,
  \aap, 455, 107

\bibitem[{{F{\"o}rster Schreiber} {et~al.}(2009){F{\"o}rster Schreiber},
  {Genzel}, {Bouch{\'e}}, {Cresci}, {Davies}, {Buschkamp}, {Shapiro},
  {Tacconi}, {Hicks}, {Genel}, {Shapley}, {Erb}, {Steidel}, {Lutz},
  {Eisenhauer}, {Gillessen}, {Sternberg}, {Renzini}, {Cimatti}, {Daddi},
  {Kurk}, {Lilly}, {Kong}, {Lehnert}, {Nesvadba}, {Verma}, {McCracken},
  {Arimoto}, {Mignoli}, \& {Onodera}}]{FS09}
{F{\"o}rster Schreiber}, N.~M., {Genzel}, R., {Bouch{\'e}}, N., {et~al.} 2009,
  \apj, 706, 1364

\bibitem[{{Garc{\'{\i}}a-Mar{\'{\i}}n}
  {et~al.}(2007){Garc{\'{\i}}a-Mar{\'{\i}}n}, {Colina}, \& {Arribas}}]{GM07}
{Garc{\'{\i}}a-Mar{\'{\i}}n}, M., {Colina}, L., \& {Arribas}, S. 2007, phD
  thesis, Universidad Autonoma de Madrid, Spain

\bibitem[{{Garc{\'{\i}}a-Mar{\'{\i}}n}
  {et~al.}(2006){Garc{\'{\i}}a-Mar{\'{\i}}n}, {Colina}, {Arribas},
  {Alonso-Herrero}, \& {Mediavilla}}]{GM06}
{Garc{\'{\i}}a-Mar{\'{\i}}n}, M., {Colina}, L., {Arribas}, S.,
  {Alonso-Herrero}, A., \& {Mediavilla}, E. 2006, \apj, 650, 850

\bibitem[{{Garc{\'{\i}}a-Mar{\'{\i}}n}
  {et~al.}(2009){Garc{\'{\i}}a-Mar{\'{\i}}n}, {Colina}, {Arribas}, \&
  {Monreal-Ibero}}]{GM09}
{Garc{\'{\i}}a-Mar{\'{\i}}n}, M., {Colina}, L., {Arribas}, S., \&
  {Monreal-Ibero}, A. 2009, \aap, 505, 1319

\bibitem[{{Genzel} {et~al.}(2001){Genzel}, {Tacconi}, {Rigopoulou}, {Lutz}, \&
  {Tecza}}]{genzel01}
{Genzel}, R., {Tacconi}, L.~J., {Rigopoulou}, D., {Lutz}, D., \& {Tecza}, M.
  2001, \apj, 563, 527

\bibitem[{{Gon{\c c}alves} {et~al.}(2010){Gon{\c c}alves}, {Basu-Zych},
  {Overzier}, {Martin}, {Law}, {Schiminovich}, {Wyder}, {Mallery}, {Rich}, \&
  {Heckman}}]{Gon10}
{Gon{\c c}alves}, T.~S., {Basu-Zych}, A., {Overzier}, R., {et~al.} 2010, \apj,
  724, 1373

\bibitem[{{Heckman} {et~al.}(1990){Heckman}, {Armus}, \& {Miley}}]{heck90}
{Heckman}, T.~M., {Armus}, L., \& {Miley}, G.~K. 1990, \apjs, 74, 833

\bibitem[{{Heckman} {et~al.}(2005){Heckman}, {Hoopes}, {Seibert}, {Martin},
  {Salim}, {Rich}, {Kauffmann}, {Charlot}, {Barlow}, {Bianchi}, {Byun},
  {Donas}, {Forster}, {Friedman}, {Jelinsky}, {Lee}, {Madore}, {Malina},
  {Milliard}, {Morrissey}, {Neff}, {Schiminovich}, {Siegmund}, {Small},
  {Szalay}, {Welsh}, \& {Wyder}}]{heck05}
{Heckman}, T.~M., {Hoopes}, C.~G., {Seibert}, M., {et~al.} 2005, \apjl, 619,
  L35

\bibitem[{{Hinz} \& {Rieke}(2006)}]{H06}
{Hinz}, J.~L. \& {Rieke}, G.~H. 2006, \apj, 646, 872

\bibitem[{{James} {et~al.}(2004){James}, {Shane}, {Beckman}, {Cardwell},
  {Collins}, {Etherton}, {de Jong}, {Fathi}, {Knapen}, {Peletier}, {Percival},
  {Pollacco}, {Seigar}, {Stedman}, \& {Steele}}]{james04}
{James}, P.~A., {Shane}, N.~S., {Beckman}, J.~E., {et~al.} 2004, \aap, 414, 23

\bibitem[{{Kennicutt}(1998)}]{Ken98}
{Kennicutt}, Jr., R.~C. 1998, \apj, 498, 541

\bibitem[{{Kennicutt} {et~al.}(1994){Kennicutt}, {Tamblyn}, \&
  {Congdon}}]{Ken94}
{Kennicutt}, Jr., R.~C., {Tamblyn}, P., \& {Congdon}, C.~E. 1994, \apj, 435, 22

\bibitem[{{Kim} {et~al.}(2002){Kim}, {Veilleux}, \& {Sanders}}]{Kim02}
{Kim}, D.-C., {Veilleux}, S., \& {Sanders}, D.~B. 2002, \apjs, 143, 277

\bibitem[{{Krajnovi{\'c}} {et~al.}(2006){Krajnovi{\'c}}, {Cappellari}, {de
  Zeeuw}, \& {Copin}}]{K06}
{Krajnovi{\'c}}, D., {Cappellari}, M., {de Zeeuw}, P.~T., \& {Copin}, Y. 2006,
  \mnras, 366, 787

\bibitem[{{Kroupa}(2008)}]{Kroupa08}
{Kroupa}, P. 2008, in Astronomical Society of the Pacific Conference Series,
  Vol. 390, Pathways Through an Eclectic Universe, ed. J.~H. {Knapen}, T.~J.
  {Mahoney}, \& A.~{Vazdekis}, 3

\bibitem[{{Lagache} {et~al.}(2005){Lagache}, {Puget}, \& {Dole}}]{lagache05}
{Lagache}, G., {Puget}, J.-L., \& {Dole}, H. 2005, \araa, 43, 727

\bibitem[{{Law} {et~al.}(2007){Law}, {Steidel}, {Erb}, {Larkin}, {Pettini},
  {Shapley}, \& {Wright}}]{Law07a}
{Law}, D.~R., {Steidel}, C.~C., {Erb}, D.~K., {et~al.} 2007, \apj, 669, 929

\bibitem[{{Law} {et~al.}(2009){Law}, {Steidel}, {Erb}, {Larkin}, {Pettini},
  {Shapley}, \& {Wright}}]{Law09}
{Law}, D.~R., {Steidel}, C.~C., {Erb}, D.~K., {et~al.} 2009, \apj, 697, 2057

\bibitem[{{Le F{\`e}vre} {et~al.}(2003){Le F{\`e}vre}, {Saisse}, {Mancini},
  {Brau-Nogue}, {Caputi}, {Castinel}, {D'Odorico}, {Garilli}, {Kissler-Patig},
  {Lucuix}, {Mancini}, {Pauget}, {Sciarretta}, {Scodeggio}, {Tresse}, \&
  {Vettolani}}]{Lfevre03}
{Le F{\`e}vre}, O., {Saisse}, M., {Mancini}, D., {et~al.} 2003, in Presented at
  the Society of Photo-Optical Instrumentation Engineers (SPIE) Conference,
  Vol. 4841, Society of Photo-Optical Instrumentation Engineers (SPIE)
  Conference Series, ed. {M.~Iye \& A.~F.~M.~Moorwood}, 1670--1681

\bibitem[{{Le Floc'h} {et~al.}(2005){Le Floc'h}, {Papovich}, {Dole}, {Bell},
  {Lagache}, {Rieke}, {Egami}, {P{\'e}rez-Gonz{\'a}lez}, {Alonso-Herrero},
  {Rieke}, {Blaylock}, {Engelbracht}, {Gordon}, {Hines}, {Misselt}, {Morrison},
  \& {Mould}}]{LF05}
{Le Floc'h}, E., {Papovich}, C., {Dole}, H., {et~al.} 2005, \apj, 632, 169

\bibitem[{{L{\'{\i}}pari} {et~al.}(2003){L{\'{\i}}pari}, {Terlevich},
  {D{\'{\i}}az}, {Taniguchi}, {Zheng}, {Tsvetanov}, {Carranza}, \&
  {Dottori}}]{Lipari03}
{L{\'{\i}}pari}, S., {Terlevich}, R., {D{\'{\i}}az}, R.~J., {et~al.} 2003,
  \mnras, 340, 289

\bibitem[{{Ma{\'{\i}}z-Apell{\'a}niz}(2004)}]{MA04}
{Ma{\'{\i}}z-Apell{\'a}niz}, J. 2004, \pasp, 116, 859

\bibitem[{{Monreal-Ibero} {et~al.}(2006){Monreal-Ibero}, {Arribas}, \&
  {Colina}}]{MI06}
{Monreal-Ibero}, A., {Arribas}, S., \& {Colina}, L. 2006, \apj, 637, 138

\bibitem[{{Monreal-Ibero} {et~al.}(2010){Monreal-Ibero}, {Arribas}, {Colina},
  {Rodr{\'{\i}}guez-Zaur{\'{\i}}n}, {Alonso-Herrero}, \&
  {Garc{\'{\i}}a-Mar{\'{\i}}n}}]{MI10}
{Monreal-Ibero}, A., {Arribas}, S., {Colina}, L., {et~al.} 2010, \aap, 517,
  A28+

\bibitem[{{Nordon} {et~al.}(2012){Nordon}, {Lutz}, {Genzel}, {Berta}, {Wuyts},
  {Magnelli}, {Altieri}, {Andreani}, {Aussel}, {Bongiovanni}, {Cepa},
  {Cimatti}, {Daddi}, {Fadda}, {F{\"o}rster Schreiber}, {Lagache}, {Maiolino},
  {P{\'e}rez Garc{\'{\i}}a}, {Poglitsch}, {Popesso}, {Pozzi}, {Rodighiero},
  {Rosario}, {Saintonge}, {Sanchez-Portal}, {Santini}, {Sturm}, {Tacconi},
  {Valtchanov}, \& {Yan}}]{nordon12}
{Nordon}, R., {Lutz}, D., {Genzel}, R., {et~al.} 2012, \apj, 745, 182

\bibitem[{{Nordon} {et~al.}(2010){Nordon}, {Lutz}, {Shao}, {Magnelli}, {Berta},
  {Altieri}, {Andreani}, {Aussel}, {Bongiovanni}, {Cava}, {Cepa}, {Cimatti},
  {Daddi}, {Dominguez}, {Elbaz}, {F{\"o}rster Schreiber}, {Genzel}, {Grazian},
  {Magdis}, {Maiolino}, {P{\'e}rez Garc{\'{\i}}a}, {Poglitsch}, {Popesso},
  {Pozzi}, {Riguccini}, {Rodighiero}, {Saintonge}, {Sanchez-Portal}, {Santini},
  {Sturm}, {Tacconi}, {Valtchanov}, {Wetzstein}, \& {Wieprecht}}]{nordon10}
{Nordon}, R., {Lutz}, D., {Shao}, L., {et~al.} 2010, \aap, 518, L24

\bibitem[{{Overzier} {et~al.}(2008){Overzier}, {Heckman}, {Kauffmann},
  {Seibert}, {Rich}, {Basu-Zych}, {Lotz}, {Aloisi}, {Charlot}, {Hoopes},
  {Martin}, {Schiminovich}, \& {Madore}}]{over08}
{Overzier}, R.~A., {Heckman}, T.~M., {Kauffmann}, G., {et~al.} 2008, \apj, 677,
  37

\bibitem[{{Papovich} {et~al.}(2007){Papovich}, {Rudnick}, {Le Floc'h}, {van
  Dokkum}, {Rieke}, {Taylor}, {Armus}, {Gawiser}, {Huang}, {Marcillac}, \&
  {Franx}}]{papov07}
{Papovich}, C., {Rudnick}, G., {Le Floc'h}, E., {et~al.} 2007, \apj, 668, 45

\bibitem[{{Peng} {et~al.}(2010){Peng}, {Ho}, {Impey}, \& {Rix}}]{peng10}
{Peng}, C.~Y., {Ho}, L.~C., {Impey}, C.~D., \& {Rix}, H.-W. 2010, \aj, 139,
  2097

\bibitem[{{P{\'e}rez-Gonz{\'a}lez} {et~al.}(2005){P{\'e}rez-Gonz{\'a}lez},
  {Rieke}, {Egami}, {Alonso-Herrero}, {Dole}, {Papovich}, {Blaylock}, {Jones},
  {Rieke}, {Rigby}, {Barmby}, {Fazio}, {Huang}, \& {Martin}}]{PG05}
{P{\'e}rez-Gonz{\'a}lez}, P.~G., {Rieke}, G.~H., {Egami}, E., {et~al.} 2005,
  \apj, 630, 82

\bibitem[{{P{\'e}rez-Gonz{\'a}lez} {et~al.}(2008){P{\'e}rez-Gonz{\'a}lez},
  {Rieke}, {Villar}, {Barro}, {Blaylock}, {Egami}, {Gallego}, {Gil de Paz},
  {Pascual}, {Zamorano}, \& {Donley}}]{PPG08}
{P{\'e}rez-Gonz{\'a}lez}, P.~G., {Rieke}, G.~H., {Villar}, V., {et~al.} 2008,
  \apj, 675, 234

\bibitem[{{Piqueras L{\'o}pez} {et~al.}(2013){Piqueras L{\'o}pez}, {Colina},
  {Arribas}, \& {Alonso-Herrero}}]{Piq13}
{Piqueras L{\'o}pez}, J., {Colina}, L., {Arribas}, S., \& {Alonso-Herrero}, A.
  2013, ArXiv e-prints

\bibitem[{{Piqueras L{\'o}pez} {et~al.}(2012){Piqueras L{\'o}pez}, {Davies},
  {Colina}, \& {Orban de Xivry}}]{Piq12}
{Piqueras L{\'o}pez}, J., {Davies}, R., {Colina}, L., \& {Orban de Xivry}, G.
  2012, \apj, 752, 47

\bibitem[{{Pope} {et~al.}(2006){Pope}, {Scott}, {Dickinson}, {Chary},
  {Morrison}, {Borys}, {Sajina}, {Alexander}, {Daddi}, {Frayer}, {MacDonald},
  \& {Stern}}]{pope06}
{Pope}, A., {Scott}, D., {Dickinson}, M., {et~al.} 2006, \mnras, 370, 1185

\bibitem[{{Puech} {et~al.}(2007){Puech}, {Hammer}, {Lehnert}, \&
  {Flores}}]{puech07}
{Puech}, M., {Hammer}, F., {Lehnert}, M.~D., \& {Flores}, H. 2007, \aap, 466,
  83

\bibitem[{{Rodr{\'{\i}}guez-Zaur{\'{\i}}n}
  {et~al.}(2011){Rodr{\'{\i}}guez-Zaur{\'{\i}}n}, {Arribas}, {Monreal-Ibero},
  {Colina}, {Alonso-Herrero}, \& {Alfonso-Garz{\'o}n}}]{RZ11}
{Rodr{\'{\i}}guez-Zaur{\'{\i}}n}, J., {Arribas}, S., {Monreal-Ibero}, A.,
  {et~al.} 2011, \aap, 527, A60+

\bibitem[{{Rodr{\'{\i}}guez Zaur{\'{\i}}n} {et~al.}(2010){Rodr{\'{\i}}guez
  Zaur{\'{\i}}n}, {Tadhunter}, \& {Gonz{\'a}lez Delgado}}]{RZ10}
{Rodr{\'{\i}}guez Zaur{\'{\i}}n}, J., {Tadhunter}, C.~N., \& {Gonz{\'a}lez
  Delgado}, R.~M. 2010, \mnras, 403, 1317

\bibitem[{{Rothberg} \& {Fischer}(2010)}]{Rothberg10}
{Rothberg}, B. \& {Fischer}, J. 2010, \apj, 712, 318

\bibitem[{{Rothberg} {et~al.}(2013){Rothberg}, {Fischer}, {Rodrigues}, \&
  {Sanders}}]{Rothberg13}
{Rothberg}, B., {Fischer}, J., {Rodrigues}, M., \& {Sanders}, D.~B. 2013, \apj,
  767, 72

\bibitem[{{Rupke} \& {Veilleux}(2013)}]{rupke13}
{Rupke}, D.~S.~N. \& {Veilleux}, S. 2013, ArXiv e-prints

\bibitem[{{Salpeter}(1955)}]{Sal55}
{Salpeter}, E.~E. 1955, \apj, 121, 161

\bibitem[{{Sanders} {et~al.}(2003){Sanders}, {Mazzarella}, {Kim}, {Surace}, \&
  {Soifer}}]{sanders03}
{Sanders}, D.~B., {Mazzarella}, J.~M., {Kim}, D.-C., {Surace}, J.~A., \&
  {Soifer}, B.~T. 2003, \aj, 126, 1607

\bibitem[{{Sanders} \& {Mirabel}(1996)}]{SM96}
{Sanders}, D.~B. \& {Mirabel}, I.~F. 1996, \araa, 34, 749

\bibitem[{{Scoville} {et~al.}(2000){Scoville}, {Evans}, {Thompson}, {Rieke},
  {Hines}, {Low}, {Dinshaw}, {Surace}, \& {Armus}}]{SCo00}
{Scoville}, N.~Z., {Evans}, A.~S., {Thompson}, R., {et~al.} 2000, \aj, 119, 991

\bibitem[{{Shapiro} {et~al.}(2010){Shapiro}, {Falc{\'o}n-Barroso}, {van de
  Ven}, {de Zeeuw}, {Sarzi}, {Bacon}, {Bolatto}, {Cappellari}, {Croton},
  {Davies}, {Emsellem}, {Fakhouri}, {Krajnovi{\'c}}, {Kuntschner}, {McDermid},
  {Peletier}, {van den Bosch}, \& {van der Wolk}}]{shapiro10}
{Shapiro}, K.~L., {Falc{\'o}n-Barroso}, J., {van de Ven}, G., {et~al.} 2010,
  \mnras, 402, 2140

\bibitem[{{Shapiro} {et~al.}(2008){Shapiro}, {Genzel}, {F{\"o}rster Schreiber},
  {Tacconi}, {Bouch{\'e}}, {Cresci}, {Davies}, {Eisenhauer}, {Johansson},
  {Krajnovi{\'c}}, {Lutz}, {Naab}, {Arimoto}, {Arribas}, {Cimatti}, {Colina},
  {Daddi}, {Daigle}, {Erb}, {Hernandez}, {Kong}, {Mignoli}, {Onodera},
  {Renzini}, {Shapley}, \& {Steidel}}]{S08}
{Shapiro}, K.~L., {Genzel}, R., {F{\"o}rster Schreiber}, N.~M., {et~al.} 2008,
  \apj, 682, 231

\bibitem[{{Shapiro} {et~al.}(2009){Shapiro}, {Genzel}, {Quataert}, {F{\"o}rster
  Schreiber}, {Davies}, {Tacconi}, {Armus}, {Bouch{\'e}}, {Buschkamp},
  {Cimatti}, {Cresci}, {Daddi}, {Eisenhauer}, {Erb}, {Genel}, {Hicks}, {Lilly},
  {Lutz}, {Renzini}, {Shapley}, {Steidel}, \& {Sternberg}}]{S09}
{Shapiro}, K.~L., {Genzel}, R., {Quataert}, E., {et~al.} 2009, \apj, 701, 955

\bibitem[{{Skrutskie} {et~al.}(2006){Skrutskie}, {Cutri}, {Stiening},
  {Weinberg}, {Schneider}, {Carpenter}, {Beichman}, {Capps}, {Chester},
  {Elias}, {Huchra}, {Liebert}, {Lonsdale}, {Monet}, {Price}, {Seitzer},
  {Jarrett}, {Kirkpatrick}, {Gizis}, {Howard}, {Evans}, {Fowler}, {Fullmer},
  {Hurt}, {Light}, {Kopan}, {Marsh}, {McCallon}, {Tam}, {Van Dyk}, \&
  {Wheelock}}]{Skru06}
{Skrutskie}, M.~F., {Cutri}, R.~M., {Stiening}, R., {et~al.} 2006, \aj, 131,
  1163

\bibitem[{{Swinbank} {et~al.}(2012){Swinbank}, {Sobral}, {Smail}, {Geach},
  {Best}, {McCarthy}, {Crain}, \& {Theuns}}]{swin12}
{Swinbank}, A.~M., {Sobral}, D., {Smail}, I., {et~al.} 2012, \mnras, 426, 935

\bibitem[{{Tacconi} {et~al.}(2002){Tacconi}, {Genzel}, {Lutz}, {Rigopoulou},
  {Baker}, {Iserlohe}, \& {Tecza}}]{T02}
{Tacconi}, L.~J., {Genzel}, R., {Lutz}, D., {et~al.} 2002, \apj, 580, 73

\bibitem[{{Takagi} {et~al.}(2010){Takagi}, {Ohyama}, {Goto}, {Matsuhara},
  {Oyabu}, {Wada}, {Pearson}, {Lee}, {Im}, {Lee}, {Shim}, {Hanami}, {Ishigaki},
  {Imai}, {White}, {Serjeant}, \& {Malkan}}]{TAK10}
{Takagi}, T., {Ohyama}, Y., {Goto}, T., {et~al.} 2010, \aap, 514, A5

\bibitem[{{Taylor} {et~al.}(2010){Taylor}, {Franx}, {Brinchmann}, {van der
  Wel}, \& {van Dokkum}}]{Taylor10}
{Taylor}, E.~N., {Franx}, M., {Brinchmann}, J., {van der Wel}, A., \& {van
  Dokkum}, P.~G. 2010, \apj, 722, 1

\bibitem[{{Tecza} {et~al.}(2000){Tecza}, {Genzel}, {Tacconi}, {Anders},
  {Tacconi-Garman}, \& {Thatte}}]{tecza00}
{Tecza}, M., {Genzel}, R., {Tacconi}, L.~J., {et~al.} 2000, \apj, 537, 178

\bibitem[{{U} {et~al.}(2012){U}, {Sanders}, {Mazzarella}, {Evans}, {Howell},
  {Surace}, {Armus}, {Iwasawa}, {Kim}, {Casey}, {Vavilkin}, {Dufault},
  {Larson}, {Barnes}, {Chan}, {Frayer}, {Haan}, {Inami}, {Ishida},
  {Kartaltepe}, {Melbourne}, \& {Petric}}]{vivian12}
{U}, V., {Sanders}, D.~B., {Mazzarella}, J.~M., {et~al.} 2012, \apjs, 203, 9

\bibitem[{{Veilleux} {et~al.}(2002){Veilleux}, {Kim}, \& {Sanders}}]{V02}
{Veilleux}, S., {Kim}, D.-C., \& {Sanders}, D.~B. 2002, \apjs, 143, 315

\bibitem[{{Westmoquette} {et~al.}(2012){Westmoquette}, {Clements}, {Bendo}, \&
  {Khan}}]{West12}
{Westmoquette}, M.~S., {Clements}, D.~L., {Bendo}, G.~J., \& {Khan}, S.~A.
  2012, \mnras, 424, 416

\bibitem[{{Williams} {et~al.}(2010){Williams}, {Bureau}, \&
  {Cappellari}}]{Williams10}
{Williams}, M.~J., {Bureau}, M., \& {Cappellari}, M. 2010, \mnras, 409, 1330

\bibitem[{{Wolf} {et~al.}(2010){Wolf}, {Martinez}, {Bullock}, {Kaplinghat},
  {Geha}, {Mu{\~n}oz}, {Simon}, \& {Avedo}}]{wolf10}
{Wolf}, J., {Martinez}, G.~D., {Bullock}, J.~S., {et~al.} 2010, \mnras, 406,
  1220

\bibitem[{{Wright}(2006)}]{Wri06}
{Wright}, E.~L. 2006, \pasp, 118, 1711

\bibitem[{{Wu} {et~al.}(2005){Wu}, {Cao}, {Hao}, {Liu}, {Wang}, {Xia}, {Deng},
  \& {Young}}]{Wu05}
{Wu}, H., {Cao}, C., {Hao}, C.-N., {et~al.} 2005, \apjl, 632, L79

\bibitem[{{Yun} {et~al.}(2001){Yun}, {Reddy}, \& {Condon}}]{Yun01}
{Yun}, M.~S., {Reddy}, N.~A., \& {Condon}, J.~J. 2001, \apj, 554, 803

\end{thebibliography}

\clearpage

\appendix
\section {Kinematic maps}
\label{App_maps}

In this appendix we present the H$\alpha$ flux, velocity field, and the velocity dispersion maps of the different kinematic components for the galaxies of the sample. When available, the HST image is also included. The spatial offsets between the peaks of the continuum and the H$\alpha$ flux emission are shown in Table \ref{OFFSETS}.

\begin{figure*}
\includegraphics[width=1\textwidth, height=1\textwidth]{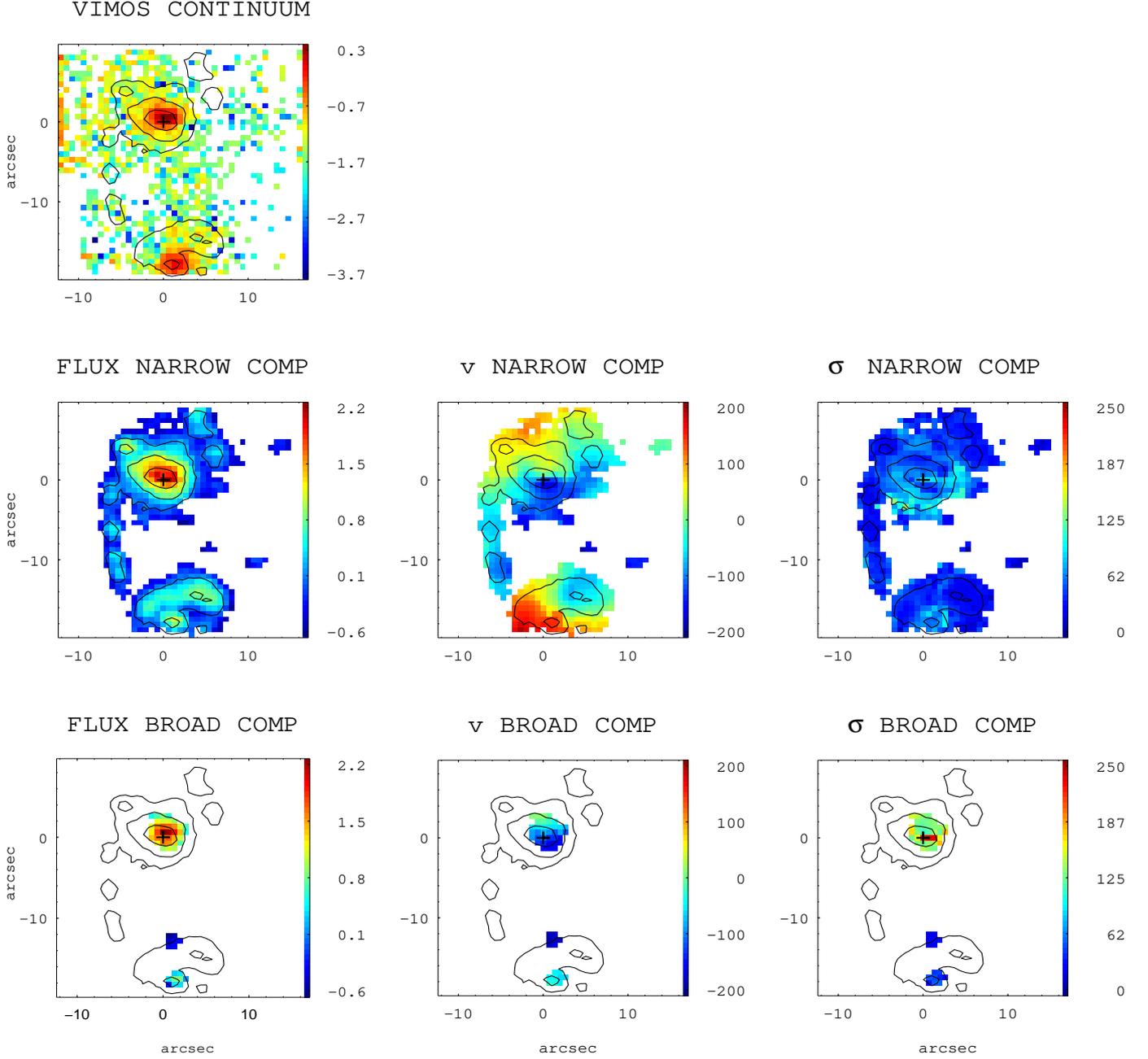}
\vspace{2mm}
\caption{\small {\it Top panel:} VIMOS continuum (6390-6490) \AA\ image within the rest-frame wavelength range. When available, the HST continuum image is also shown. {\it Middle panel:} the H$\alpha$ flux intensity, velocity field, $v$ (km s$^{-1}$), and velocity dispersion, $\sigma$ (km s$^{-1}$), for the narrow component. {\it Bottom panel:} similar maps for the broad component. The flux intensity maps are represented in logarithmic scale (applying a factor of -13) in units of erg s$^{-1}$ cm$^{-2}$ for the H$\alpha$ flux maps and erg s$^{-1}$ cm$^{-2}$ \AA$^{-1}$ for the continuum map. The center (0,0) is identified with the H$\alpha$ flux intensity peak and the iso-contours of the H$\alpha$ flux map are overplotted. North is at the top and East to the left in all the panels. }
\label{all_panels}
\vspace{3mm}
IRAS F01159-4443 (ESO 244-G012): this is an interacting pair with a nuclear separation $\sim$ 8.4 kpc, where the northern galaxy shows the brightest nuclear emission in both the H$\alpha$ and continuum maps. The two galaxies show regular velocity fields in the narrow component. The scale is 0.462 kpc/$^{\prime\prime}$. 
\end{figure*}

\begin{figure*}
\includegraphics[width=1.\textwidth, height=1\textwidth]{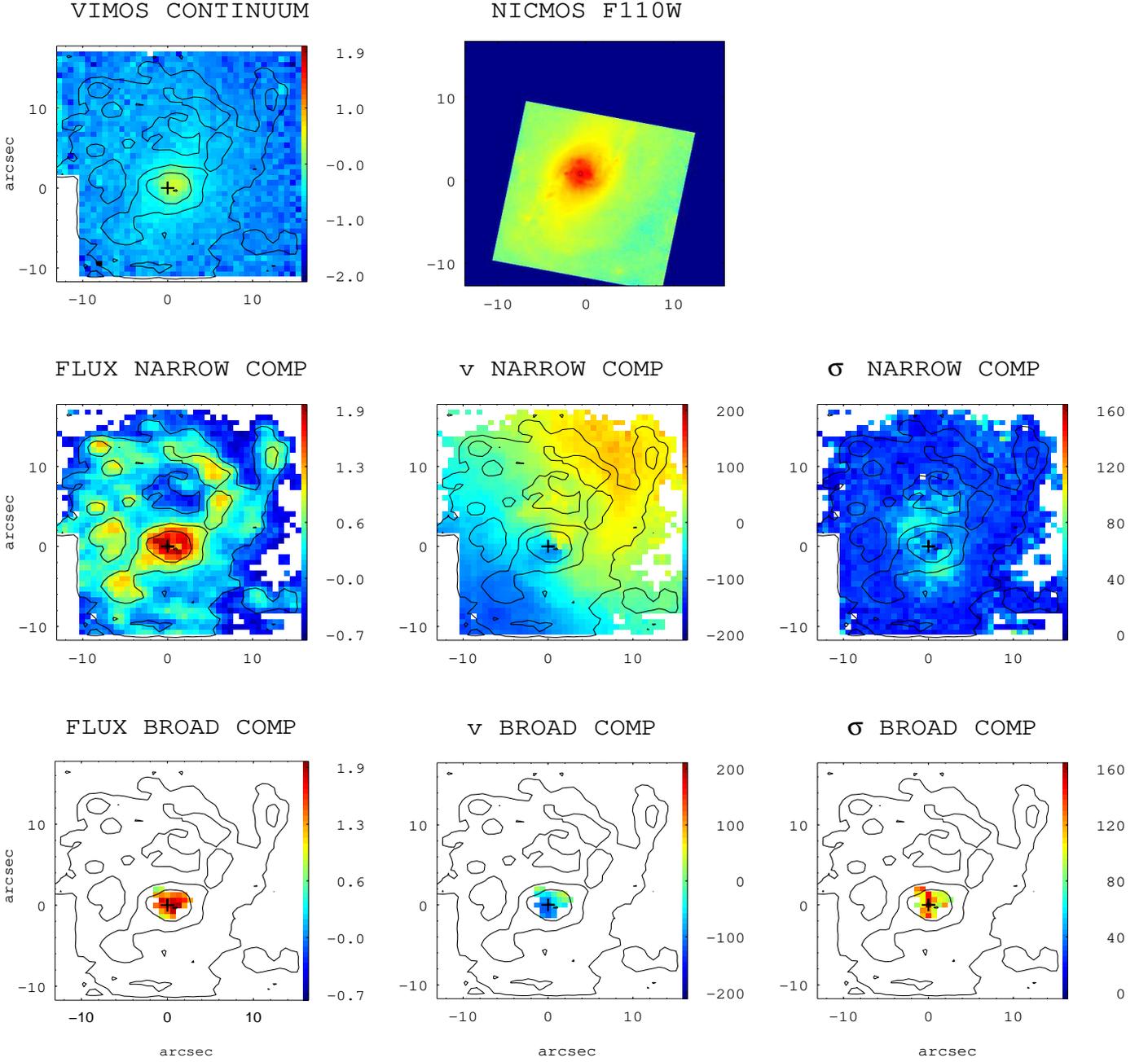}
\vspace{2mm}
\caption{(General comments about the panels as in Fig. A.1.) IRAS F01341-3735 (ESO 297-G011/G012): this system is composed of two galaxies (i.e., northern, southern) separated by $\sim$ 25 kpc, implying the need of two VIMOS pointings. The present maps correspond to the northern galaxy, which shows a regular velocity field. The velocity dispersion map has two symmetric local maxima of $\sim$ 70 km s$^{-1}$ around the nucleus, which are well fitted using one Gaussian model. These regions have low H$\alpha$ surface brightness and are associated with high excitation and high $\sigma$, as reported by \cite{MI10}. In the innermost regions (i.e., $\sim$ 1.5 kpc$^2$), the H$\alpha$-[NII] emission lines need a secondary broad component. The spatial scale is of 0.352 kpc/$^{\prime\prime}$.}
\label{all_panels}
\vspace{3mm}
\end{figure*}

\begin{figure*}
\vspace{0cm}
\includegraphics[width=1.\textwidth, height=1\textwidth]{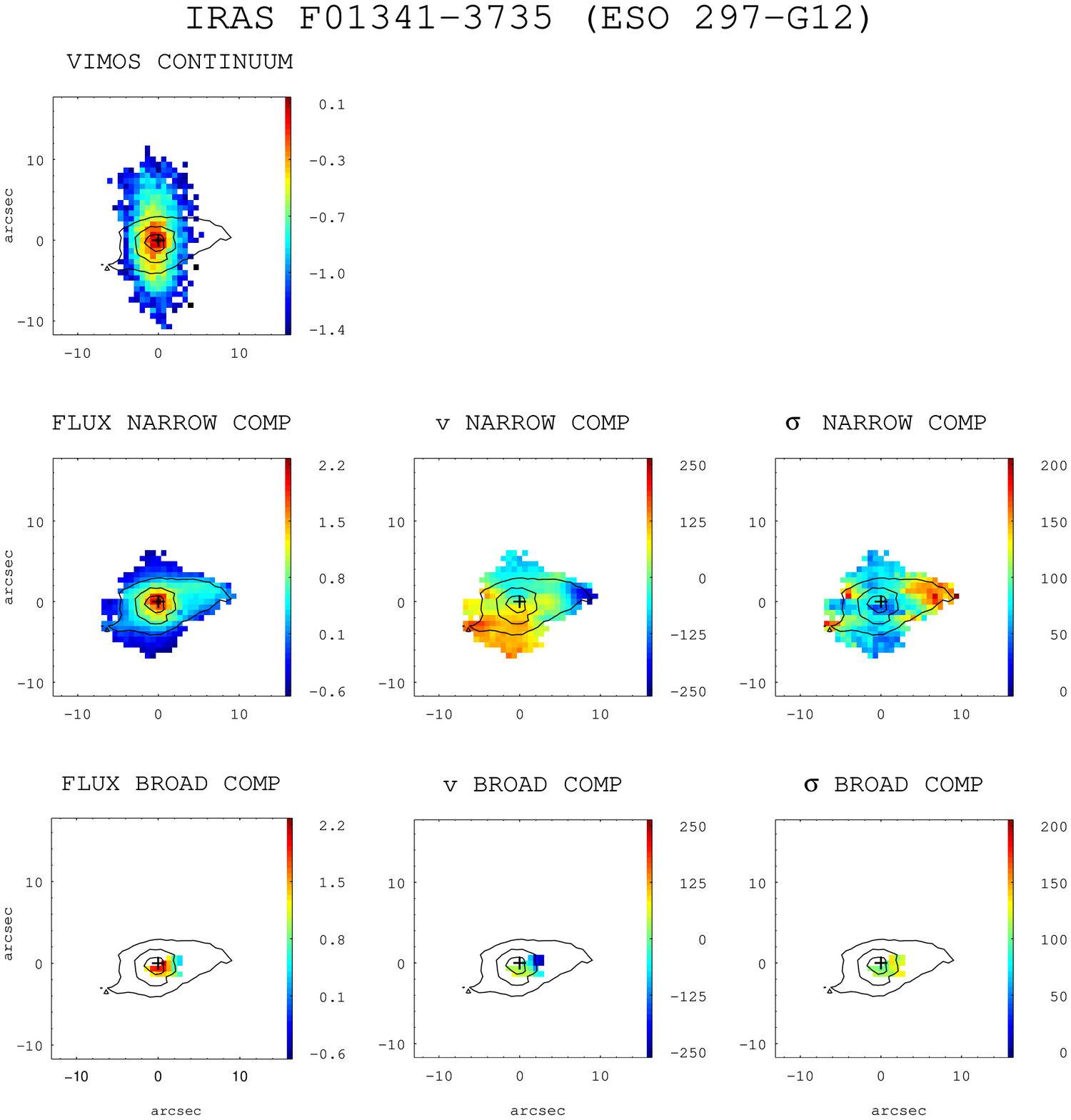}
\vspace{2mm}
\caption{(General comments about the panels as in Fig. A.1.). ESO 297-G012: this is the southern galaxy of the system IRAS F01341-3735. The H$\alpha$ emission is oriented perpendicular to the major stellar axis as traced by the continuum image. Although a rotation component is visible in the velocity field along the major photometrical axis, its structure, as well as that of the velocity dispersion map, is irregular. In an inner region ($\sim$ 1 kpc$^2$), the spectra have been fitted with two components, with the broad component blueshifted by $\sim$ 50 km s$^{-1}$. The scale is 0.352 kpc/$^{\prime\prime}$.}
\label{all_panels}
\end{figure*}

\begin{figure*}
\vspace{0cm}
\includegraphics[width=1.\textwidth, height=1\textwidth]{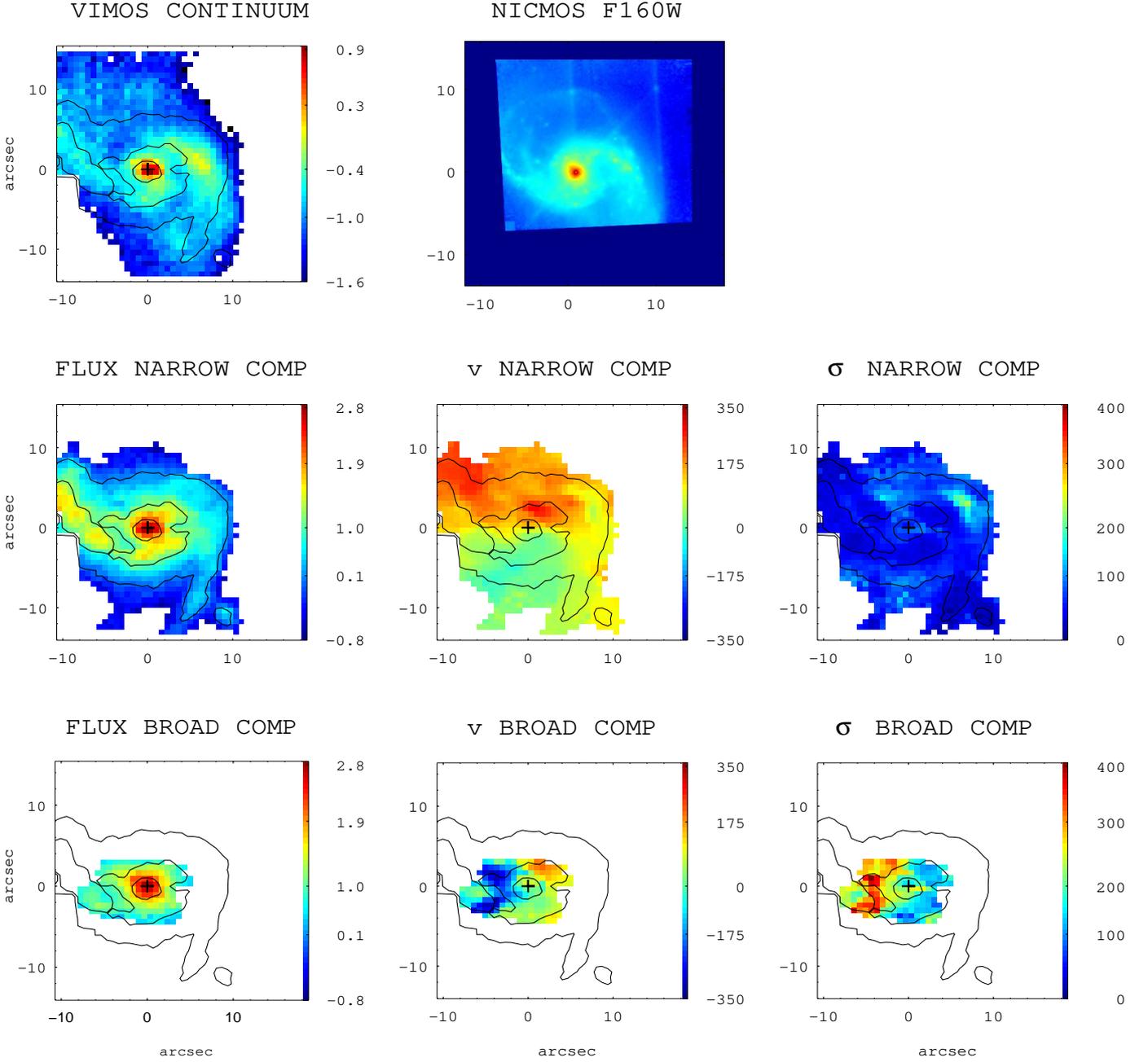}
\vspace{2mm}
\caption{(General comments about the panels as in Fig. A.1.) IRAS F04315-0840 (NGC 1614): this is a well-studied, post-coalescence late merger, with a bright, spiral structure of a length scale of few kpcs. The velocity field of the narrow component is somewhat distorted and chaotic with an amplitude of 324 km s$^{-1}$. Its velocity dispersion map shows an offset peak of $\sim$ 220 km s$^{-1}$ at around 2.4 kpc from the nucleus in the western arm. The spectra in the inner regions are complex, and a secondary broad component covers a relatively large area of about 8.6 kpc$^2$. The projection of the kinematic major axes of the narrow (main) and broad components differs by $\sim$ 90$^\circ$. The blueshifted region of the velocity field of the broad component has the largest velocity dispersion (i.e., $\sim$400 km s$^{-1}$). All this supports the hypothesis of a dusty outflow, where the receding components (which are behind the disk) are obscured, making the whole profile relatively narrow with respect to the approaching component. This object has been analyzed in \cite{YO2012}. The spatial scale is 0.325 kpc/$^{\prime\prime}$.}
\label{all_panels}
\end{figure*}

\begin{figure*}
\vspace{0cm}
\includegraphics[width=1.\textwidth, height=1\textwidth]{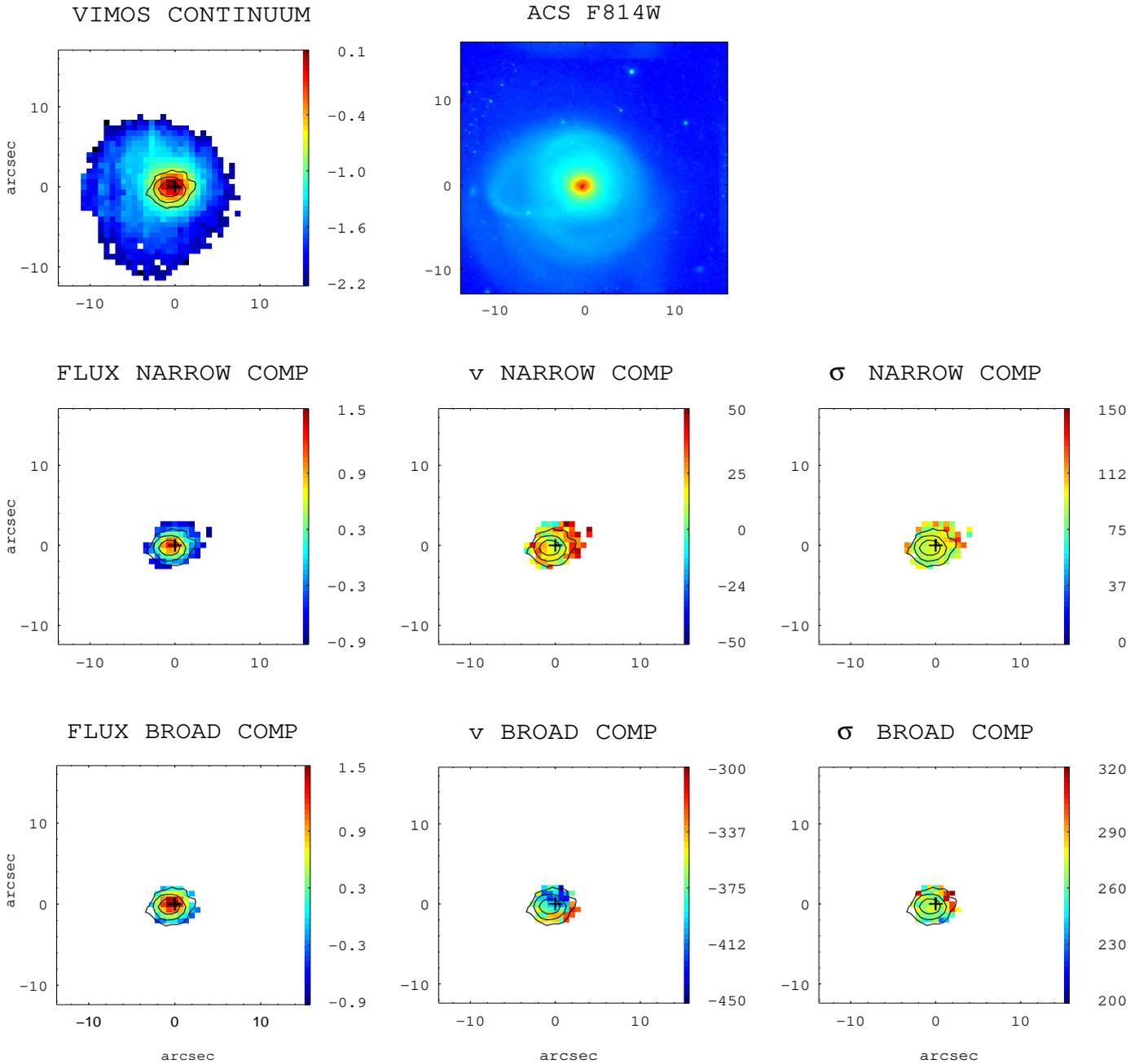} 
\vspace{2mm}
\caption{(General comments about the panels as in Fig. A.1.) IRAS F05189-2524: this galaxy shows a compact H$\alpha$ emission, both in the narrow and broad components.Their velocity fields and velocity dispersion maps do not show regular patterns. The broad component is blueshifted up to $\sim$ 400 km s$^{-1}$ with respect to the narrow one and is dominant in terms of flux in the innermost spaxels. The scale is of 0.839 kpc/$^{\prime\prime}$.}
\label{all_panels}
\end{figure*}

\begin{figure*}
\vspace{0cm}
\includegraphics[width=1.\textwidth, height=1\textwidth]{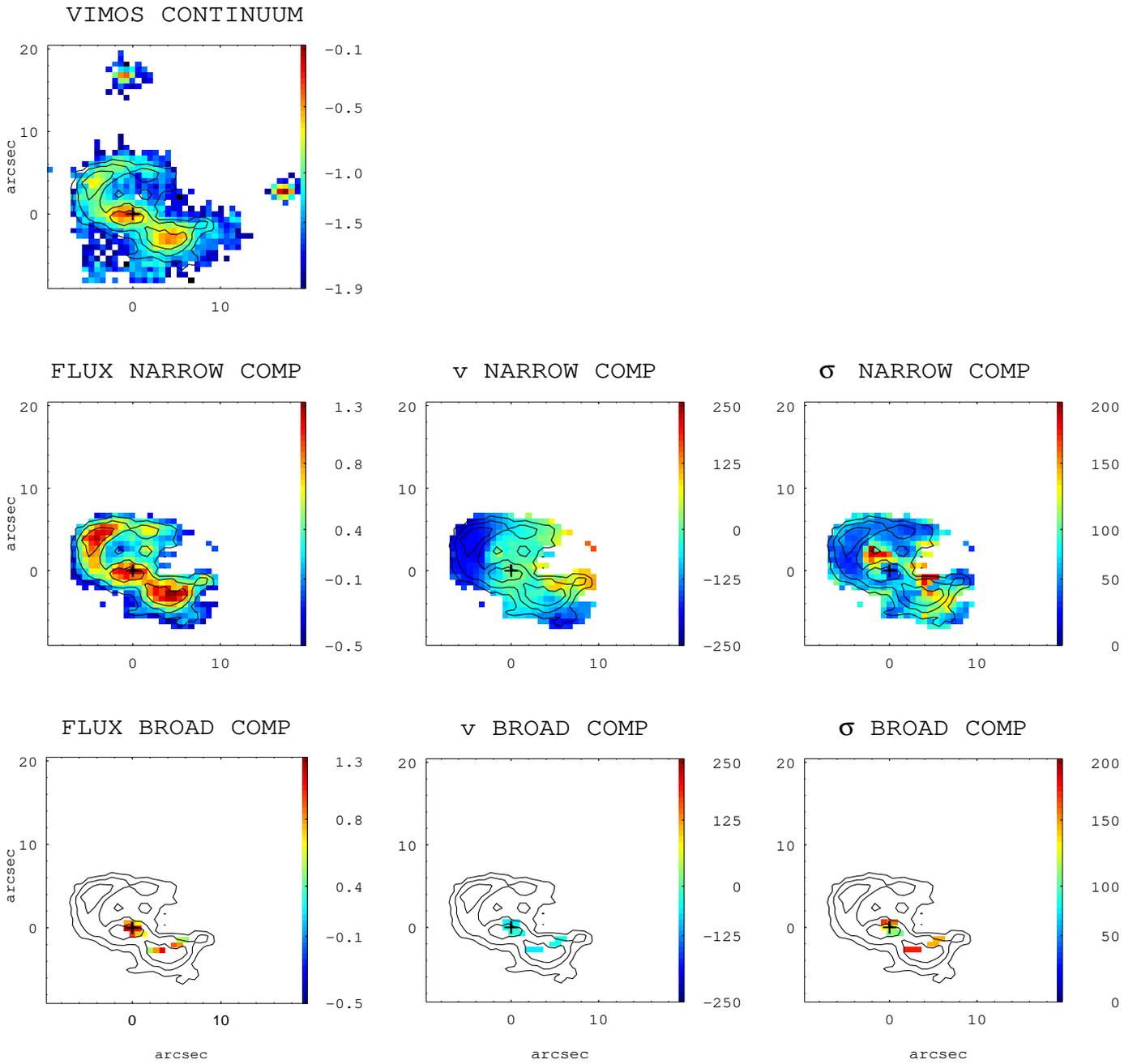}
\vspace{2mm}
\caption{(General comments about the panels as in Fig. A.1.) IRAS F06035-7102: this system consists of two pre-coalescence galaxies with a nuclear separation of $\sim$ 8 kpc. The velocity dispersion map of the narrow component clearly shows an irregular pattern, in which the two local maxima correspond to low H$\alpha$ surface brightness regions. The spatial scale is 1.5 kpc/$^{\prime\prime}$.} 
\label{all_panels}
\end{figure*}

\begin{figure*}
\vspace{0cm}
\includegraphics[width=1.\textwidth, height=1\textwidth]{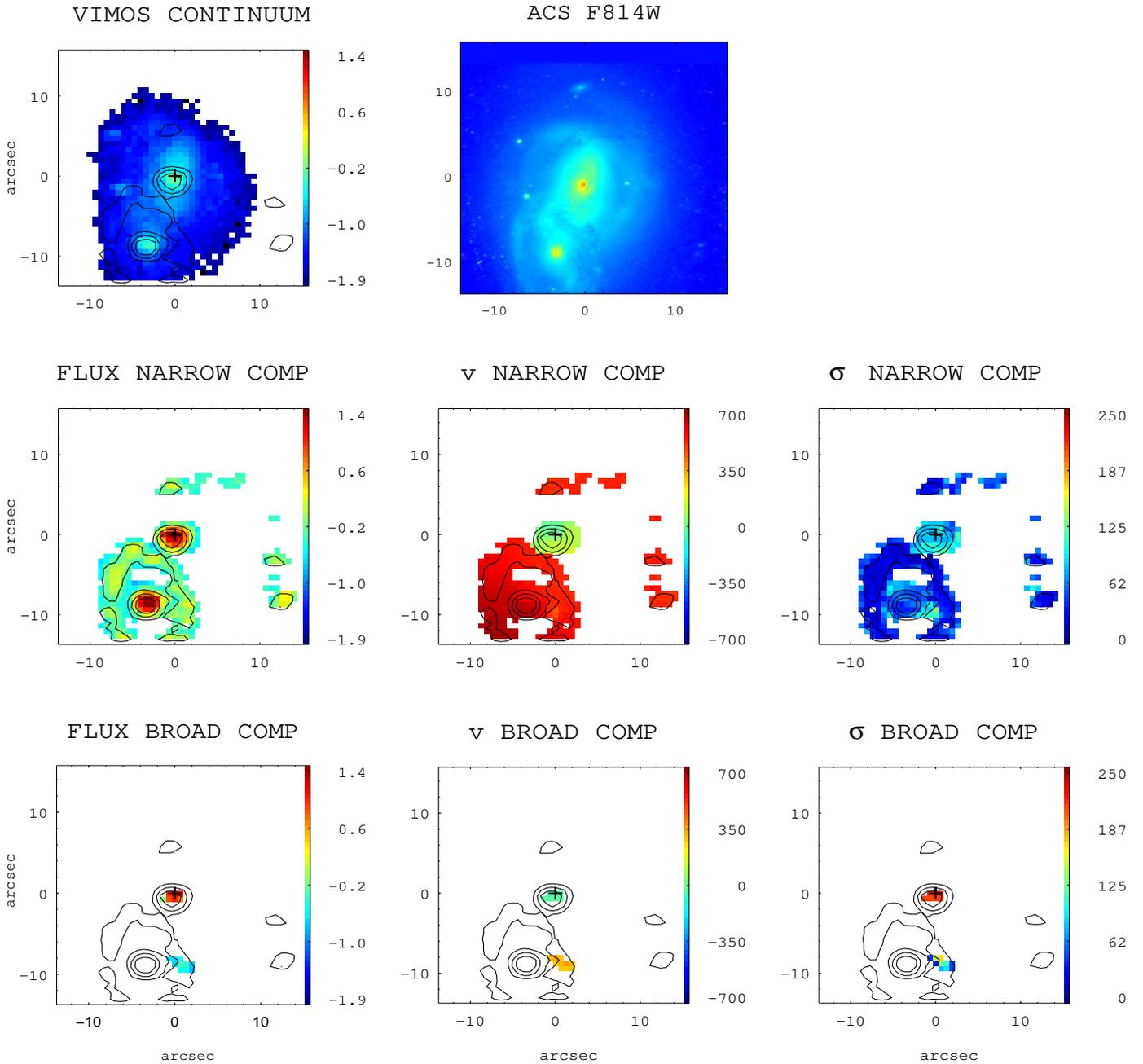}
\vspace{2mm}
\caption{(General comments about the panels as in Fig. A.1.) IRAS F06076-2139: this system consists of two galaxies in interaction with a rather complex morphology. The H$\alpha$ image shows a rather different structure than that shown by the VIMOS continuum and the HST-ACS image. The southern galaxy is characterized by a ring and a tidal tail-like structures extending towards the west and northwest part from the nucleus. The faint broad component found in a small region of the ring in the southern object is likely due to the superposition of the emission of the two galaxies along the line-of-sight. This system was already studied in \cite{A08}. The scale is 0.743 kpc/$^{\prime\prime}$. }
\label{all_panels}
\end{figure*}

\begin{figure*}
\vspace{0cm}
\includegraphics[width=1.\textwidth, height=1\textwidth]{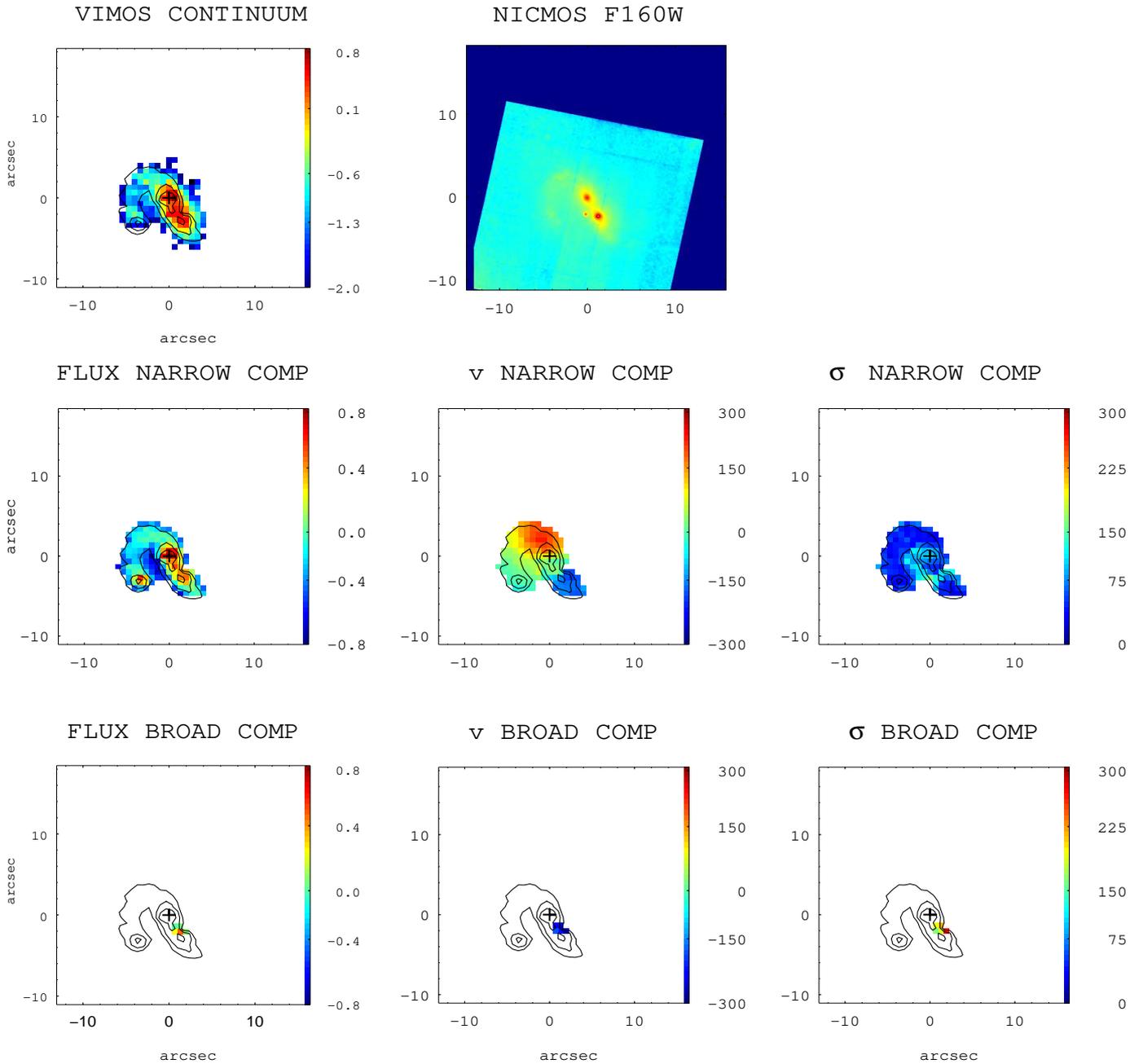}
\vspace{2mm}
\caption{(General comments about the panels as in Fig. A.1.) IRAS F06206-6315: this system has a double nuclei separated by $\sim$ 4 kpc, as clearly seen in the the Near Infrared Camera and Multi-Object Spectrometer (NICMOS/HST) image. There is a tidal tail starting in the north and bending towards the southeast, which contains a local peak of H$\alpha$ emission (i.e., a possible tidal dwarf galaxy candidate). The brightest nucleus seems to be in positional agreement with the kinematic center. The velocity field is regular, and most of the spectra are well fitted using one component. The scale is of 1.72 kpc/$^{\prime\prime}$.}
\label{all_panels}
\end{figure*}

\begin{figure*}
\vspace{0cm}
\includegraphics[width=1.\textwidth, height=1\textwidth]{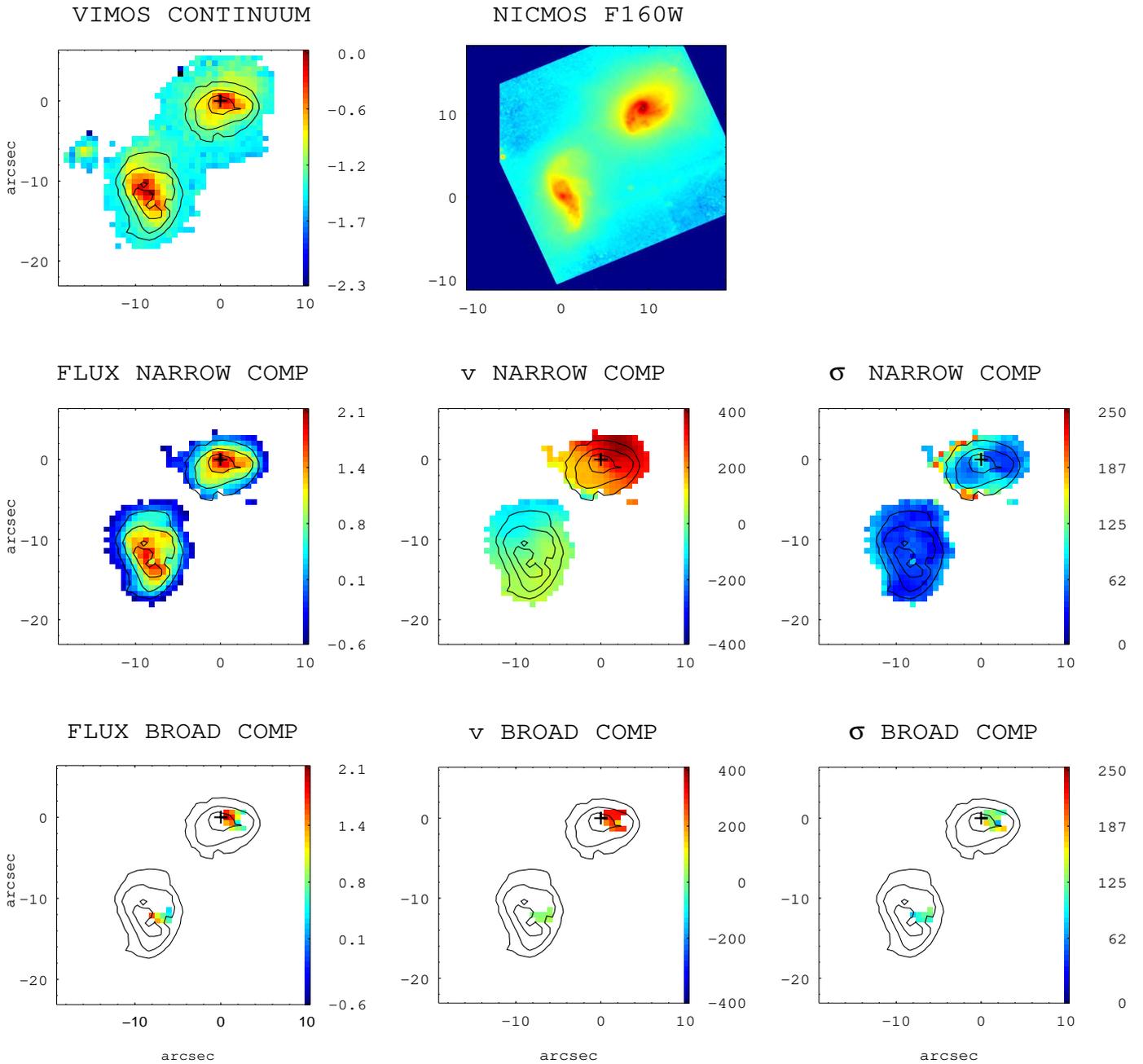}
\vspace{2mm}
\caption{(General comments about the panels as in Fig. A.1.) IRAS F06259-4708 (ESO 255-IG007): two VIMOS pointings were used during the observation of this triple system. The two brightest galaxies, the northern and the central ones, are separated by a distance of $\sim$11 kpc, while the third one is located at $\sim$ 13 kpc to the south from the central galaxy (see next panel). The present panels correspond to the northern and central galaxies, which show a regular velocity field pattern. The scale is of 0.769 kpc/$^{\prime\prime}$.}
\label{all_panels}
\end{figure*}

\begin{figure*}
\vspace{0cm}
\includegraphics[width=1.\textwidth, height=1\textwidth]{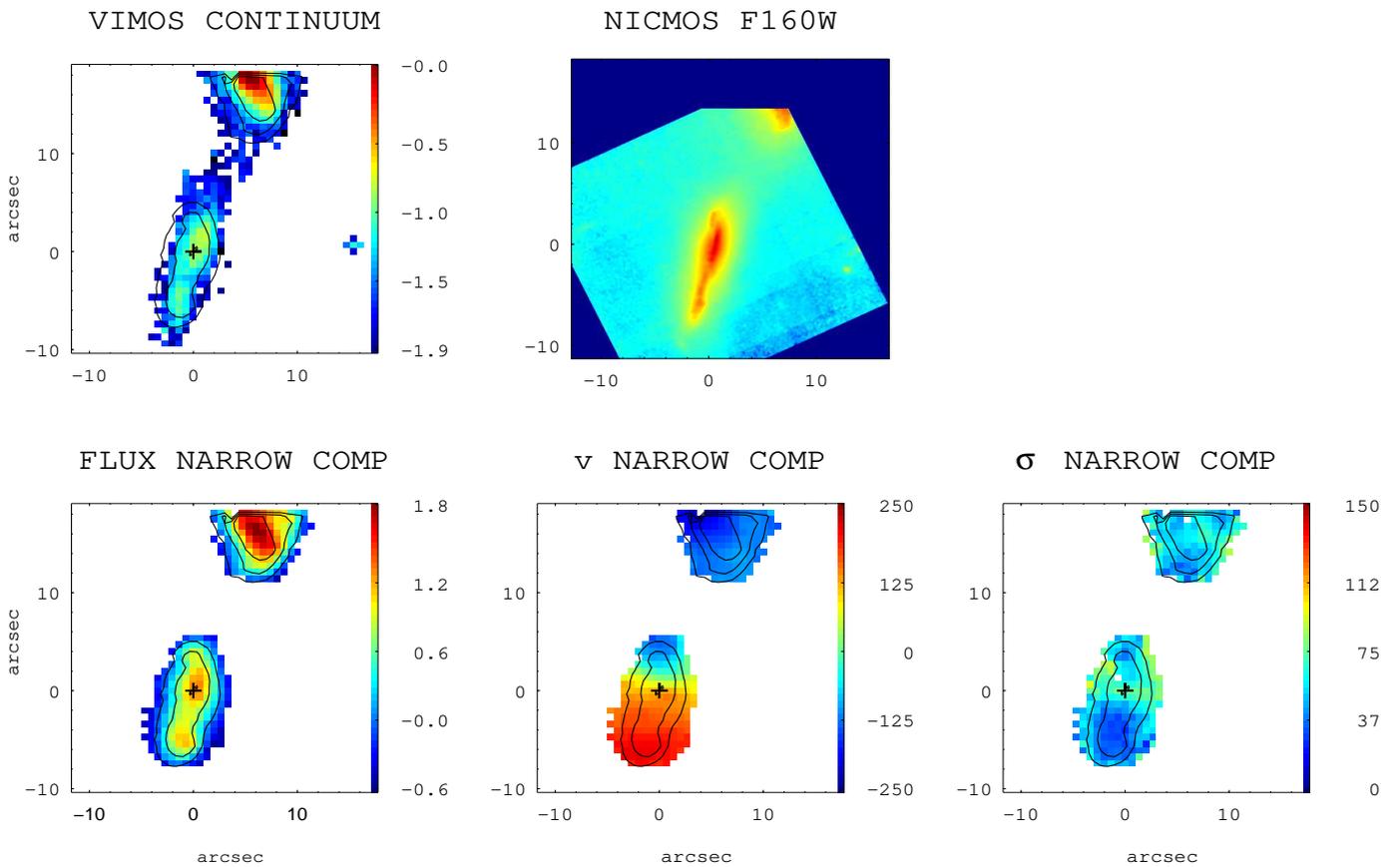}
\vspace{2mm}
\caption{(General comments about the panels as in Fig. A.1.) ESO 255-IG007: this is the southern galaxy of the system IRAS F06259-4708. It has a regular velocity field and the velocity dispersion map. One single Gaussian component is adequate to fit the H$\alpha$-[NII] emission lines of this galaxy. The scale is 0.769 kpc/$^{\prime\prime}$.}
\label{all_panels}
\end{figure*}

\begin{figure*}
\vspace{0cm}
\includegraphics[width=1.\textwidth, height=1\textwidth]{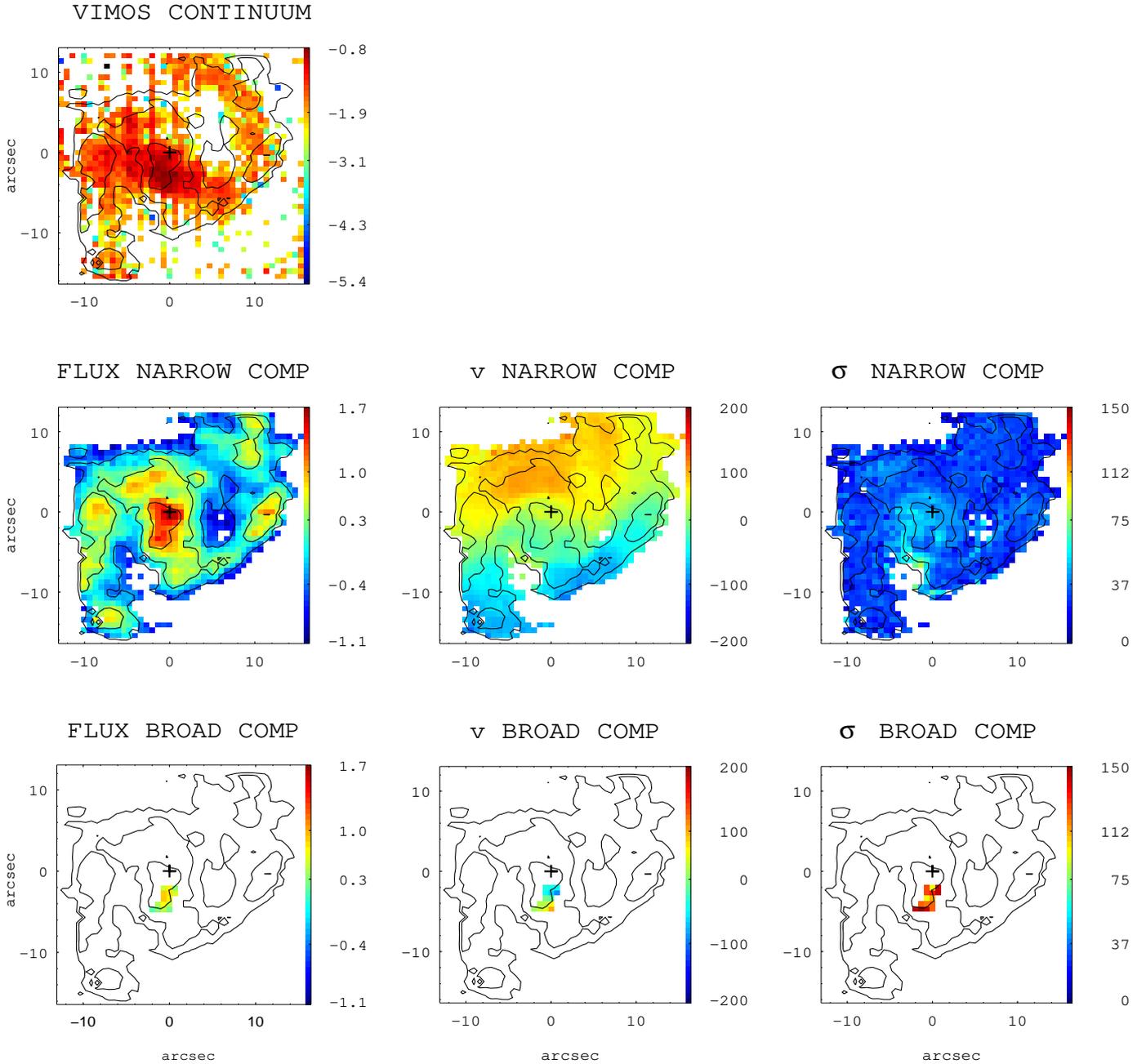}
\vspace{2mm}
\caption{(General comments about the panels as in Fig. A.1.) IRAS F06295-1735 (ESO 557-G002): the continuum image of this barred spiral shows vertical strips, which were not possible to remove during the reduction process (see \citealt{RZ11}). However, the H$\alpha$ maps are not affected by this problem. Interestingly, neither the arms nor the bar in the H$\alpha$ image coincide with those in the continuum image. The velocity field shows a regular structure, while its velocity dispersion map reaches the highest values (i.e., $\sigma \sim$ 70-80 km s$^{-1}$) at about 2 kpc to the south of the nucleus, in a region of relatively low emission. The scale is of 0.431 kpc/$^{\prime\prime}$.}
\label{all_panels}
\end{figure*}

\begin{figure*}
\vspace{0cm}
\includegraphics[width=1.\textwidth, height=1\textwidth]{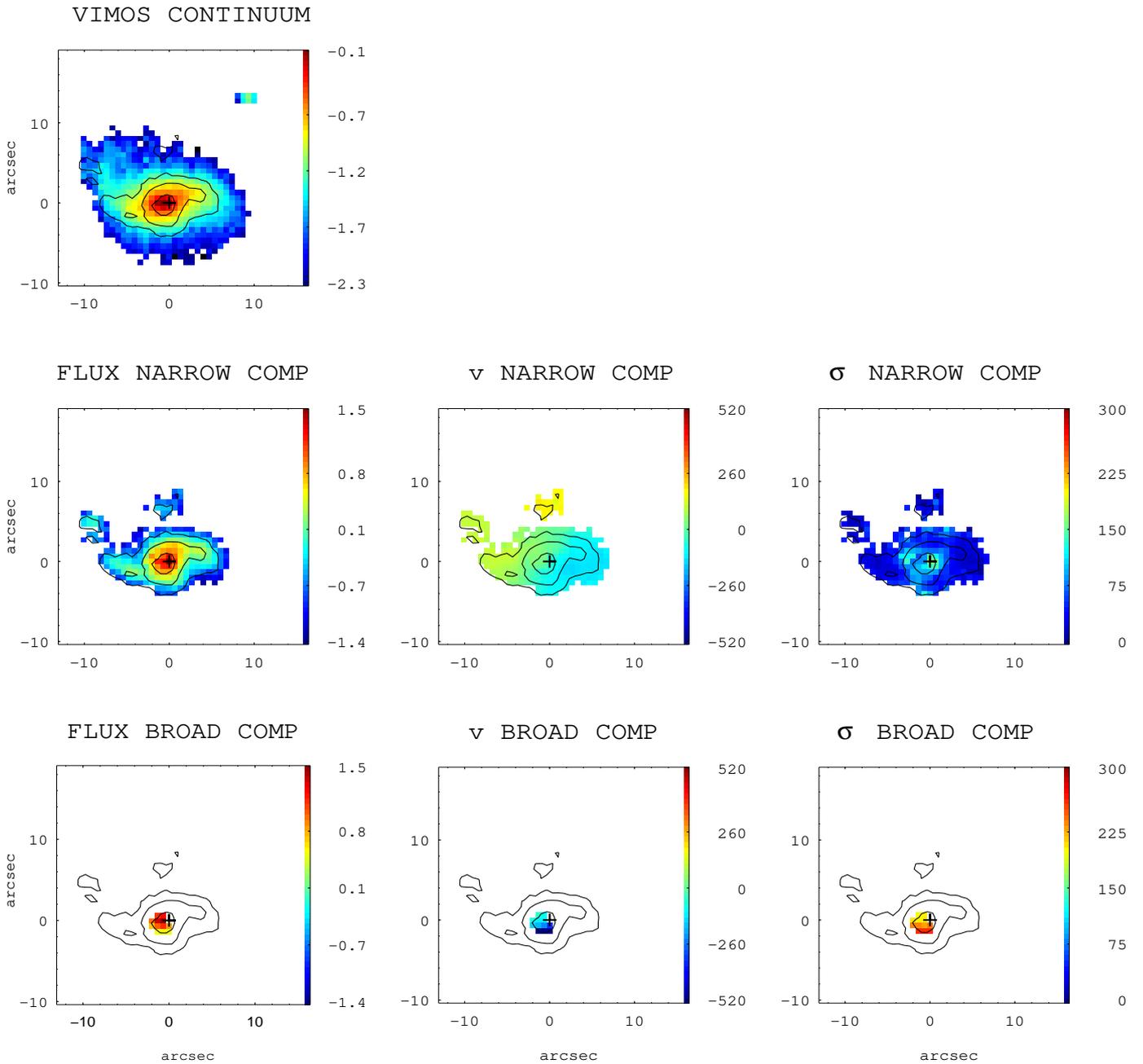}
\vspace{2mm}
\caption{(General comments about the panels as in Fig. A.1.) IRAS 06592-6313: this spiral galaxy shows a minor kinematic axis not aligned with the corresponding photometric one. Its velocity field shows a regular pattern with an amplitude of 160 km s$^{-1}$, while the velocity dispersion map shows some structure around the nucleus. A broad component blueshifted up to $\sim$ 300 km s$^{-1}$ has been found in the inner regions. The scale is of 0.464 kpc/$^{\prime\prime}$. }
\label{all_panels}
\end{figure*}

\clearpage

\begin{figure*}
\vspace{0cm}
\includegraphics[width=1.\textwidth, height=1\textwidth]{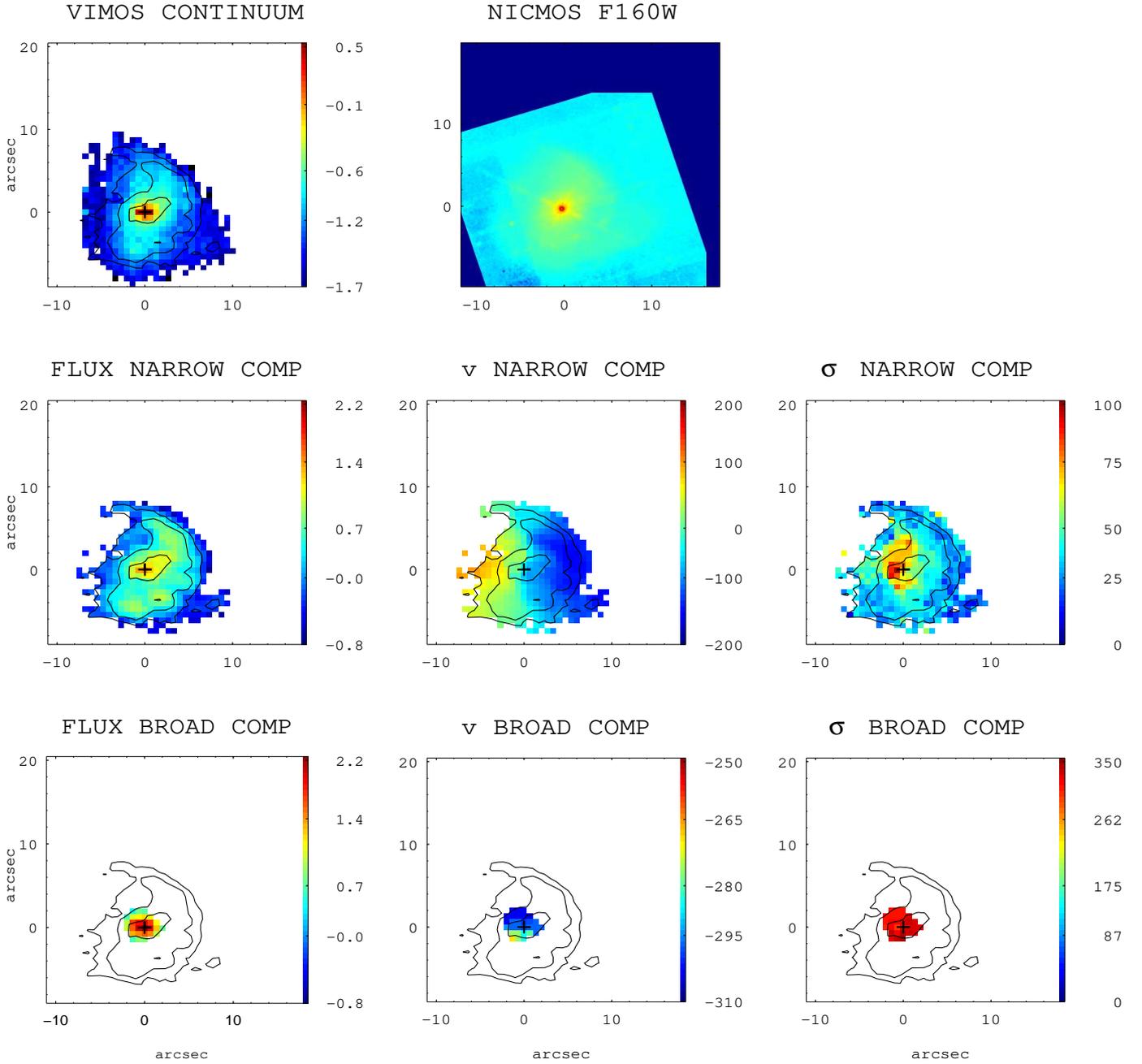}
\vspace{2mm}
\caption{(General comments about the panels as in Fig. A.1.) IRAS F07027-6011 (AM 0702-601): this system consists of two galaxies separated by $\sim$ 54 kpc. The present maps correspond to the northern galaxy. This object has a regular velocity field, while the velocity dispersion map shows a local maximum offset by $\sim$ 0.4 kpc with respect to the nucleus (or H$\alpha$ flux peak). The secondary component found in the inner part shows a $\sigma_{mean}$ of about 320 km s$^{-1}$, which can be explained by the fact that this galaxy possibly hosts an AGN (see \citealt{A12}). The spatial scale is of 0.626 kpc/$^{\prime\prime}$. }
\label{all_panels}
\end{figure*}

\begin{figure*}
\vspace{0cm}
\includegraphics[width=1.\textwidth, height=1\textwidth]{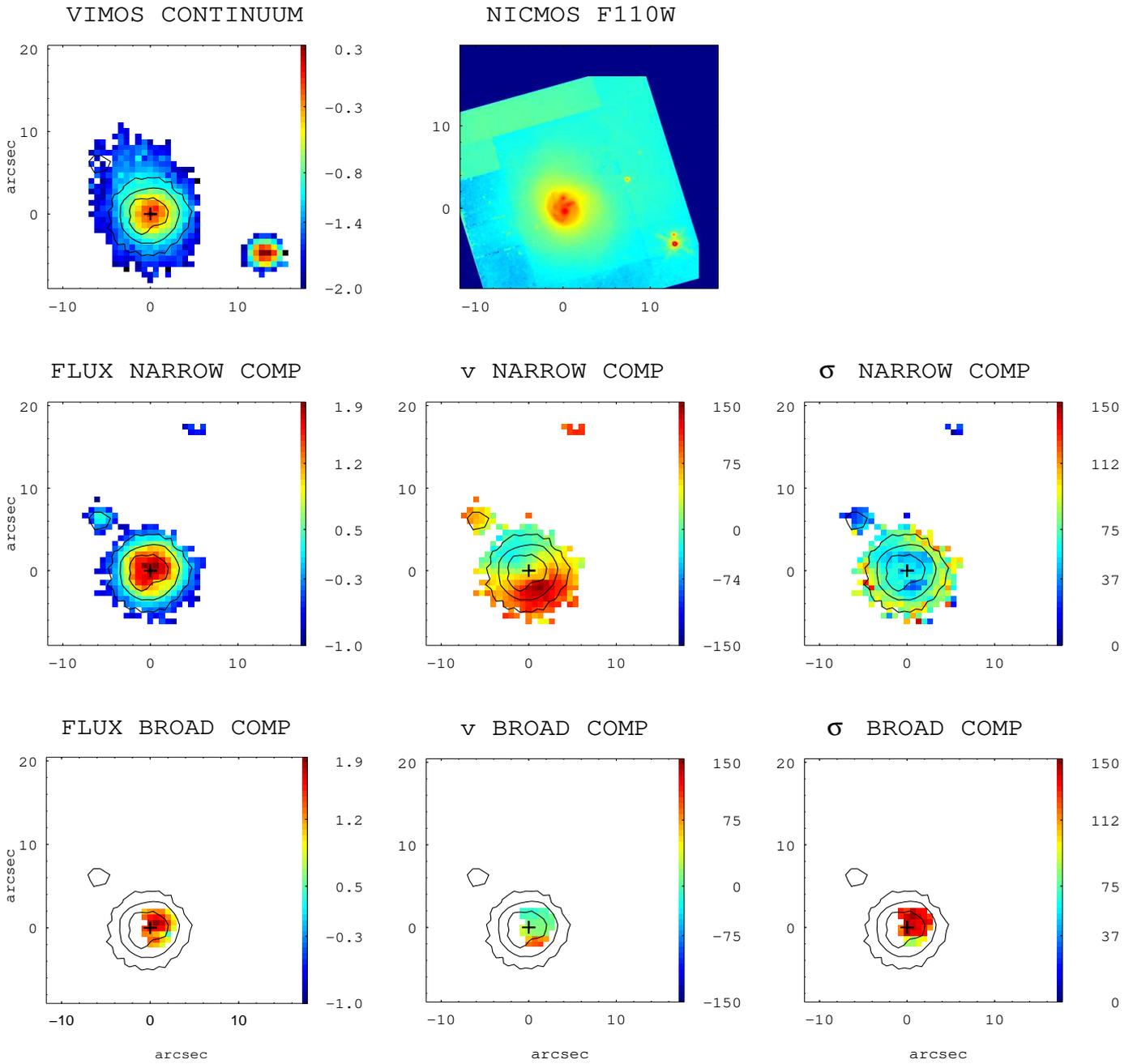}
\vspace{2mm}
\caption{(General comments about the panels as in Fig. A.1.) IRAS F07027-6011 (AM 0702-601): these maps correspond to the southern galaxy of the system. The velocity field of the narrow component is rather regular. In the velocity dispersion map a local maximum is found in correspondence of the nucleus (or H$\alpha$ peak). The spatial scale is of 0.626 kpc/$^{\prime\prime}$.}
\label{all_panels}
\end{figure*}

\begin{figure*}
\vspace{0cm}
\includegraphics[width=1.\textwidth, height=1\textwidth]{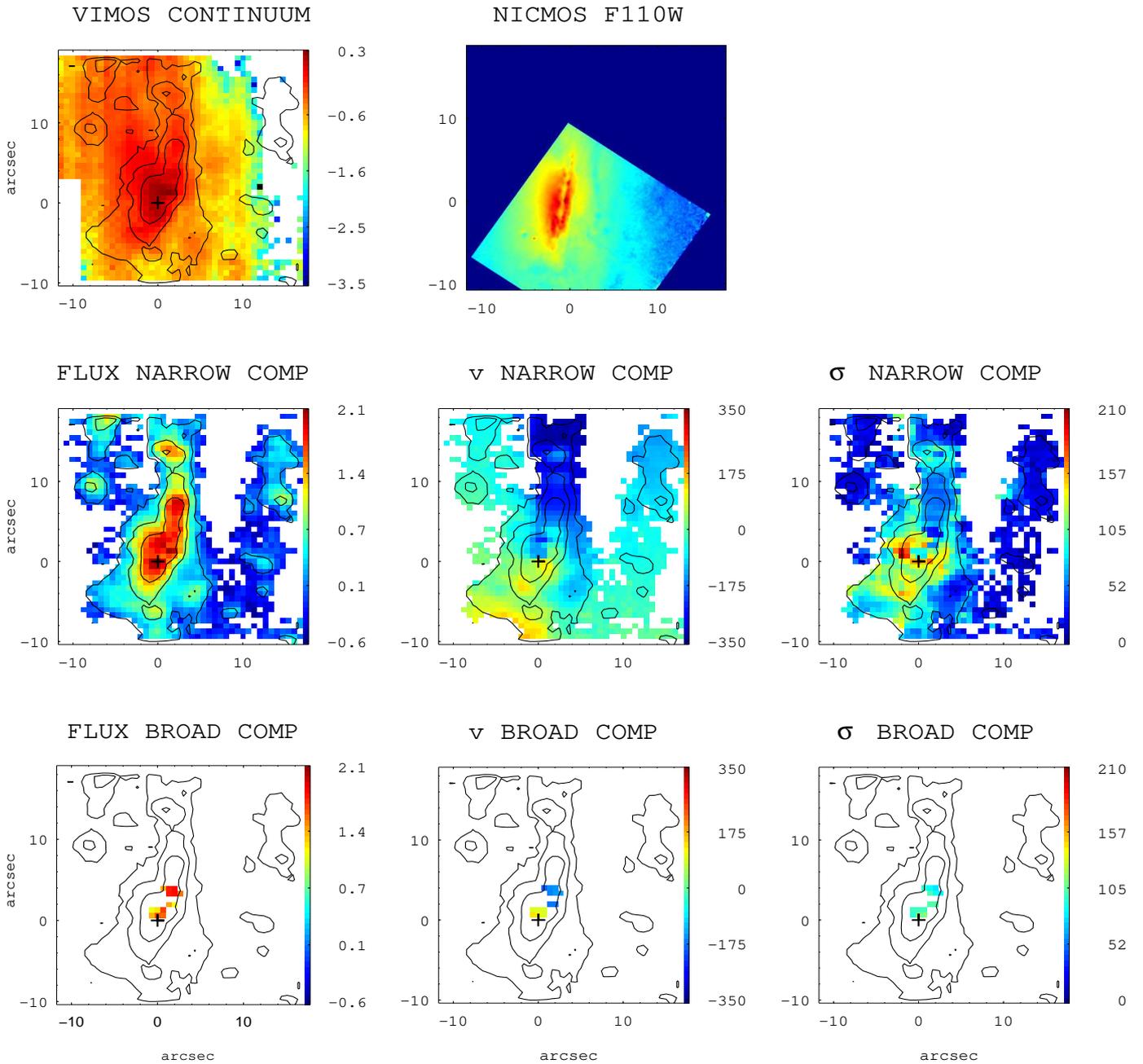}
\vspace{2mm}
\caption{(General comments about the panels as in Fig. A.1.) IRAS F07160-6215 (NGC 2369): this is an edge-on dusty galaxy, whose structure could strongly affect the pattern of its velocity field and velocity dispersion map. Indeed, the velocity field has a clear rotation component with some irregularities, and the velocity dispersion map shows distorted inner ($\sim$ 1 kpc) regions, with values ranging between 120 and 200 km s$^{-1}$, where a broad component has been considered to properly fit the spectra. The scale is 0.221 kpc/$^{\prime\prime}$.}
\label{all_panels}
\end{figure*}

\begin{figure*}
\includegraphics[width=1.\textwidth, height=1\textwidth]{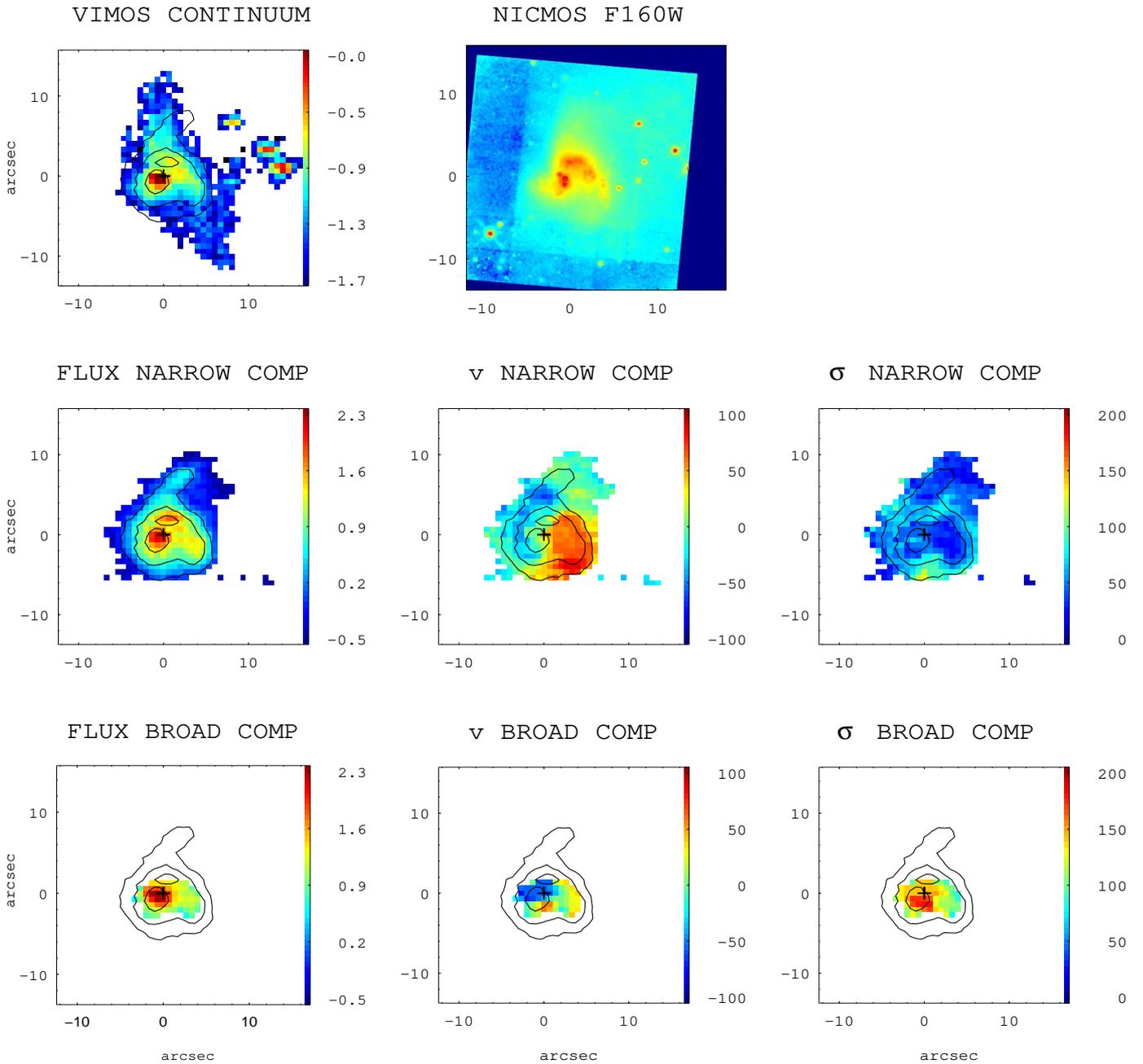}
\vspace{2mm}
\caption{(General comments about the panels as in Fig. A.1.) IRAS 08355-4944: this galaxy is morphologically classified as a post-coalescence merger. It has a quite regular velocity pattern and a relatively disturbed structure in the velocity dispersion map. Most of the spectra in the inner region have been fitted applying 2-Gaussian fit. The spatial scale is 0.521 kpc/$^{\prime\prime}$.}
\label{all_panels}
\end{figure*}

\begin{figure*}
\vspace{0cm}
\includegraphics[width=1.\textwidth, height=1\textwidth]{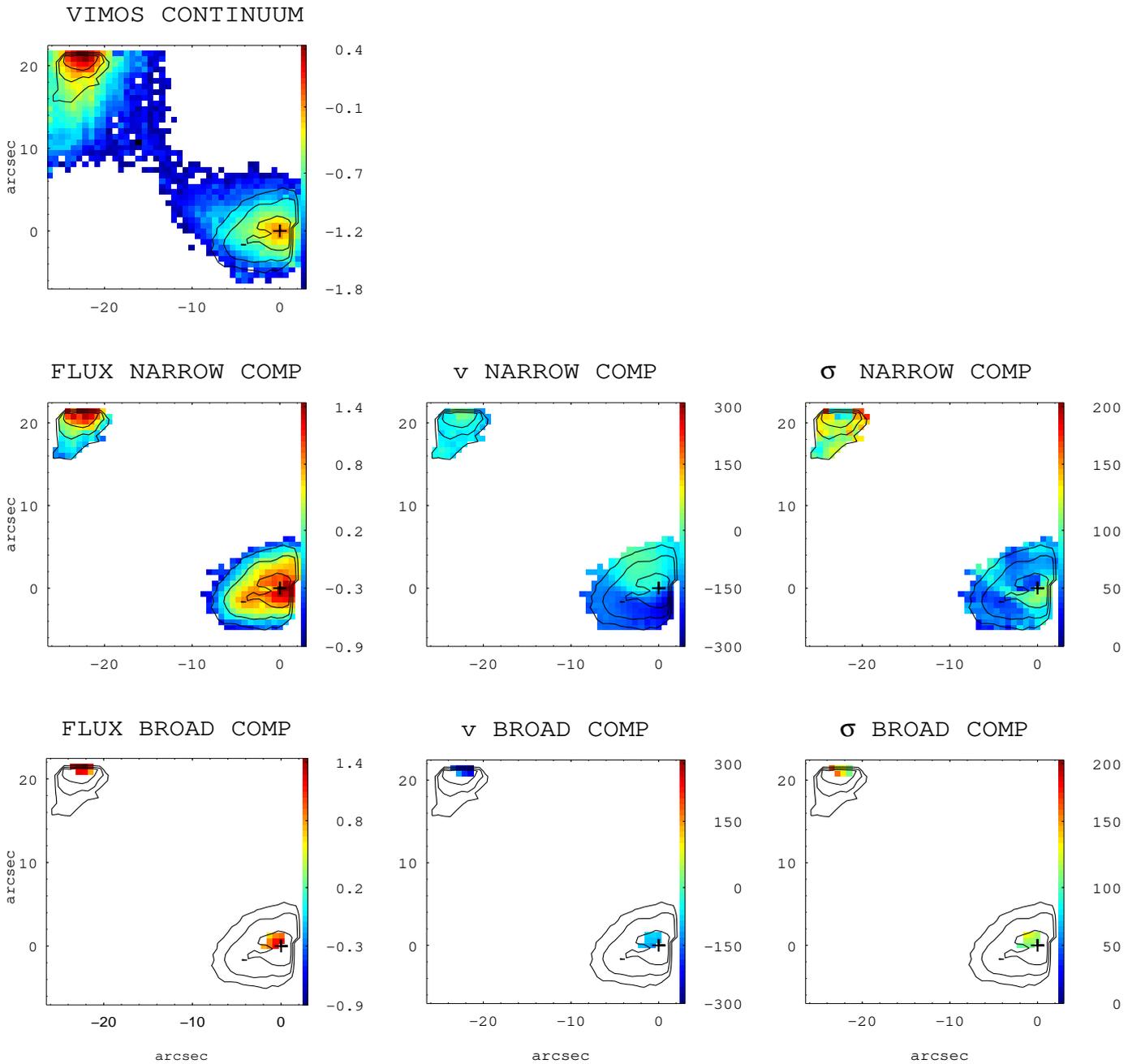}
\vspace{2mm}
\caption{(General comments about the panels as in Fig. A.1.) IRAS 08424-3130 (ESO 432-IG006): the Digital Sky Survey (DSS) image shows a pair of spiral galaxies (nuclear separation of $\sim$ 9 kpc) in interaction. Only part of the nuclear region of both galaxies is covered by our VIMOS FoV. The area covered by the H$\alpha$ emission line for the northern galaxy is very small, so that its kinematical classification was not possible. The southern object reveals a relatively regular velocity field with an amplitude of 133 km s$^{-1}$ and an almost centrally peaked velocity dispersion map. The spatial scale is 0.329 kpc/$^{\prime\prime}$.}  
\label{all_panels}
\end{figure*}

\begin{figure*}
\vspace{0cm}
\includegraphics[width=1.\textwidth, height=1\textwidth]{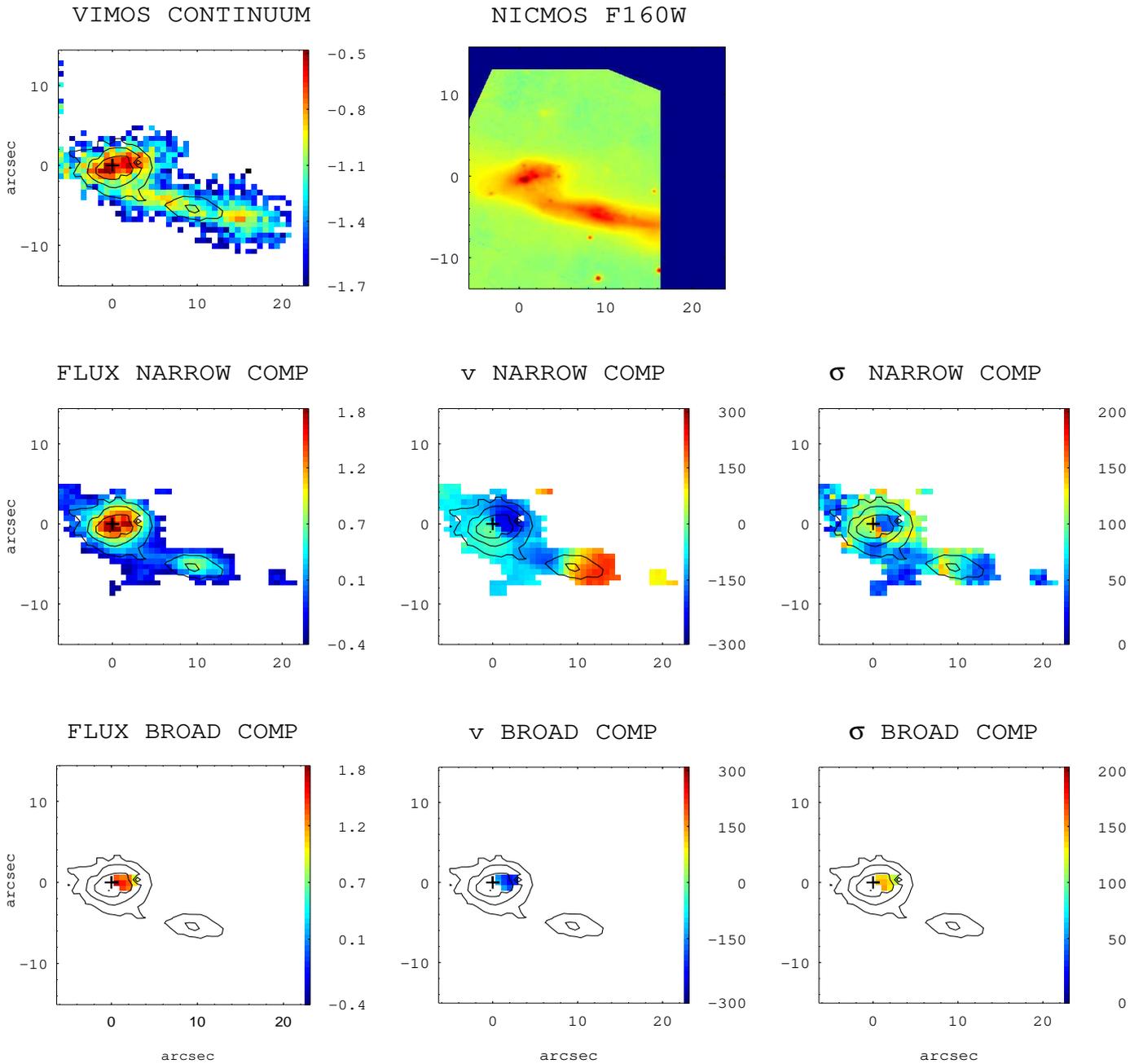}
\vspace{2mm}
\caption{(General comments about the panels as in Fig. A.1.) IRAS F08520-6850 (ESO 60-IG016): this system is composed of two disk galaxies in interaction with a nuclear separation of 10 kpc. The H$\alpha$ flux peak and the kinematic center of the eastern source are in positional agreement. The velocity fields of the respective galaxies are regular. The spatial scale is 0.909 kpc/$^{\prime\prime}$. }
\label{all_panels}
\end{figure*}

\clearpage

\begin{figure*}
\vspace{0cm}
\includegraphics[width=1.\textwidth, height=1\textwidth]{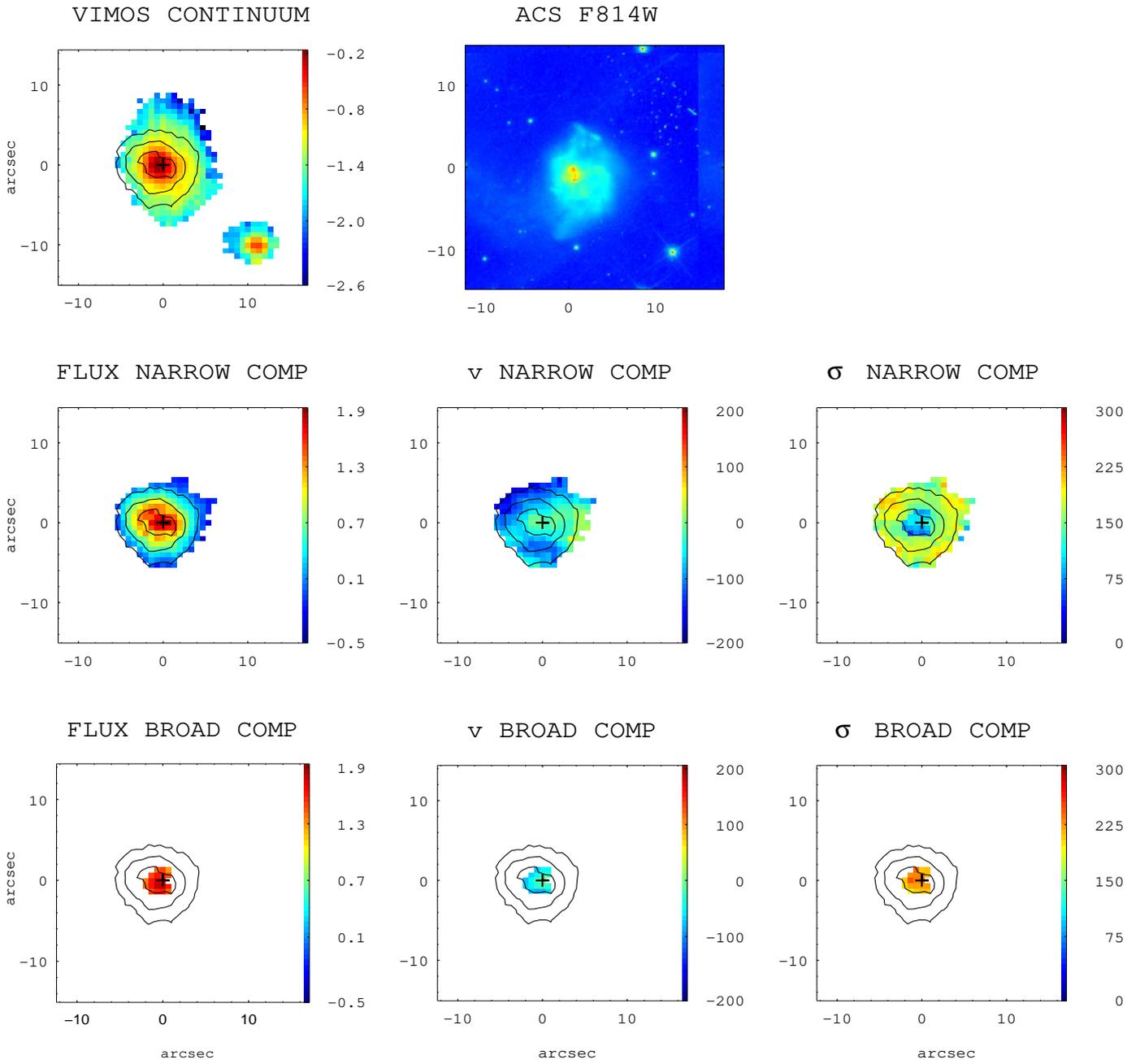}
\vspace{2mm}
\caption{(General comments about the panels as in Fig. A.1.) IRAS F09022-3615: the velocity field is quite distorted and irregular with poorly defined kinematic axes. The ring of high-velocity dispersion around the nucleus has several local peaks. A broad component is present in the inner parts. The spatial scale is of 1.153 kpc/$^{\prime\prime}$.}
\label{all_panels}
\end{figure*}

\begin{figure*}
\vspace{0cm}
\includegraphics[width=1\textwidth, height=0.78\textwidth]{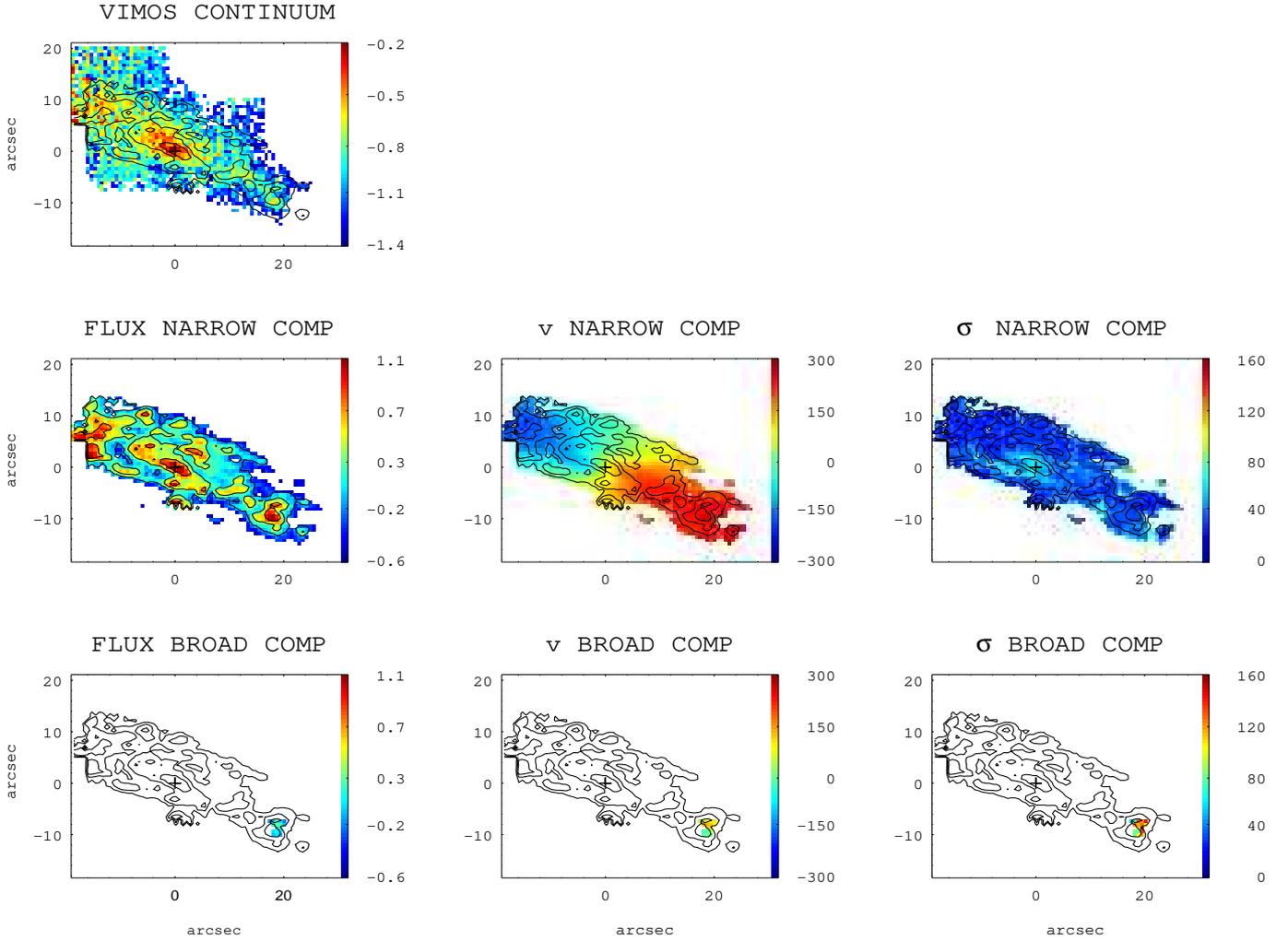}
\vspace{5mm}
\caption{(General comments about the panels as in Fig. A.1.) IRAS F09437+0317 (IC 564/IC 563): this system is formed by two galaxies (north: IC564/ south: IC563) with a nuclear separation of about 39 kpc. Three VIMOS pointings were carried out for observing this system. For the northern galaxy (IC 564), two pointings sample the northeast (NE) and the southwest (SW) parts of the galaxy (see Fig.1 Paper III); both of them  have been used to derived the mean kinematic values (Table \ref{NARROW}). These panels (i.e., $\sim$ 50$^{\prime\prime}$ $\times$ 40$^{\prime\prime}$) were generated by combining the NE and NW pointings. A faint broad component was found in correspondence of an offset H$\alpha$ peak in the SW direction, possibly due to the presence of a star-forming region. The velocity field is regular, and the kinematic center seems to be in positional agreement with the continuum and H$\alpha$ flux peaks. The spatial scale is 0.415 kpc/$^{\prime\prime}$.}
\label{all_panels}
\end{figure*}

\begin{figure*}
\vspace{0cm}
\includegraphics[width=1.\textwidth, height=1\textwidth]{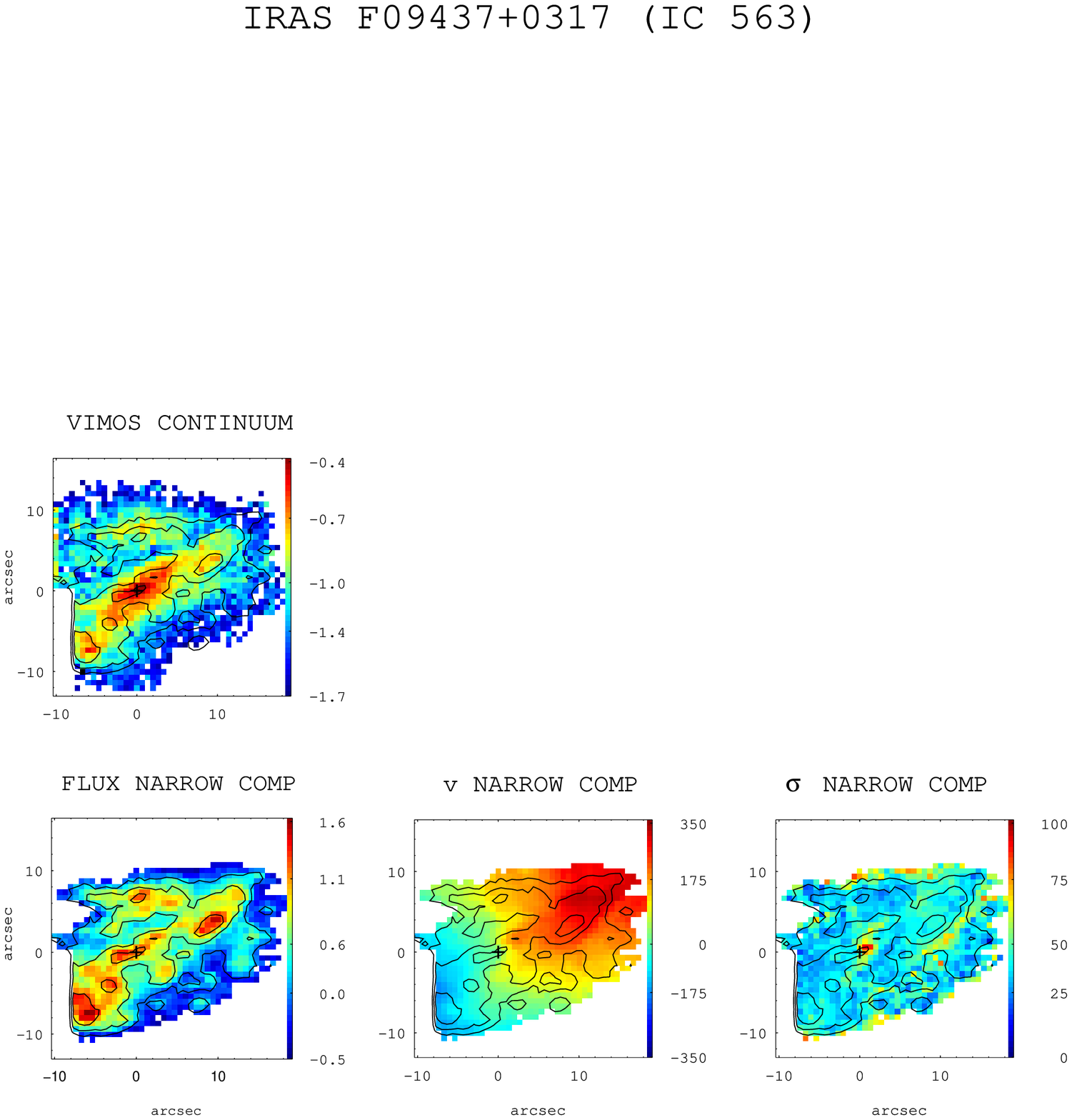}
\vspace{2mm}
\caption{(General comments about the panels as in Fig. A.1.) IRAS F09437+0317 (IC 563): this object shows a very regular velocity field and a centrally peaked velocity dispersion map. The VIMOS continuum image is considered the center of the image since the H$\alpha$ peak is offset due to a knot of star formation and does not properly define the center. The spatial scale is 0.415 kpc/$^{\prime\prime}$.}
\label{all_panels}
\end{figure*}

\begin{figure*}
\vspace{0cm}
\includegraphics[width=1.\textwidth, height=1\textwidth]{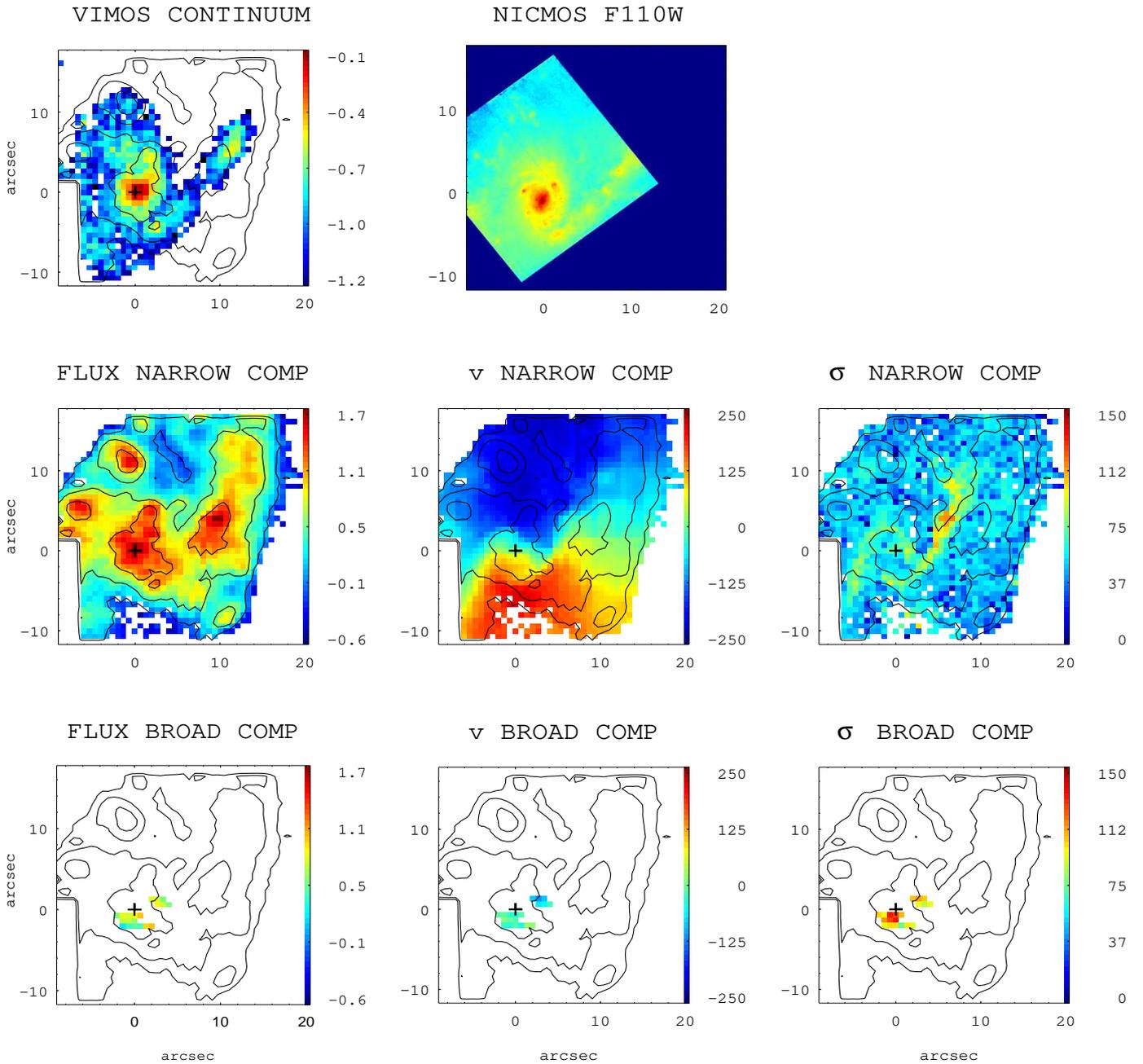}
\vspace{2mm}
\caption{(General comments about the panels as in Fig. A.1.) IRAS F10015-0614 (NGC 3110): this galaxy shows two well-defined spiral arms in NICMOS/HST image. The velocity field and the velocity dispersion maps reproduce the spiral structure of this galaxy. Broad-profile spectra are found in regions of low-H$\alpha$ surface brightness round the nucleus.The spatial scale is 0.343 kpc/$^{\prime\prime}$. }
\label{all_panels}
\end{figure*}

\begin{figure*}
\vspace{0cm}
\includegraphics[width=1.\textwidth, height=1\textwidth]{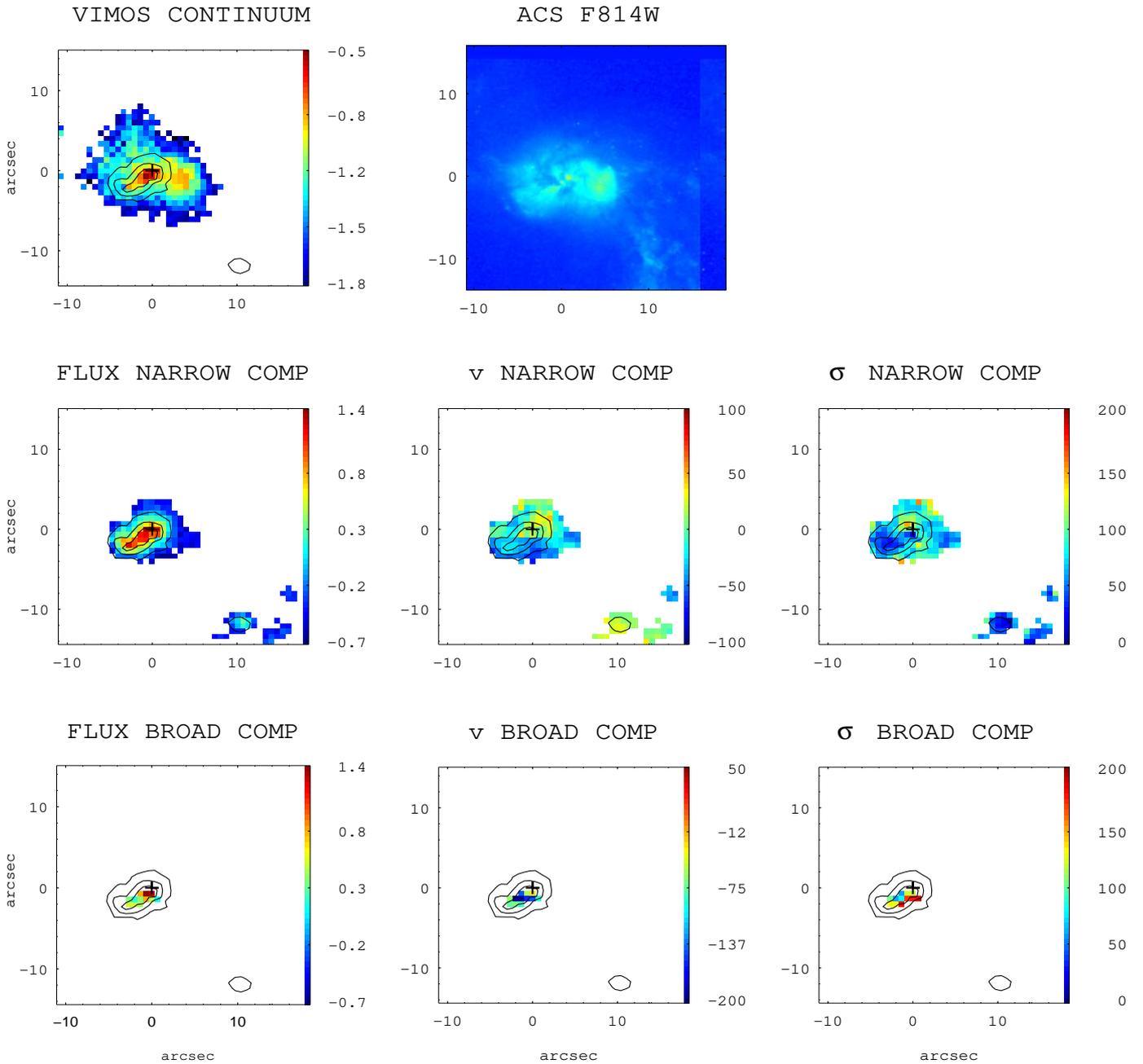}
\vspace{2mm}
\caption{(General comments about the panels as in Fig. A.1.) IRAS F10038-3338 (IC 2545): the VIMOS continuum image shows two nuclei separated by $\sim$ 3 kpc. The H$\alpha$ peak corresponds to the eastern nucleus. The velocity field and velocity dispersion maps are quite disturbed. The broad component is blueshifted by 100 km s$^{-1}$. The scale is of 0.679 kpc/$^{\prime\prime}$.}
\label{all_panels}
\end{figure*}

\begin{figure*}
\vspace{0cm}
\includegraphics[width=1.\textwidth, height=1\textwidth]{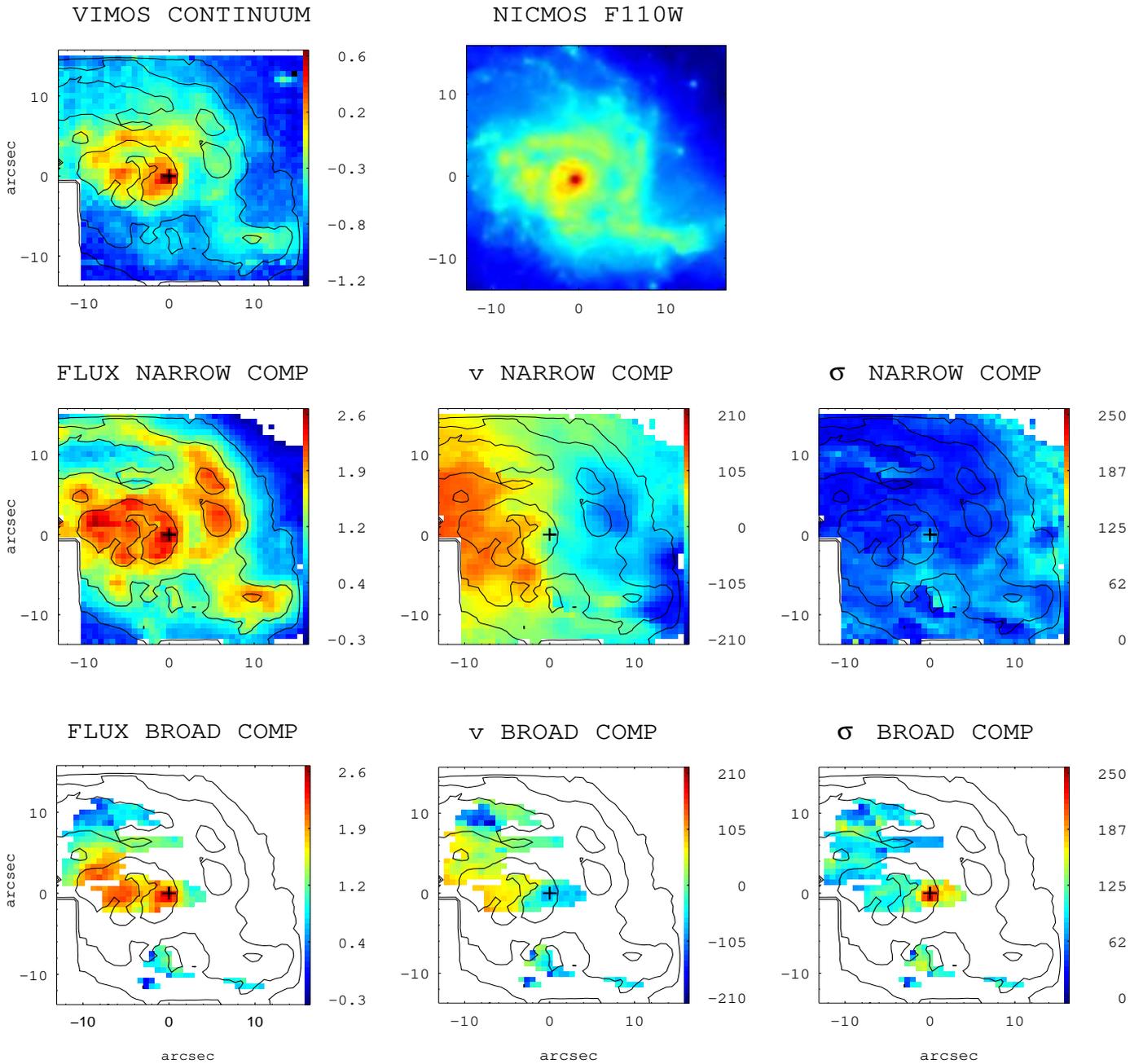}
\vspace{2mm}
\caption{(General comments about the panels as in Fig. A.1.) IRAS F10257-4338 (NGC 3256): this galaxy has a somewhat complex morphological structure. It shows a clumpy pattern in the H$\alpha$ flux intensity map. The velocity field of the narrow component has the kinematic axes not well defined, though a rotation component with an amplitude of 170 km s$^{-1}$ can be identified. The velocity dispersion map shows high values in regions where high ionized gas is found (see \citealt{MI10}). The broad component has been found in a quite large and irregular area. The scale is of 0.192 kpc/$^{\prime\prime}$. }
\label{all_panels}
\end{figure*}

\begin{figure*}
\vspace{0cm}
\includegraphics[width=1.\textwidth, height=1\textwidth]{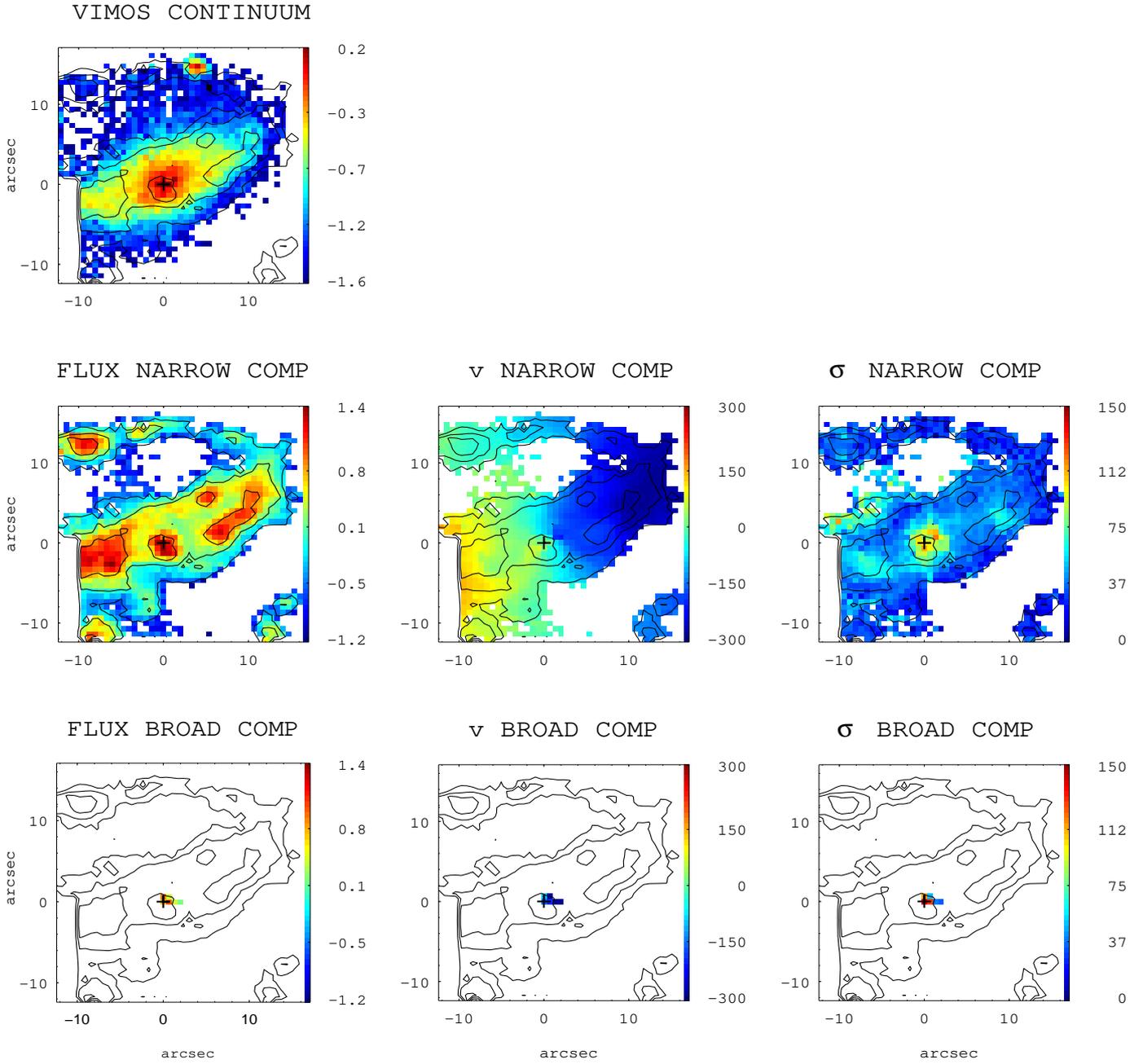}
\vspace{2mm}
\caption{(General comments about the panels as in Fig. A.1.) IRAS F10409-4556 (ESO 264-G036): this is an isolated barred spiral galaxy morphologically classified as type 0 (i.e., {\it isolated disk}). Its velocity field is very regular and the velocity dispersion map is centrally peaked. The spatial scale is 0.425 kpc/$^{\prime\prime}$.}
\label{all_panels}
\end{figure*}

\begin{figure*}
\vspace{0cm}
\includegraphics[width=1.\textwidth, height=1\textwidth]{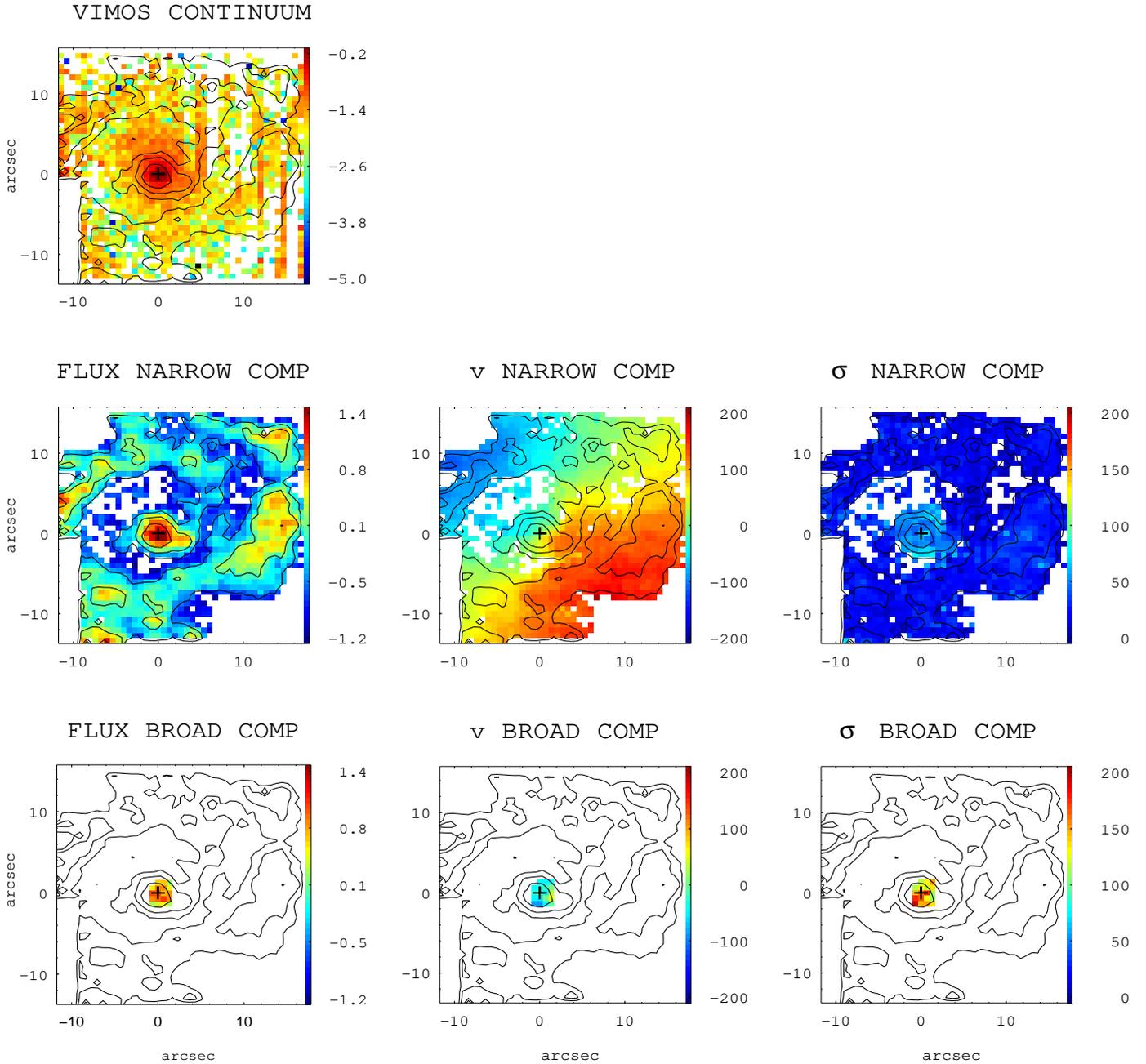}
\vspace{2mm}
\caption{(General comments about the panels as in Fig. A.1.) IRAS F10567-4310 (ESO 264-G057): the continuum image shows vertical patterns, which were not possible to remove during the reduction process (see \citealt{RZ11}). However, the H$\alpha$ maps are not affected by this problem. The narrow component shows a very regular velocity field and a centrally peaked velocity dispersion map. This object has been analyzed in \cite{YO2012}. The scale is 0.35 kpc/$^{\prime\prime}$.}
\label{all_panels}
\end{figure*}

\begin{figure*}
\vspace{0cm}
\includegraphics[width=1.\textwidth, height=1\textwidth]{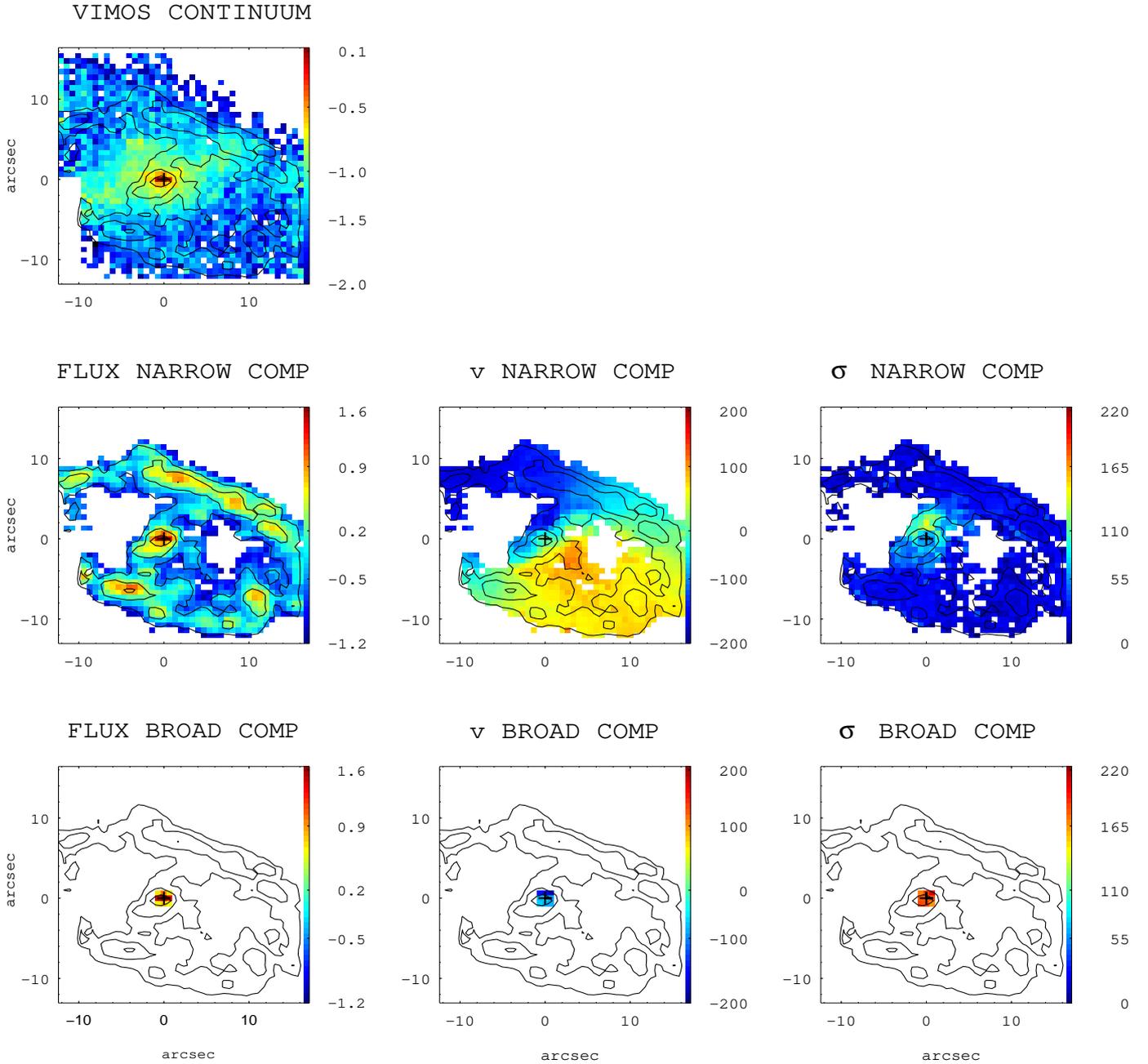}
\vspace{2mm}
\caption{(General comments about the panels as in Fig. A.1.) IRAS F11255-4120 (ESO 319-G022): this is a barred spiral with a circumnuclear ring structure that is at $\sim$ 4 kpc from the nuclear region in the H$\alpha$ image but not observed in the continuum images. Interestingly, the orientation of the bar seen in continuum emission (PA $\sim$ 110$^\circ$) is different from that of the ionized gas emission (PA $\sim$ 150$^\circ$). The projected orientation of the rotation axis (minor kinematic axis) seems to be different by about 30$^\circ$ from that of the minor photometric axis (and the bar in the H$\alpha$ flux intensity map) but is well aligned with the bar in the continuum. The kinematic center closely agrees with that of the H$\alpha$ flux and the continuum peaks. The velocity field is very regular and the velocity dispersion map has an almost centrally peaked structure with a local maximum of 120 km s$^{-1}$ in the bar structure. This object has been previously studied in \cite{YO2012}. The scale is of 0.333 kpc/$^{\prime\prime}$.}
\label{all_panels}
\end{figure*}

\begin{figure*}
\vspace{0cm}
\includegraphics[width=1.\textwidth, height=1\textwidth]{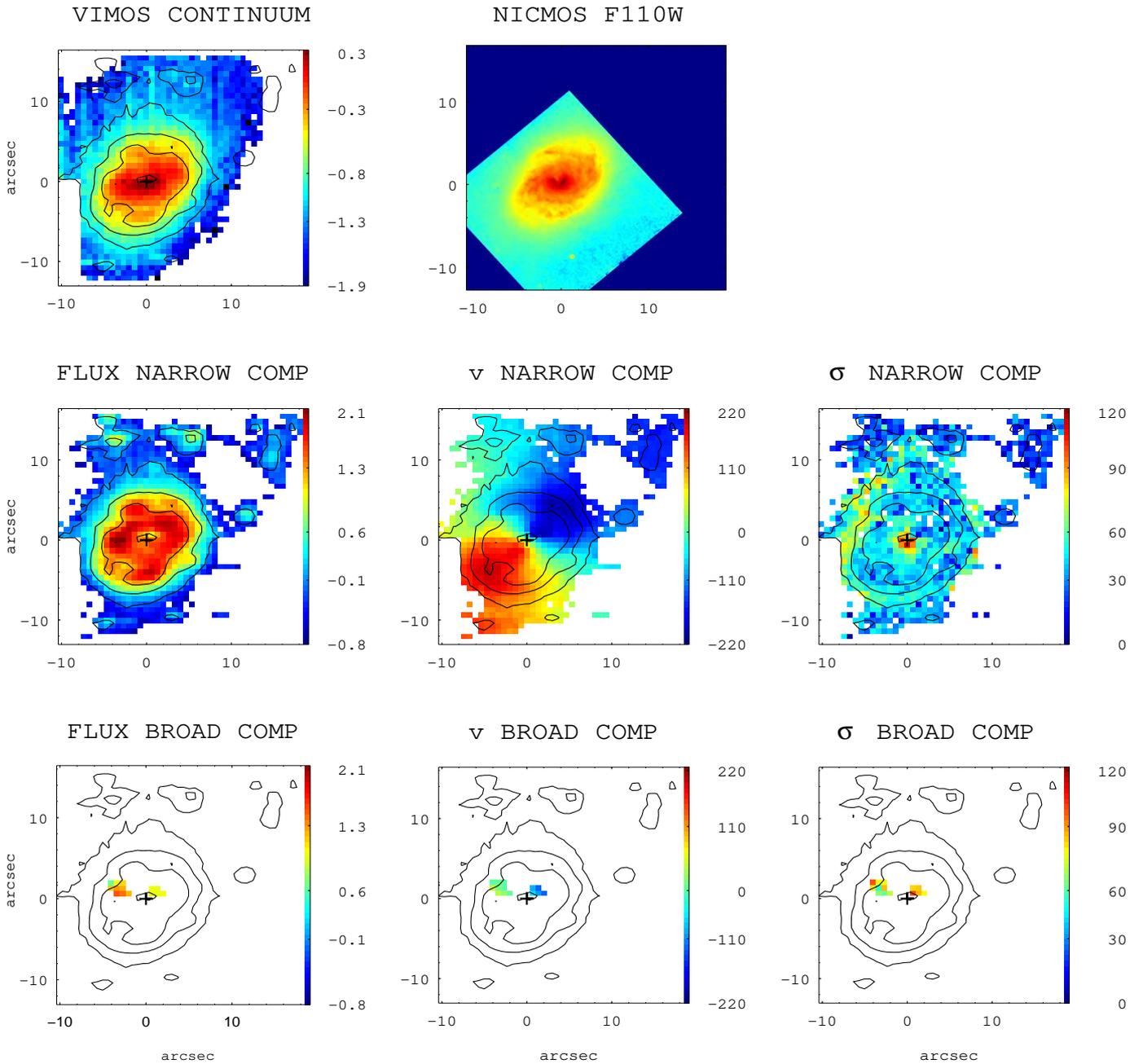}
\vspace{2mm}
\caption{(General comments about the panels as in Fig. A.1.) IRAS F11506-3851 (ESO 320-G030): a circumnuclear ring structure is detected in the H$\alpha$ image. The velocity field is extremely regular and the velocity dispersion map is centrally peaked (i.e., $\sigma_c$ = 81 km s$^{-1}$) with high values (i.e., $\sim$ 60 km s$^{-1}$) outwards the ring structure. The spatial scale is of 0.221 kpc/$^{\prime\prime}$.}
\label{all_panels}
\end{figure*}

\begin{figure*}
\vspace{0cm}
\includegraphics[width=1.\textwidth, height=1\textwidth]{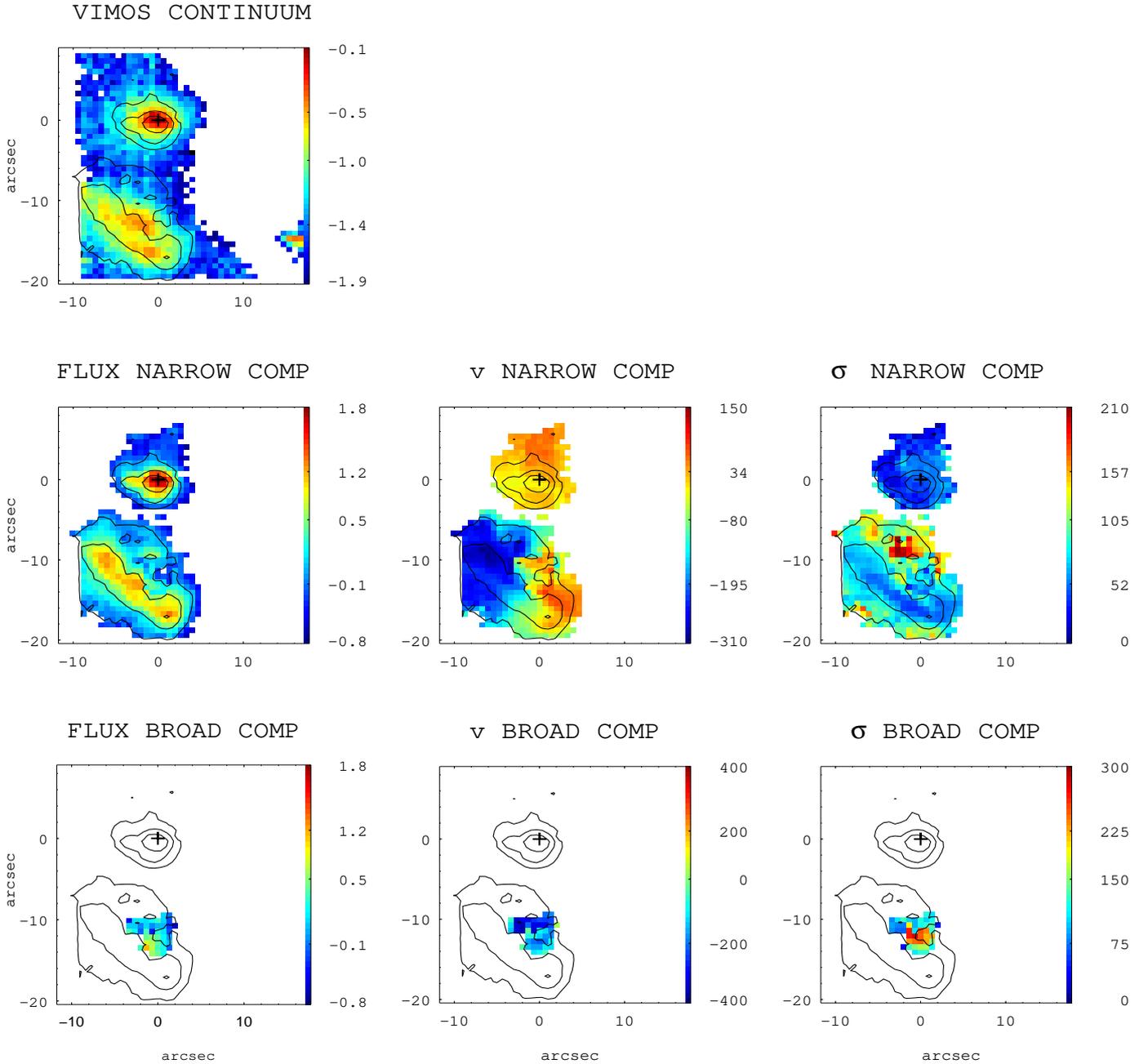}
\vspace{2mm}
\caption{(General comments about the panels as in Fig. A.1.) IRAS 12043-3140 (ESO 440-IG058): this system consists of two galaxies with a nuclear separation of $\sim$ 6 kpc. The northern galaxy is compact, while the southern source presents several knots in the H$\alpha$ map. The velocity field of the southern galaxy has two relatively well-defined approaching and receding parts, with the kinematic and photometric axes in relatively good agreement. Its velocity dispersion map reaches the highest values (i.e., $\sigma\sim$ 250 km s$^{-1}$) in the northern part, in correspondence of a local maximum of the H$\alpha$ intensity map. Around that region, a secondary broad component has been found in some spectra . The stellar distribution (i.e., continuum map) of the northern galaxy suggests that it is almost face-on: this could possibly explain the derived relatively low velocity shear (i.e., $\sim$ 42 km s$^{-1}$). Its velocity field shows a somewhat irregular pattern. The scale is of 0.468 kpc/$^{\prime\prime}$.} 
\label{all_panels}
\end{figure*}

\begin{figure*}
\vspace{0cm}
\includegraphics[width=1.\textwidth, height=1\textwidth]{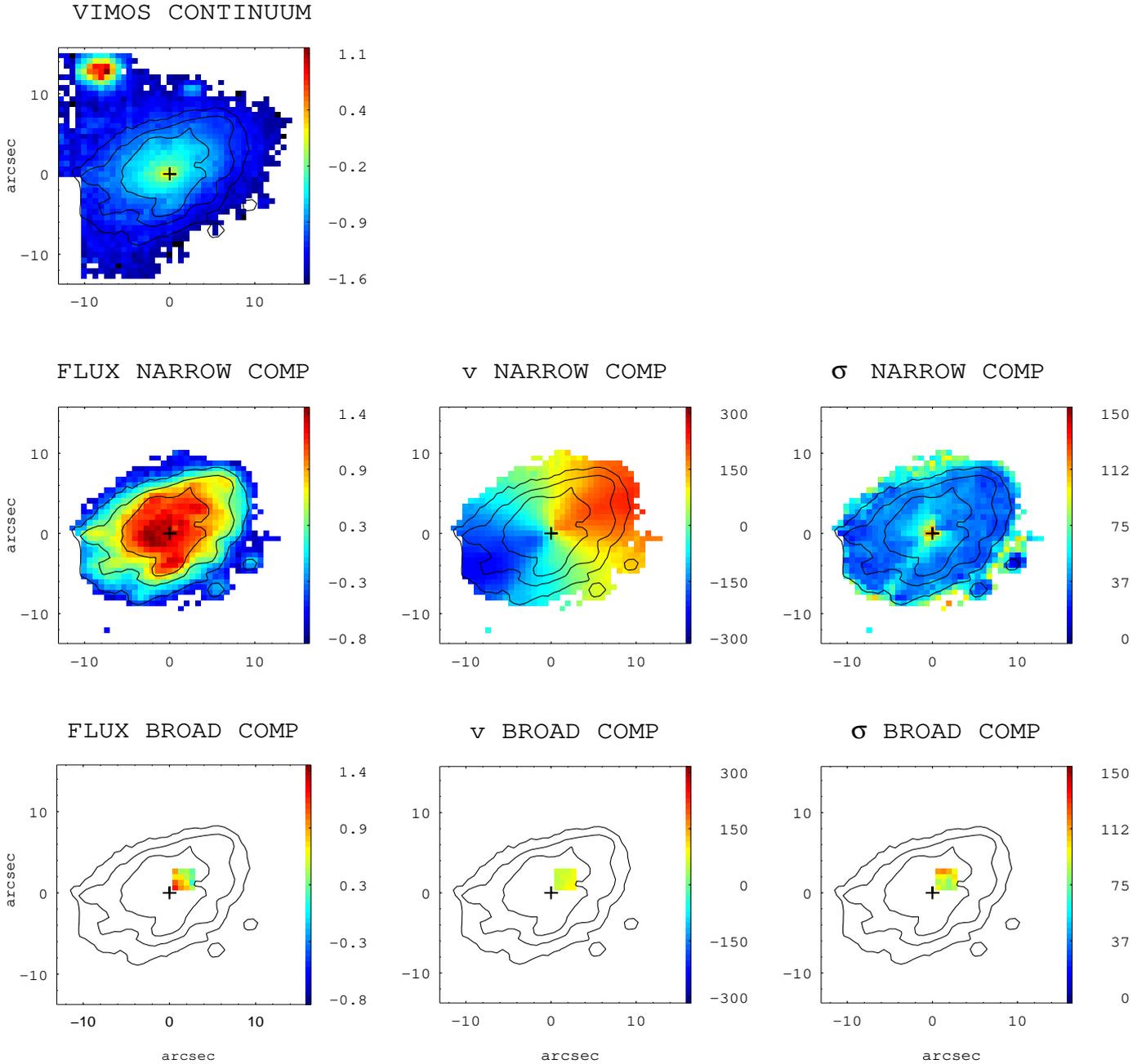}
\vspace{2mm}
\caption{(General comments about the panels as in Fig. A.1.) IRAS 12115-4656 (ESO 267-G030): the H$\alpha$ image shows a peak displaced by 0.8 kpc in the southwest direction with respect to the nucleus (continuum maximum). The kinematic center is in positional agreement with the nucleus. The velocity field and velocity dispersion map are those typical of a rotating $disk$-like object. At about $\sim$ 3 kpc northwest from the nucleus, the spectra show the presence of two kinematically distinct components. The scale is of 0.375 kpc/$^{\prime\prime}$. }
\label{all_panels}
\end{figure*}

\begin{figure*}
\vspace{0cm}
\includegraphics[width=1.\textwidth, height=1\textwidth]{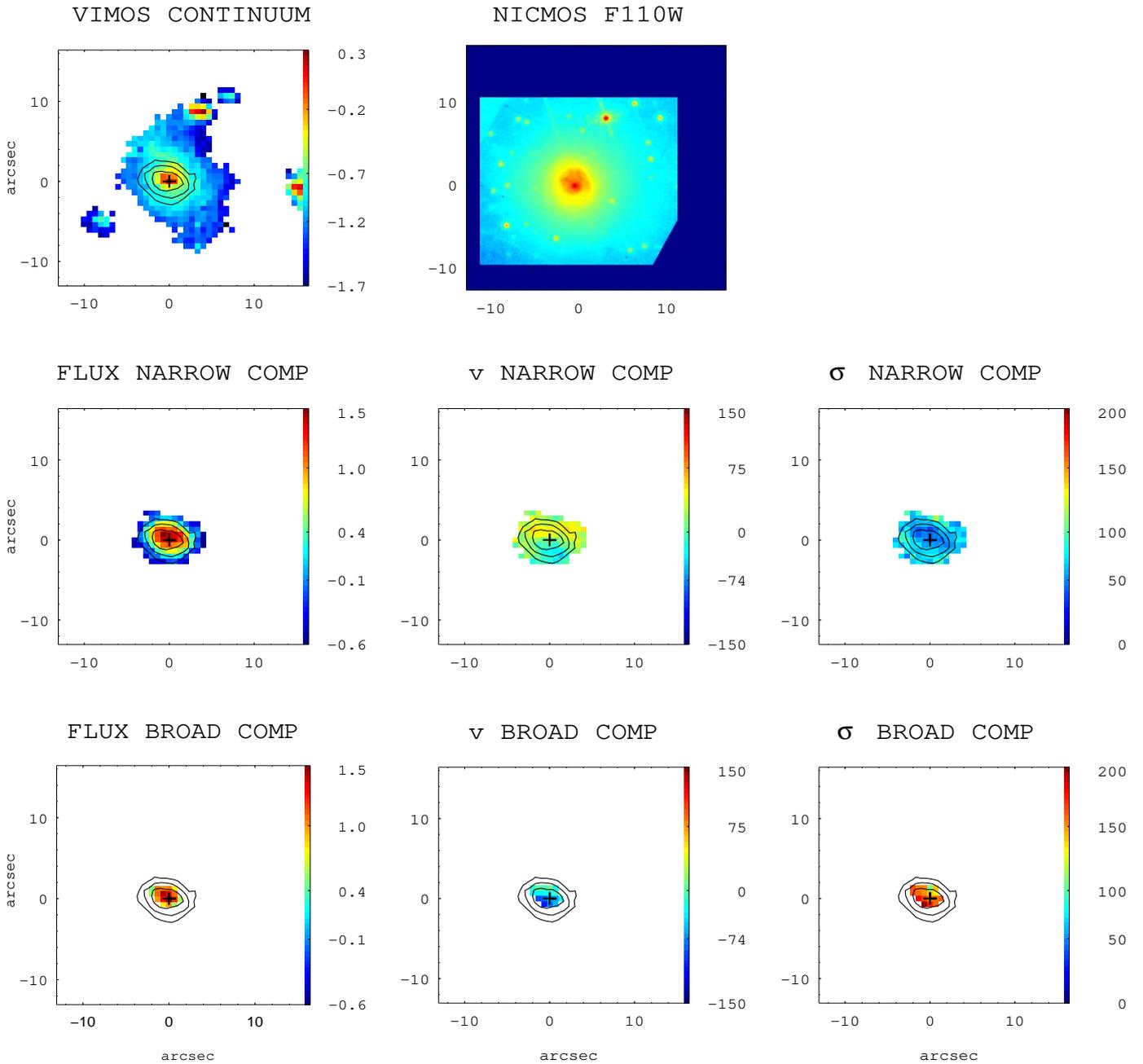}
\vspace{2mm}
\caption{(General comments about the panels as in Fig. A.1.) IRAS 12116-5615: this is one of the few objects for which the morphological classification is uncertain (class 2 or 0). It shows a very compact H$\alpha$ emission. The H$\alpha$ velocity field has a typical rotation pattern, but the velocity dispersion map is not centrally peaked. The latter shows higher values in the outer parts, which also have high excitation and low-H$\alpha$ surface brightness (see \citealt{MI10}). The scale is of 0.545 kpc/$^{\prime\prime}$. }
\label{all_panels}
\end{figure*}

\begin{figure*}
\vspace{0cm}
\includegraphics[width=1.\textwidth, height=1\textwidth]{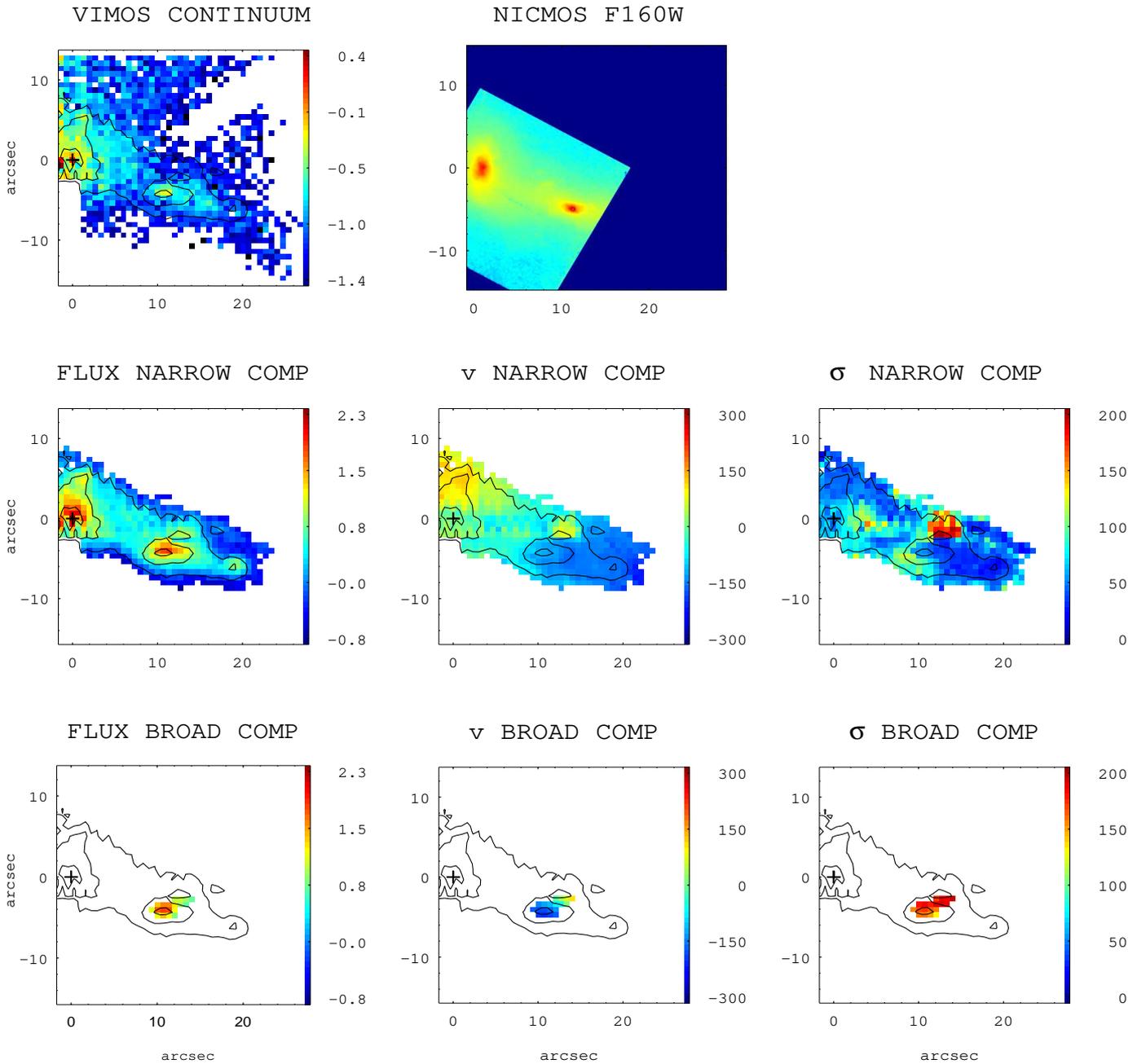}
\vspace{2mm}
\caption{(General comments about the panels as in Fig. A.1.) IRAS 12596-1529 (MGC-02-33-098): this system is interacting with MCG-02-33-099, at $\sim$ 37 kpc in the southeast direction. The HST image clearly shows two distinct galaxies in interaction, still visible in the VIMOS H$\alpha$ flux intensity map, where the nuclei are separated by $\sim$ 5 kpc. For this system it was not possible to reliably separate the contribution of each object. The two galaxies have been considered as a whole systems, deriving the mean kinematic values reported in Table \ref{NARROW}. However, the velocity field shows a quite regular pattern. The velocity dispersion map does not have a well-defined structure with a maximum of $\sim$ 200 km s$^{-1}$ in the northern part of the western galaxy: in this region low-H$\alpha$ surface brightness spectra are well fitted using one component and for this reason show a broader profile. However, no kinematic classification can be inferred from the whole system. The spatial scale is of 0.324 kpc/$^{\prime\prime}$.}
\label{all_panels}
\end{figure*}

\begin{figure*}
\vspace{0cm}
\includegraphics[width=1.\textwidth, height=1\textwidth]{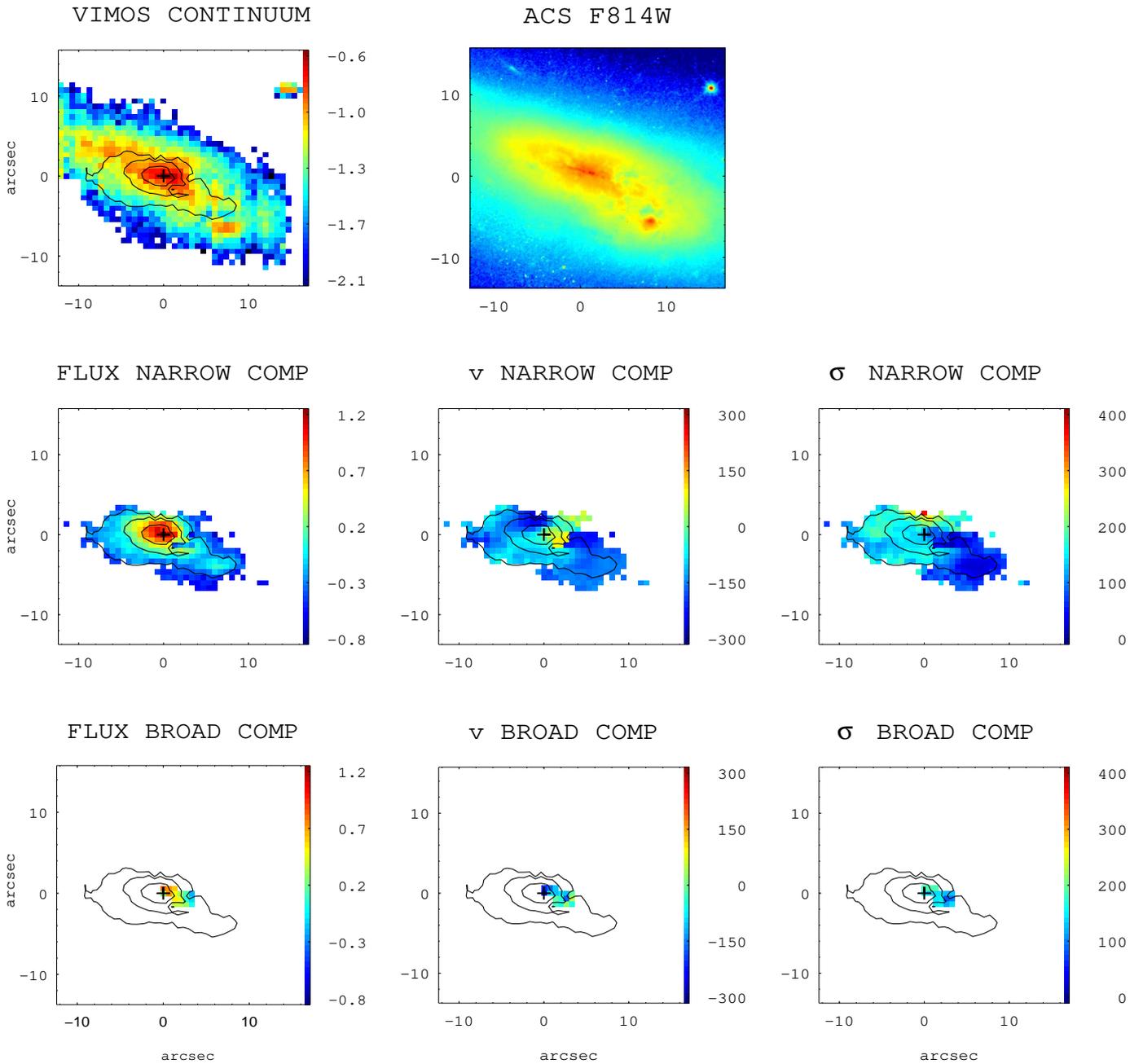}
\vspace{2mm}
\caption{(General comments about the panels as in Fig. A.1.) IRAS F13001-2339 (ESO 507-G070): the velocity field and velocity dispersion map have an irregular pattern. Some of the spectra in the northern region show a broad profile and relatively high excitation (see \citealt{MI10}). Its scale is of 0.439 kpc/$^{\prime\prime}$.}
\label{all_panels}
\end{figure*}

\begin{figure*}
\vspace{0cm}
\includegraphics[width=1.\textwidth, height=1\textwidth]{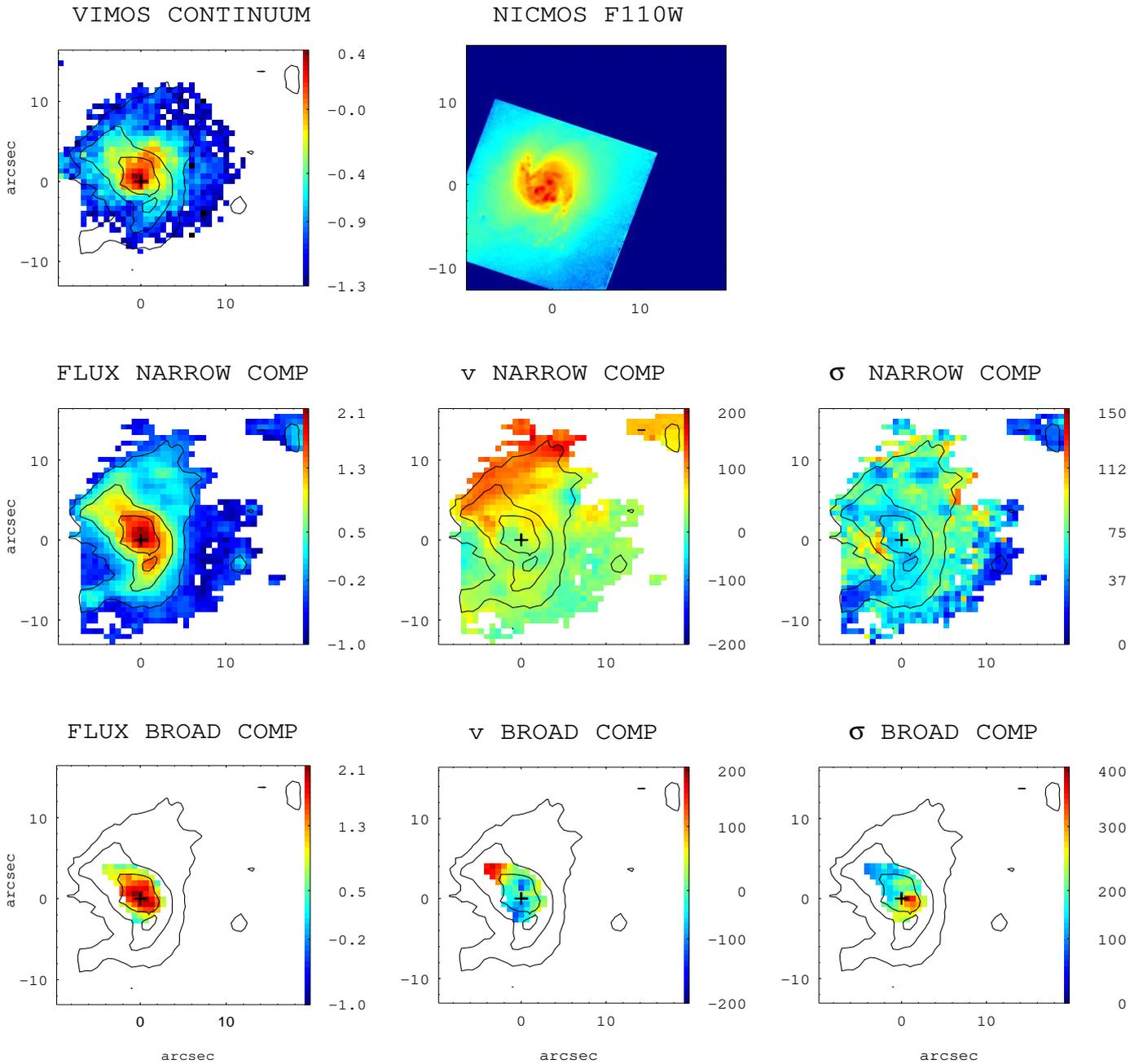}
\vspace{2mm}
\caption{(General comments about the panels as in Fig. A.1.) IRAS F13229-2934 (NGC 5135): the morphology of the ionized gas emission is substantially different from that of the continuum. Moreover, the photometric major axis of the continuum image seems to be perpendicular with respect to that of the H$\alpha$ flux intensity in the inner regions. The spatial scale is 0.28 kpc/$^{\prime\prime}$. }
\label{all_panels}
\end{figure*}

\clearpage

\begin{figure*}
\vspace{0cm}
\includegraphics[width=1.\textwidth, height=1\textwidth]{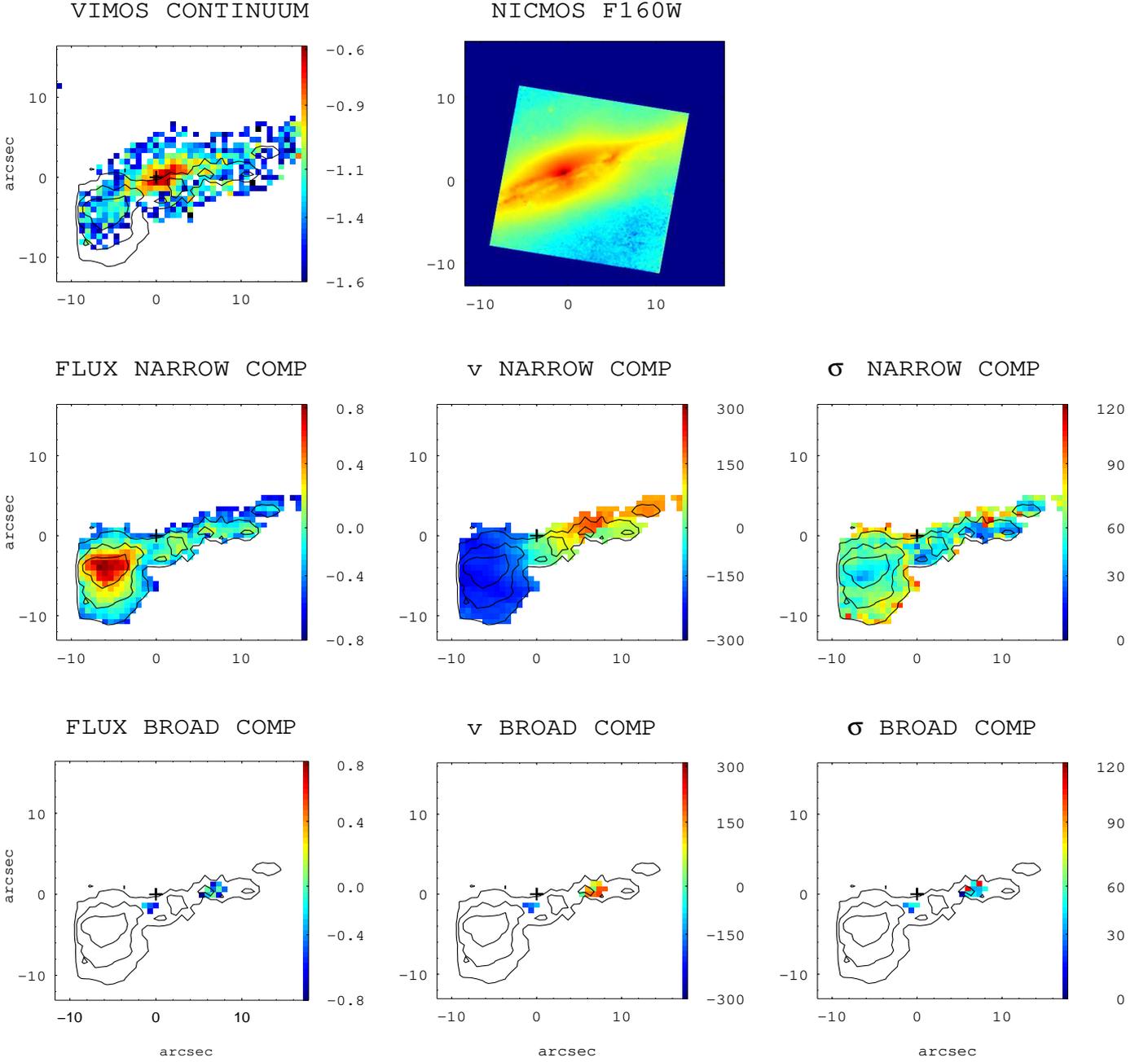}
\vspace{2mm}
\caption{(General comments about the panels as in Fig. A.1.) IRAS F14544-4255 (IC 4518): this system is formed by two galaxies with a nuclear separation $\sim$ 12 kpc, which were observed separately using two VIMOS pointings. The present panels correspond to the eastern source. This galaxy has a faint tail extended towards the northwest, as seen in the DSS image (see Paper III). The H$\alpha$ image reveals a very strong knot of star formation towards the southeast of the nucleus ($\sim$ 2.5 kpc): for this reason, the VIMOS continuum peak is used to identify the center (0,0) of the images. The velocity field is rather regular, although some deviations of pure rotation pattern are seen towards the west, probably due to the projection of star-forming knots with peculiar velocity. Its velocity dispersion map has a relatively constant value of $\sigma_{mean}$ $\sim$ 60 km s$^{-1}$ in almost the whole FoV. The broad component found in a few spectra has velocity dispersion values similar to those of the narrow component. The spatial scale is of 0.32 kpc/$^{\prime\prime}$. }  
\label{all_panels}
\end{figure*}

\begin{figure*}
\vspace{0cm}
\includegraphics[width=1.\textwidth, height=1\textwidth]{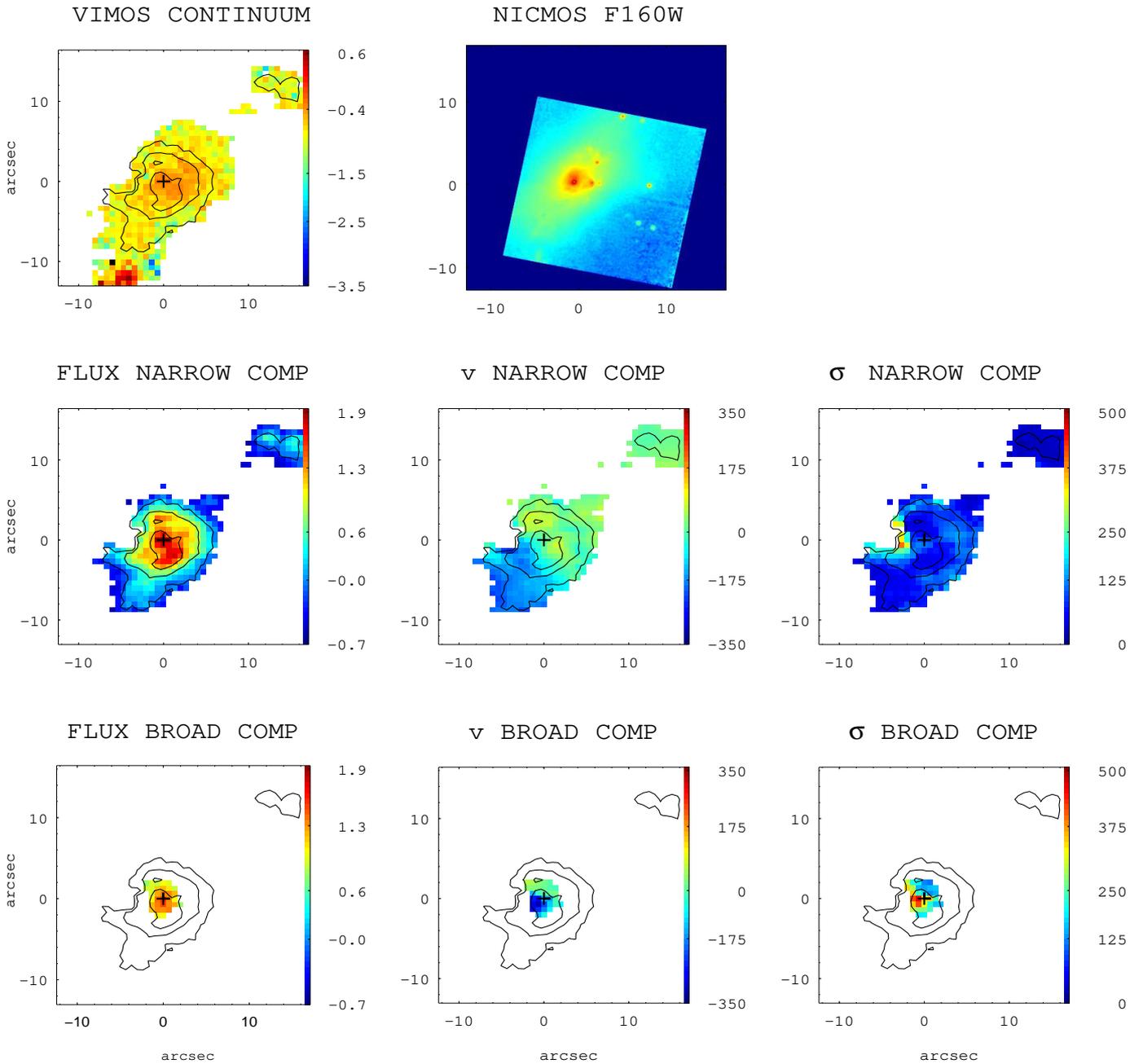}
\vspace{2mm}
\caption{(General comments about the panels as in Fig. A.1.) IRAS F14544-4255 (IC 4518): this is the western source of the system. It is more compact than the eastern one, although some extended emission is also observed towards the northwest likely associated with two star-forming knots. The velocity field has some irregularities (i.e., two maxima in the redshifted part), although a weak rotation pattern is visible. The velocity dispersion map shows a maximum of 300 km s$^{-1}$ on the eastern part of this galaxy at 0.9 kpc from the H$\alpha$ flux peak. The high values of the velocity dispersion of the narrow component correspond to low-H$\alpha$ surface brightness spectra. The spectra showing a secondary broad component in this object are quite complex. The spatial scale is 0.32 kpc/$^{\prime\prime}$.}
\label{all_panels}
\end{figure*}

\begin{figure*}
\vspace{0cm}
\includegraphics[width=1.\textwidth, height=1\textwidth]{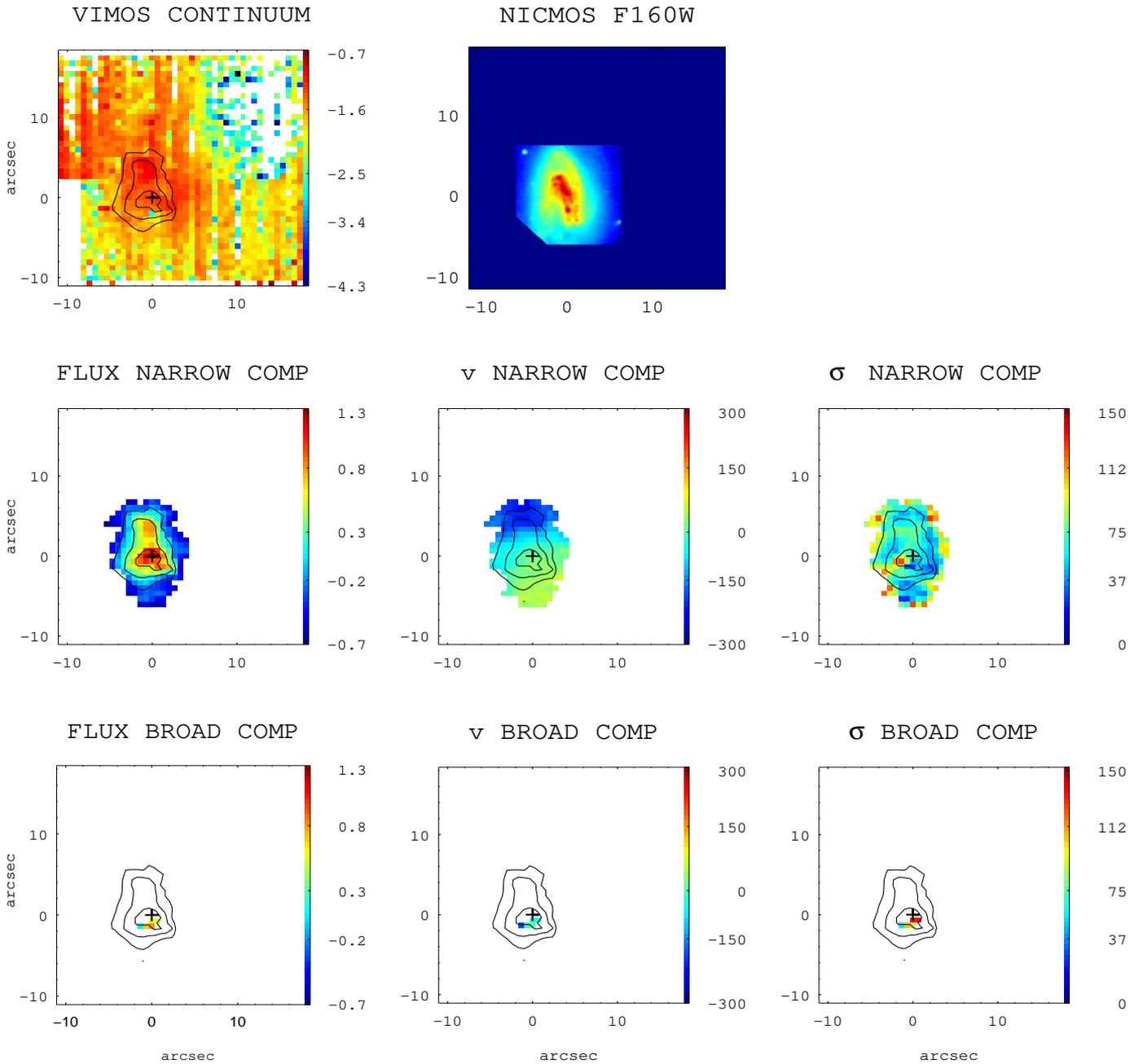}
\vspace{2mm}
\caption{(General comments about the panels as in Fig. A.1.) IRAS F17138-1017: the velocity field is regular, while the velocity dispersion map has no defined pattern. The continuum image shows vertical patterns that could not be removed during the reduction process (see \citealt{RZ11}). However, the H$\alpha$ maps (flux intensity and velocity field) are not affected by this problem. The spatial scale is 0.352 kpc/$^{\prime\prime}$.}
\label{all_panels}
\end{figure*}

\begin{figure*}
\vspace{0cm}
\includegraphics[width=1.\textwidth, height=1\textwidth]{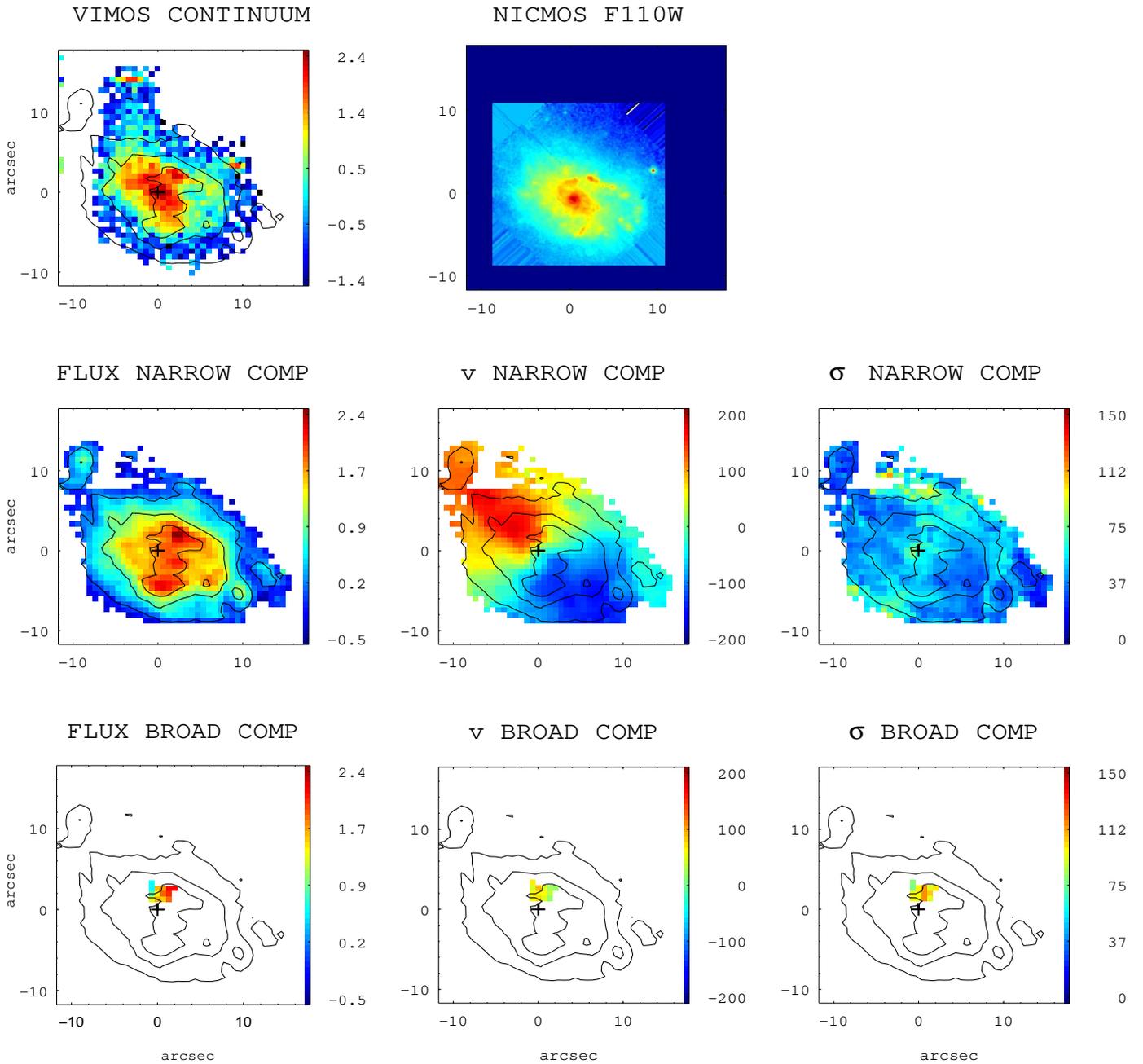}
\vspace{2mm}
\caption{(General comments about the panels as in Fig. A.1.) IRAS F18093-5744: the whole system consists of three galaxies in interaction, which have been observed separately. The nuclear separations between the northern (IC 4687) and the central galaxy (IC 4686) and between the central and the southern galaxy (IC 4689) are $\sim$ 10 kpc and $\sim$ 20 kpc, respectively. The HST (F110W) image shows a spiral-like morphology for IC 4687, with several knots and concentrations in the nuclear region.
IC 4687 shows a velocity field dominated by rotation, with the kinematic center well defined and coincident with the nucleus (continuum maximum). On the other hand, the velocity dispersion map shows an almost centrally peaked map, with higher values in the outer parts of the galaxy, corresponding to ionized regions (see \citealt{MI10}). The scale is 0.353 kpc/$^{\prime\prime}$. }
\label{all_panels}
\end{figure*}

\begin{figure*}
\vspace{0cm}
\includegraphics[width=1.\textwidth, height=1\textwidth]{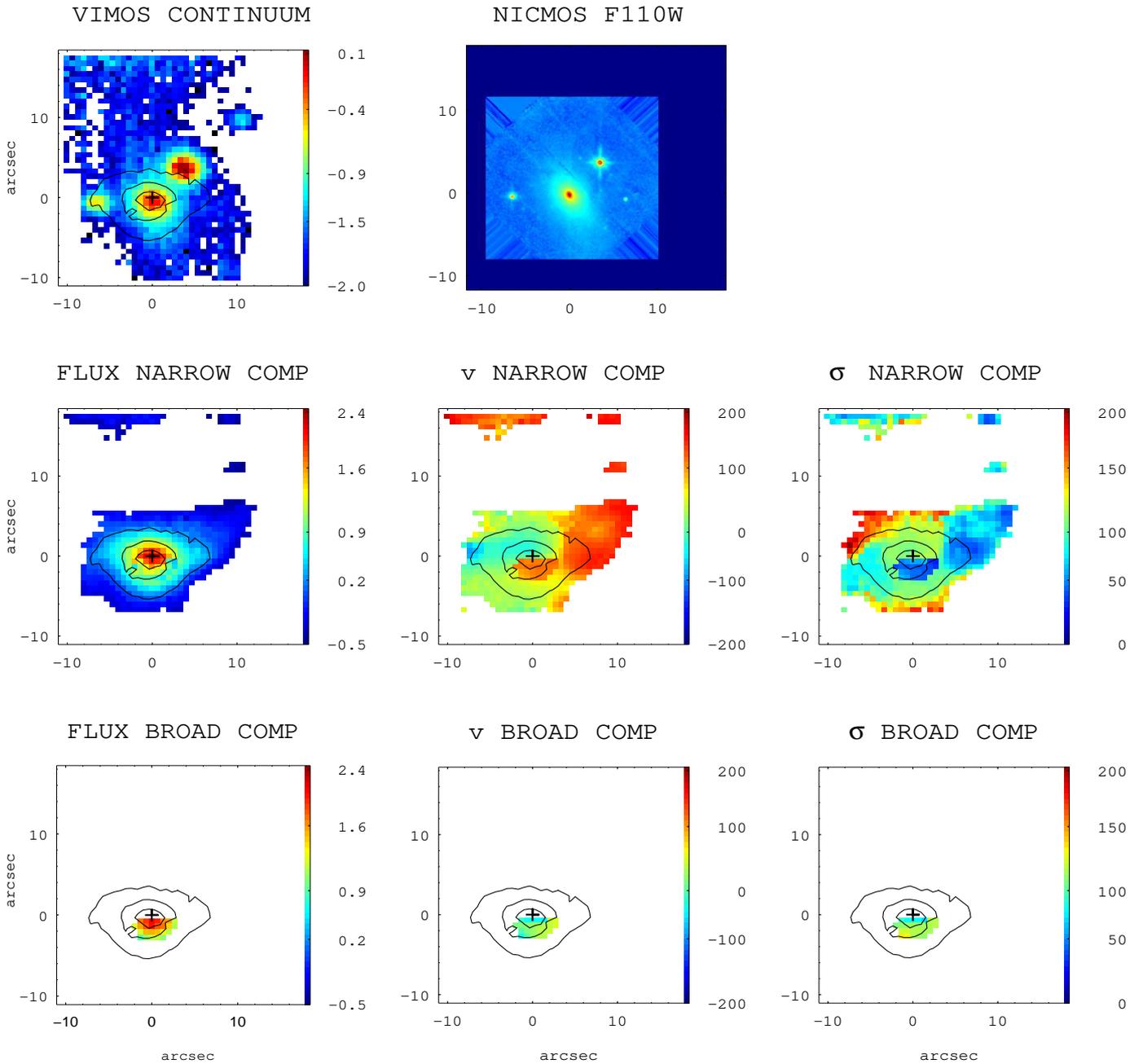}
\vspace{2mm}
\caption{(General comments about the panels as in Fig. A.1.) IRAS F18093-5744 (C): the H$\alpha$ emission from the source is highly concentrated in the nuclear region. Ionized gas emission is also observed extending towards the northwest of the nuclear region, probably as part of a tidal tail linking this galaxy with IC 4687. The VIMOS continuum emission is strongly contaminated by a star located in the northwest part with respect to the galaxy. The northern/western part of the velocity field is dominated by the kinematics of the tidal tail, which shows redshifted velocities and very low velocity dispersion: the blue part of the velocity field is not well defined, even excluding the region south of the nucleus. Due to the presence of the tidal tail, the interpretation of the broad component in the innermost region is uncertain (i.e., it may correspond to the main systemic component because it is the narrow one associated with the tidal tail). The scale is 0.353 kpc/$^{\prime\prime}$. }
\label{all_panels}
\end{figure*}

\begin{figure*}
\vspace{0cm}
\includegraphics[width=1.\textwidth, height=1\textwidth]{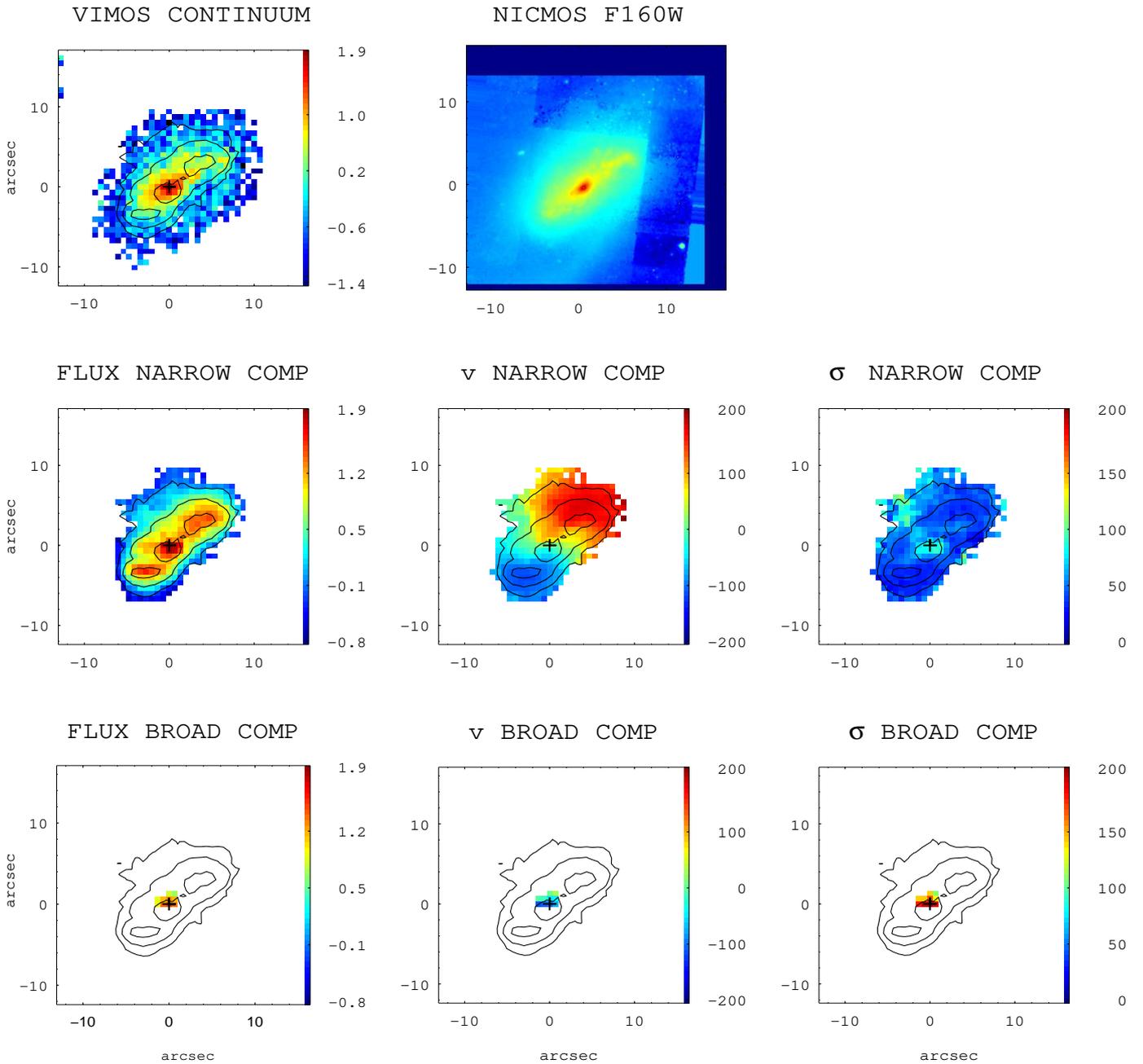}
\vspace{2mm}
\caption{(General comments about the panels as in Fig. A.1.) IRAS F18093-5744 (S) (IC 4689): this galaxy has a spiral morphology in the NICMOS/HST image, without clear evidence of strong interaction with the other galaxies of the system. The kinematic center seems to be coincident with the VIMOS continuum peak: the velocity field shows a mostly regular pattern, typical of a rotation-dominated $disk$-like object. Its velocity dispersion map shows a centrally peaked structure. It also shows regions in  the northern part with relatively large velocity dispersion ($\sim$ 100 km s$^{-1}$). A small area ($\sim$ 0.6 kpc$^2$) at the center shows the presence of a broad component, blueshifted by $\sim$ 85 km s$^{-1}$. The scale is 0.353 kpc/$^{\prime\prime}$. }
\label{all_panels}
\end{figure*}

\begin{figure*}
\vspace{0cm}
\includegraphics[width=1.\textwidth, height=1\textwidth]{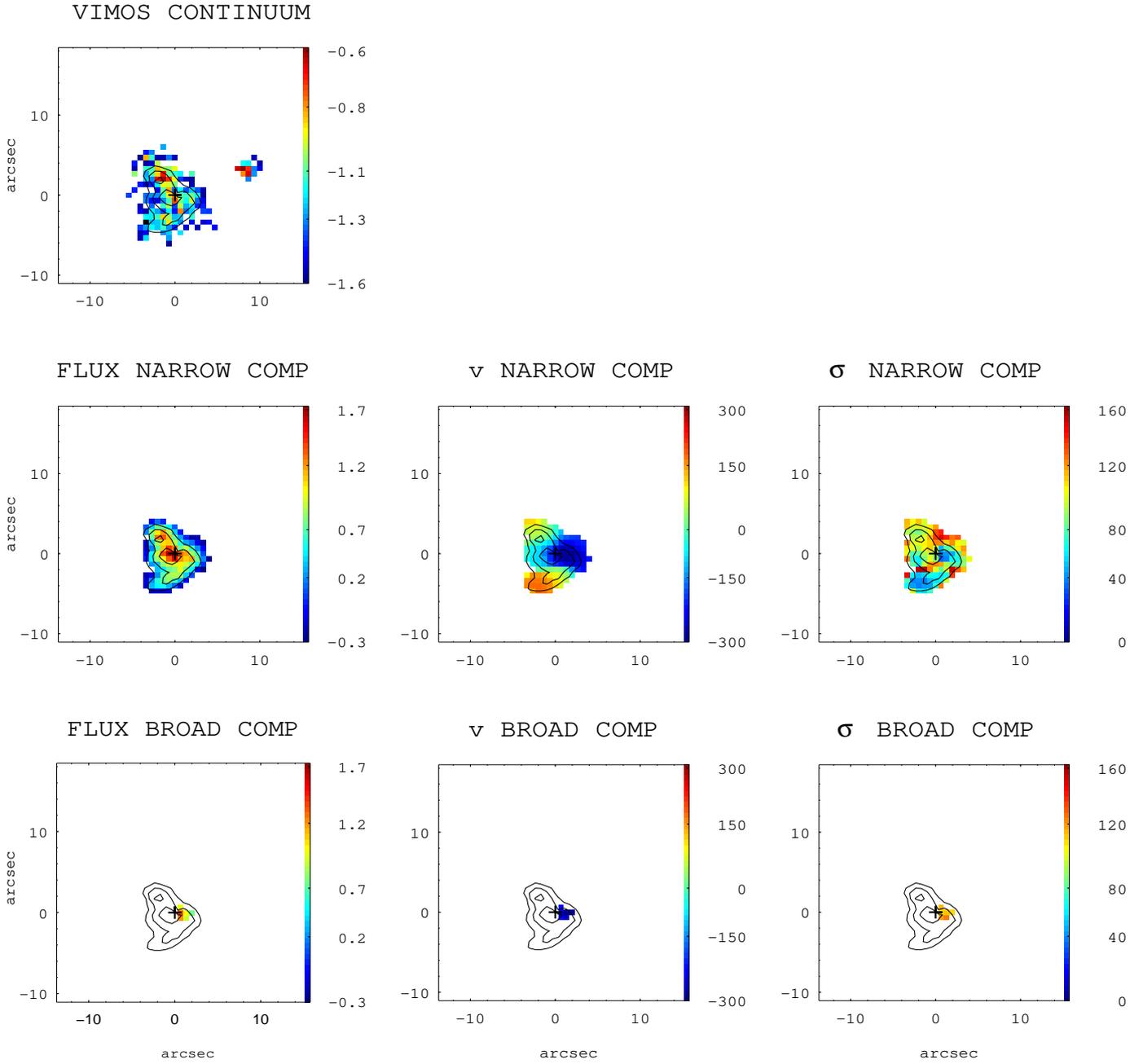}
\vspace{2mm}
\caption{(General comments about the panels as in Fig. A.1.) IRAS F21130-4446: this system has a double nucleus with a separation of 5.4 kpc (i.e., \citealt{dasyra06}). The peak of the continuum emission coincides with the location of the northern (weaker) H$\alpha$ condensation. The velocity field has a quite complex structure. The velocity dispersion map shows a distorted and irregular structure. The scale is of 1.72 kpc/$^{\prime\prime}$.}
\label{all_panels}
\end{figure*}

\begin{figure*}
\vspace{0cm}
\includegraphics[width=1.\textwidth, height=1\textwidth]{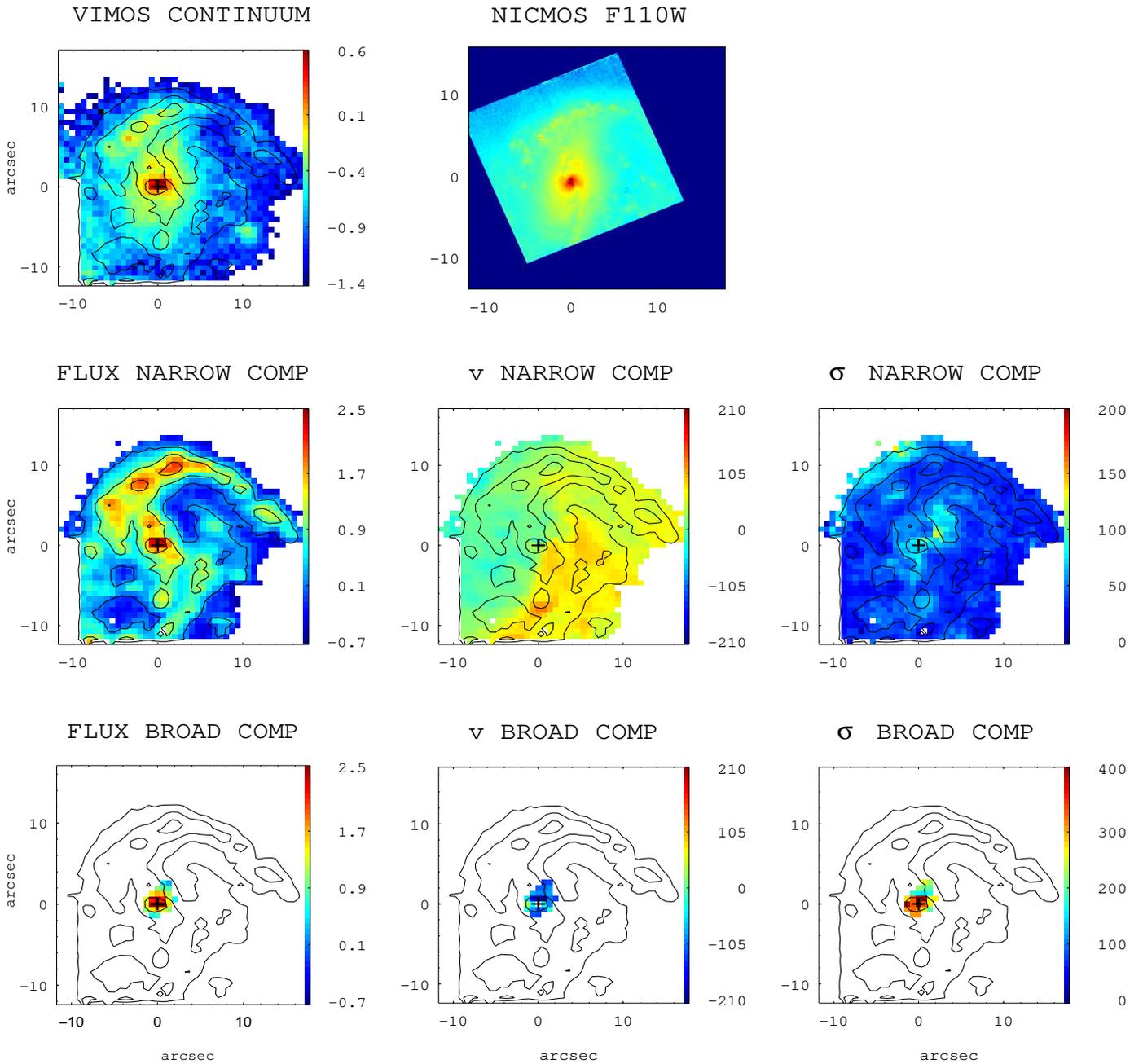}
\vspace{2mm}
\caption{(General comments about the panels as in Fig. A.1.) IRAS F21453-3511 (NGC 7130): the velocity field and velocity dispersion maps of the narrow component show asymmetric patterns. In the velocity field three main regions can be identified with the rotation axes not well-defined. The velocity dispersion map shows high values in the northern arm and its central part, with values larger than 100 km s$^{-1}$. This object has been analyzed in \cite{YO2012}. The scale is 0.329 kpc/$^{\prime\prime}$. }
\label{all_panels}
\end{figure*}

\begin{figure*}
\vspace{0cm}
\includegraphics[width=1.\textwidth, height=1\textwidth]{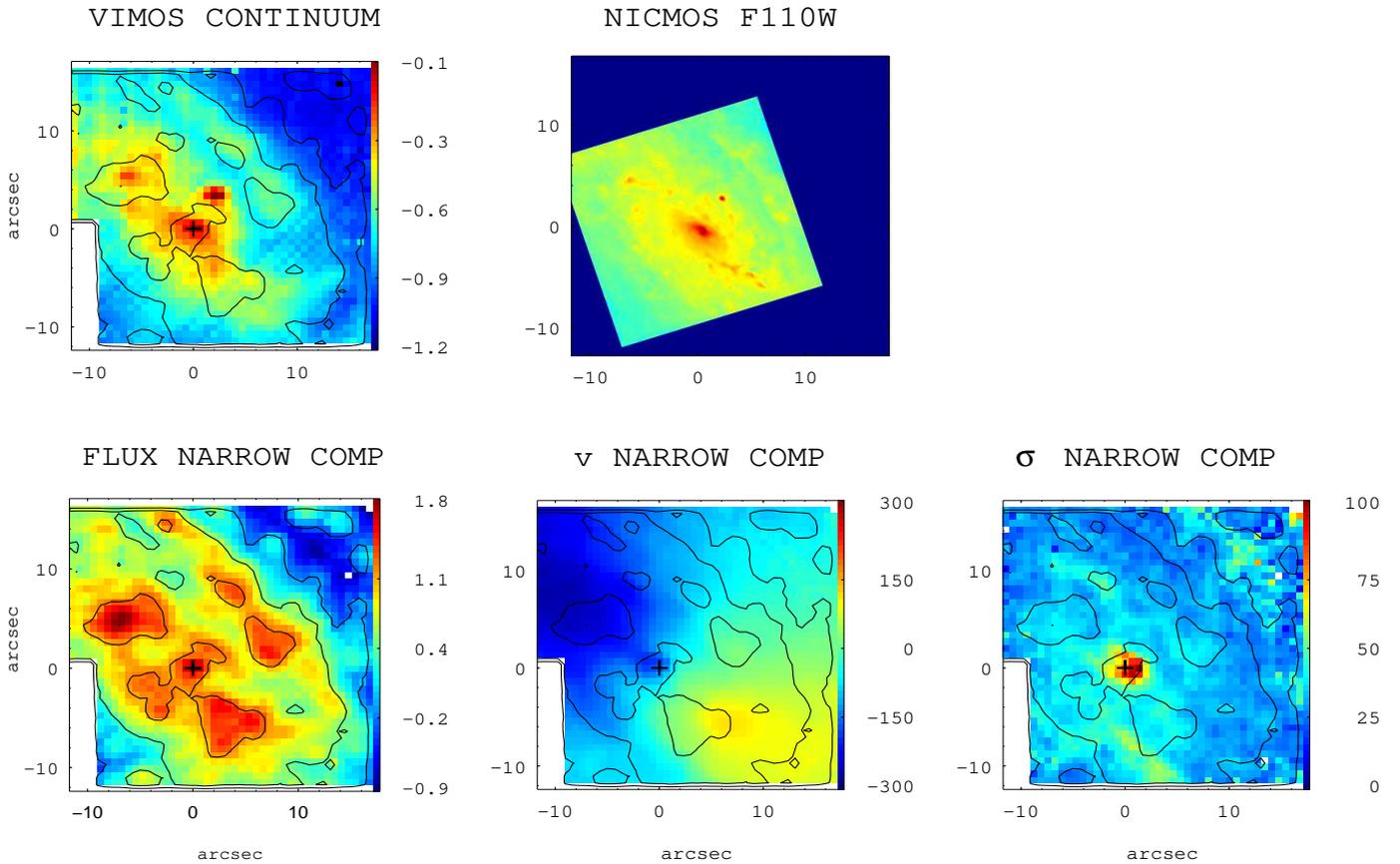}
\vspace{2mm}
\caption{(General comments about the panels as in Fig. A.1.) IRAS F22132-3705 (IC 5179): the velocity field of this source has the typical `spider-diagram' shape characterizing a rotating $disk$-like object. The kinematic center and the peak of the velocity dispersion map are well defined and coincident with the VIMOS continuum maximum and the H$\alpha$ flux peak. In the south and northwest there are regions with velocity dispersion of the order of 50 km s$^{-1}$. The spatial scale is 0.234 kpc/$^{\prime\prime}$. }
\label{all_panels}
\end{figure*}

\begin{figure*}
\vspace{0cm}
\includegraphics[width=1.\textwidth, height=1\textwidth]{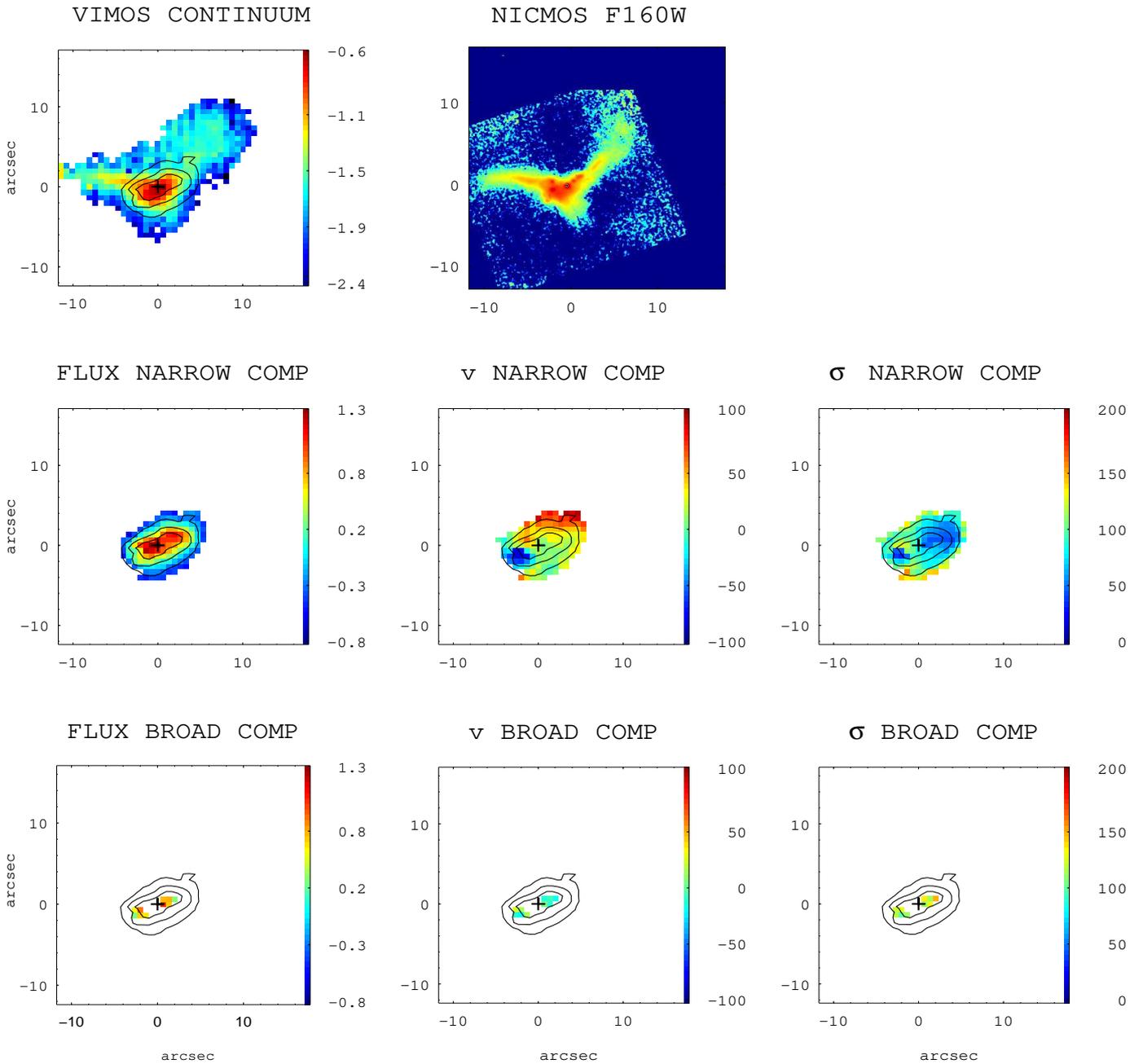}
\vspace{2mm}
\caption{(General comments about the panels as in Fig. A.1.) IRAS F22491-1808: this is a double nucleus system (separation of 2.6 kpc) with several knots and condensations located in both the nuclear region and the tidal tails observed to the east and northwest of the system. The H$\alpha$ image shows a more compact morphology than that of the continuum, with no evidence for the already mentioned tidal tails. This system seems to be kinematically disturbed, with a poorly defined minor rotation axis, although a rotation pattern is visible. The velocity dispersion map has a quite irregular structure, with an offset peak with respect to the nucleus and high values in correspondence of low H$\alpha$ surface brightness. There are two regions where the spectra profiles suggest the presence of an extra broad component. The scale is of 1.47 kpc/$^{\prime\prime}$. }
\label{all_panels}
\end{figure*}

\begin{figure*}
\vspace{0cm}
\includegraphics[width=1.\textwidth, height=1\textwidth]{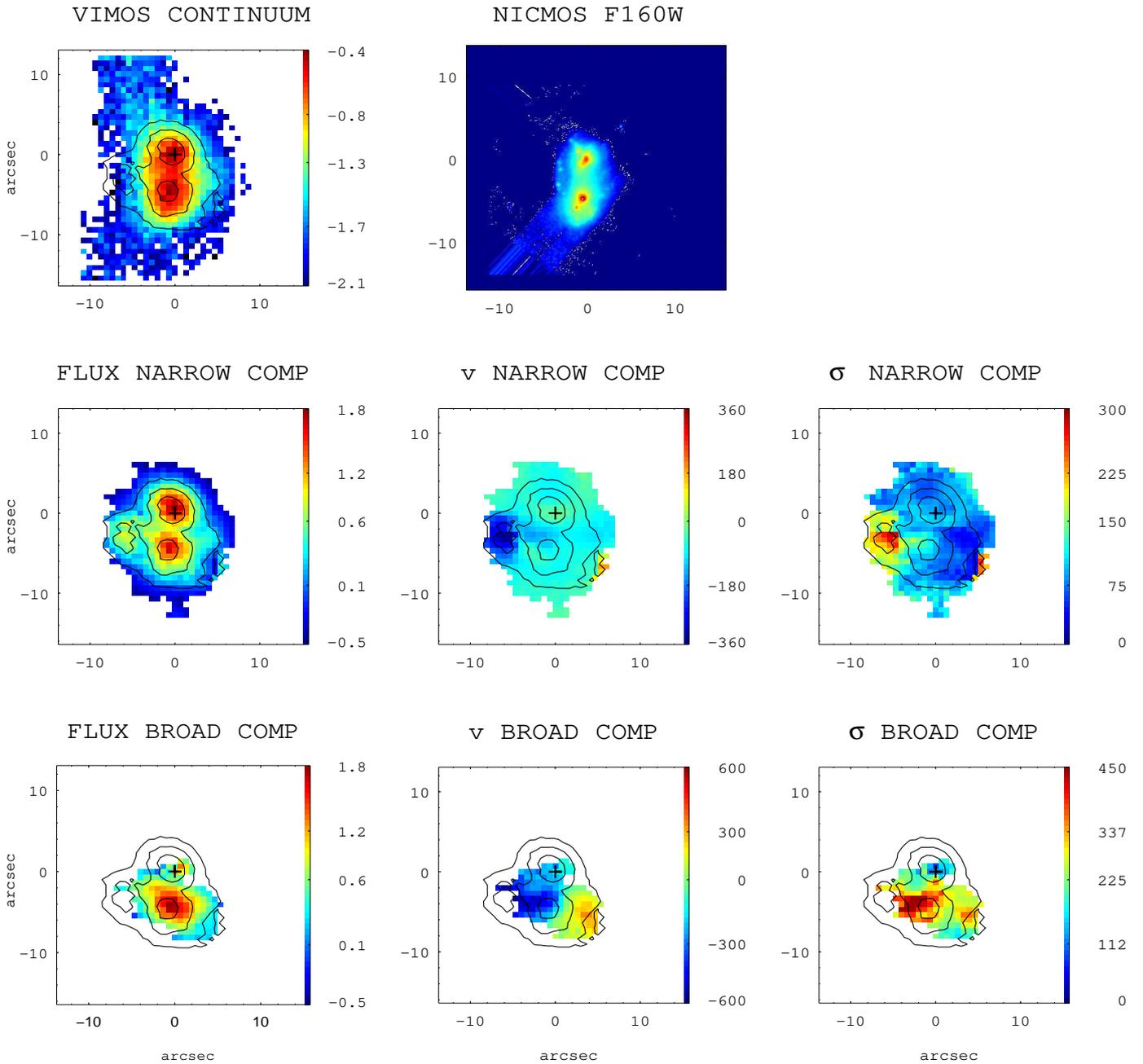}
\vspace{2mm}
\caption{(General comments about the panels as in Fig. A.1.) IRAS F23128-5919 (AM 2312-591): this is a ULIRG with a double nucleus (separation of $\sim$ 4 kpc). It has not been possible to separate reliably the contribution of each individual galaxy and, therefore, the kinematic values reported in Table \ref{NARROW} correspond to the whole system. The velocity field has a quite complex pattern, where the two galaxies merger their velocity fields. A relatively extended broad component has been found in the inner regions of this system, covering an area of 55 kpc$^2$ and with a mean blueshift of 32 km s$^{-1}$. The velocity maps of this second component suggest the presence of a strong outflow. The scale is of 0.878 kpc/$^{\prime\prime}$.}
\label{all_panels}
\end{figure*}

\clearpage

\begin{table*}
 \centering
\caption{Spatial offsets between the VIMOS continuum and H$\alpha$ flux peaks.}
\vspace{1cm}
\label{OFFSETS}
\begin{scriptsize}
   \begin{tabular}{cccc}
\hline\hline\noalign{\smallskip}
 Galaxy ID             &    RA offset  & DEC offset  &  {}Notes\\
{\smallskip}
(IRAS)          &   ($arcsec$) & ($arcsec$)     &       \\
  (1) & (2) & (3)  \\
\hline\noalign{\smallskip}
        F06295-1735          &  0.7  	& -2.5   	   \\
        F06592-6313          &  0.6  	&  0    \\
        F07027-6011 S       & -0.5  	&  0.4	      	\\
        F07027-6011 N       &  0 		&   0    \\
        F07160-6215          & -1.4 	&  1.6    	 \\
        F10015-0614          & 0 		&  0         \\
        F10409-4556          & 0 		&  0     \\
        F10567-4310          &  0 		&  0.3         \\
        F11255-4120          &  0 		& 0           \\
        F11506-3851          &  3.4 	&  0.1  & (a)                    \\
        F12115-4656          &  1.9 	& -0.7  & (a)                \\
        F13229-2934          &  0.6 	&  0.5              	    \\
        F22132-3705          & 0  		& 0                   \\
        \hline\noalign{\smallskip}
      F01159-4443 N         & 0.6		& 0.4 \\
    F01341/ESO G12        & 0   		&  0\\
    F01341/ESO G11        &-0.8  	& 0.6\\
        F06035-7102          & 0.7  	&  0	\\
      F06076-2139 N         & 0   & 0 \\
        F06206-6315          & -0.5   	& 0 \\
    F06259-4780 S           &  -1.0 	&  1.3\\
    F06259-4780 N           & -0.6  	&  0  \\
     08424-3130 S            &  0	  	& 0 \\
       F08520-6850           &  0.6   	& -0.4 \\
       F09437+0317 S       & 1.5  	& 0 & (a)\\
       F09437+0317 N       & 0.7   	& 0.6 \\
       12043-3140 N          & 0	   	&  0  \\
        F12596-1529           & 1.7   	&  -0.6 \\
          F14544-4255 E     &  -6.7 	& 3.9   &  (a)\\
          F14544-4255 W    & -0.6 	& 0\\
          F18093-5744 S     &  0  		& 0  \\
          F18093-5744 C     &  0   	& 0 \\
          F18093-5744 N     &  -2.2   &   2.0  &  (a)  \\
       F22491-1808            &  -0.7	& 0 \\
       F23128-5919            & 0 	&   -0.6\\
      \hline\noalign{\smallskip}
        F04315-0840            &  0	  	& 0 \\
        F05189-2524            &  -0.4 	& 0  \\
        08355-4944              & 0   	& 0 \\
        09022-3615              & 1.1  	&  0  \\
        F10038-3338            &  0.5 	& -0.4\\
        F10257-4339            & 0 		& 0\\
        12116-5615              & -0.5 	& -0.4  \\
        F13001-2339            & 0	  	& 0 \\
        F17138-1017            &  -	& - & (b)\\
        F21130-4446            & 1.2 	& 2.2  \\
        F21453-3511            & 0 		& 0 \\
\end{tabular}
\vskip0.2cm\hskip0.0cm
\end{scriptsize}
\begin{minipage}[h]{18cm}
\tablefoot{Spatial offsets along the right ascension (RA) and declination (DEC) directions: they are defined as the difference between the position of the H$\alpha$ flux intensity peak and that of the VIMOS continuum (rest frame 6390-6490 \AA) intensity peak (i.e., $x_{H\alpha}$ - $x_{cont}$, $y_{H\alpha}$ - $y_{cont}$). 
The coordinates ($x_i$, $y_i$) of the peaks in the maps have been derived as the flux-weighted mean position when considering 3$\times$3 spaxels around the maximum intensity value. A typical error of 0.34$^{\prime\prime}$ is estimated. For double systems having both galaxies in the VIMOS FoV, the position of the peak refers to the one used as reference to center the images.
Col (4): Notes with the following code: 
(a) The peak of the VIMOS continuum image is considered the center of the images since the H$\alpha$ peak corresponds to an extranuclear knot of star formation and does not properly define the center of the galaxy. 
(b) For this galaxy, it was not possible to identify the VIMOS continuum peak due to the S/N. }
\end{minipage}
\end{table*}

\clearpage

\section {Effective radius determinations based on near-infrared continuum imaging}
\label{App_rad}

Rest frame near-IR continuum light traces to first order the bulk of the galaxy stellar component. Here we obtain near-IR-based effective radii R$_{eff}$ for the present sample from infrared imaging obtained from the literature. Specifically, we have compiled the K$_s$-band images from the 2MASS (\citealt{Skru06}), and the existing near-IR imaging obtained with HST in H-band (K-band HST imaging is restricted to a reduced number of objects). The 2MASS images have the advantage of being available for all the galaxies in the sample and without limitations by FoV. On the other hand, they have a relatively low angular resolution (i.e., $\sim$ 2$^{\prime\prime}$), which prevents us from distinguishing kpc-scale structures for many objects in the sample. The HST imaging obviously solves this problem, but in contrast the relatively small FoV of the Near Infrared Camera and Multi-Object Spectrometer (NICMOS) imposes serious limitations for the closest systems. NICMOS/HST H-band infrared imaging existed for about 50 percent of the sample. A few cases were observed with WFC3, which combines good angular resolution and a relatively large FoV.

Two methods have been used to derive the effective radii using 2MASS and HST images: 1) GALFIT (\citealt{peng10}) and 2) the so-called A/2 method, for which the effective radius is defined as R$_{eff}$ = $\sqrt{(A/\pi)}$ and where A is the solid angle subtended by the minimum number of pixels encompassing 50 percent of the galaxy light for the images considered. GALFIT and A/2 methods agree relatively well: in general, radii smaller than $\sim$ 20\% are derived when applying the A/2 method. For those cases with clearly defined disks, the GALFIT radii were preferred to those with the A/2 method. For systems with a high ellipticity or with linear structure, GALFIT tends to interpret them as inclined or edge-on disks. This interpretation is correct in some cases, but in others the actual morphology of the system is distorted by tidal forces or it is uncertain. Also, for highly structured images (generally from HST) the A/2 measurements tend to be smaller than those from GALFIT. 

The HST-based measurements tend to be smaller ($\sim$ 10\%) than the 2MASS ones as a consequence of the relatively small FoV, which often does not cover the whole galaxy emission. Furthermore, the wavelength dependence of the radius measurements between the H- and K-bands are expected to be very small. In fact, \cite{V02} found that the mean values obtained from K-band imaging were about 38 percent smaller than those obtained in the optical (R-band). Assuming conservatively that this difference is mainly due to reddening, one should expect a minor change between the values inferred at H and K.

In Table \ref{latex_tabla} we present the individual results of the derived effective radii when applying the two methods to the 2MASS and HST images, as well as the adopted values and their estimated uncertainties. The table includes a code that indicates the values considered for deriving the adopted one.

\begin{table*}
\centering
\vspace{1cm}
\caption {Near-IR continuum (stellar) radii determinations.} 
\label{latex_tabla}
\begin{scriptsize}
    \begin{tabular}{l ccccccccc } 
\hline\hline\noalign{\smallskip}
IRAS     & R$_{hl}^{2MASS}$(GALFIT) & R$_{hl}^{2MASS}$(A/2)  &  Comm.& R$_{hl}^{HST}$(GALFIT) & R$_{hl}^{HST}$(A/2)  & Comm.& Adopted & code & Previous R$_{hl}$ \\
{\smallskip}
name   &           (kpc)           		&          (kpc)         		&  code 		&         (kpc)          		&         (kpc)        	& code	 &  (kpc)  		&      &  (kpc)(ref)       \\
(1)       &    (2) &           (3)            &           (4)          &   (5) &          (6)           &          (7)         & (8)  &   (9)   &  (10)        \\
 \hline\noalign{\smallskip}                                             
     01159 N &  0.81 $\pm$ 0.32	& 1.30 $\pm$ 0.44 & 1,7  &	  ...	   &	    ...        &      & 1.0 $\pm$  0.4 & 1100&  		     \\
     01159 S &  2.39 $\pm$ 0.72	& 2.09 $\pm$ 0.47 & 1   &	  ...	   &	    ...        &      & 2.2 $\pm$  0.6 & 1100&  		     \\
     01341 N &  2.56 $\pm$ 0.50	& 2.20 $\pm$ 0.56 &     & 1.55 $\pm$ 0.65  &   0.95 $\pm$ 0.15 &   c  & 2.2 $\pm$  0.6 & 1110&  		     \\
     01341 S &  1.50 $\pm$ 0.60	& 0.93 $\pm$ 0.26 & 4   &	     ...   &	    ...        &      & 1.5 $\pm$  0.6 & 1000&  		     \\
     04315   &  1.15 $\pm$ 0.35	& 0.94 $\pm$ 0.14 & 4   & 0.55 $\pm$ 0.30  &   0.36 $\pm$ 0.12 &   c  & 1.2 $\pm$  0.4 & 1000& 0.06 (b)		     \\
     05189   &   ...							&  (0.67)         & 6   &    (0.10)        &	   (0.25)      &   h  &     (0.18)     & 0000&  		     \\
     06035   &   ...							& 4.71 $\pm$ 1.35 &     &	     ...   &	    ...        &      & 4.7 $\pm$  1.4 & 0100&  		     \\
     06076 N &  3.00 $\pm$ 1.40	& 2.21 $\pm$ 0.37 & 1   & 3.23 $\pm$ 0.06  &        ...        &   d  & 3.2 $\pm$  0.6 & 1110&  		     \\
     06076 S &  3.40 $\pm$ 1.20	& 1.73 $\pm$ 0.45 & 1   & 3.11 $\pm$ 0.06  &        ...        &   d  & 3.1 $\pm$  0.6 & 1010&  		     \\  
     06206     &  (3.60)     		& (4.56)          & 5   &	    ...	   &   2.41 $\pm$ 0.60 &      & 2.4 $\pm$  0.6 & 0001&  		     \\
     06259 N  &  1.09 $\pm$ 0.20	& 1.45 $\pm$ 0.58 & 1   & 1.15 $\pm$ 0.05  &   1.02 $\pm$ 0.02 &      & 1.0 $\pm$  0.2 & 1011&  		     \\  
     06259 C  &  1.67 $\pm$ 0.12	& 1.73 $\pm$ 0.35 & 1   & 2.91 $\pm$ 0.07  &   1.16 $\pm$ 0.04 &   f  & 2.6 $\pm$  0.7 & 1010&  		     \\
     06259 S  &  2.60 $\pm$ 0.50	& 1.74 $\pm$ 0.58 & 4   &	    ...	   &   1.14 $\pm$ 0.04 &   e  & 2.6 $\pm$  0.7 & 1000&  		     \\       
     06295    &   ... 								& 3.03 $\pm$ 0.62 & 3   &	    ...	   &	    ...        &      & 3.0 $\pm$  0.6 & 0100&  		     \\ 
     06592    &  1.53 $\pm$ 0.60	& 1.17 $\pm$ 0.26 &     &	    ...	   &	    ...        &      & 1.2 $\pm$  0.4 & 1100&  		     \\ 
     07027 N  &   ... 							&  (0.79)         & 6   &	    ...	   &       (0.2 )      &   h  &     (0.2)      & 0000&  		     \\
     07027 S  &  1.61 $\pm$ 0.70	& 1.54 $\pm$ 0.69 &     & 1.75 $\pm$ 0.06  &   0.96 $\pm$ 0.02 &      & 1.5 $\pm$  0.5 & 1111&  		     \\  
     07160   &  3.64 $\pm$ 0.35	& 2.32 $\pm$ 0.35 & 4   &	    ...	   &   0.82 $\pm$ 0.03 &  a,e  & 3.6 $\pm$  0.5 & 1000&  		     \\   
     08355   &  0.91 $\pm$ 0.25	& 1.21 $\pm$ 0.34 &     & 2.09 $\pm$ 0.38  &   0.91 $\pm$ 0.18 &   f  & 1.3 $\pm$  0.8 & 1110&  		     \\   
     08424 N  &  2.50 $\pm$ 1.10	& 0.81 $\pm$ 0.07 & 1,4  &	    ...	   &	    ...        &      & 2.5 $\pm$  0.5 & 1000&  		     \\
     08424 S  &  2.20 $\pm$ 0.80	& 1.02 $\pm$ 0.10 & 1   &	    ...	   &	    ...        &      & 2.2 $\pm$  0.4 & 1000&  		     \\     
     08520 N  &  2.87 $\pm$ 0.98	& 	...	  &     & 2.36 $\pm$ 0.34  &   2.17 $\pm$ 0.25 &   g  & 2.4 $\pm$  0.6 & 1010&  		     \\
     08520 S  &  4.00 $\pm$ 0.30	& 	...	  & 4   & 5.95 $\pm$ 0.22  &   2.64 $\pm$ 0.30 &   g  & 5.3 $\pm$  1.4 & 1010&  		     \\     
     09022   &  1.96 $\pm$ 0.45	& 1.95 $\pm$ 0.40 &     & 3.25 $\pm$ 0.85  &   2.40 $\pm$ 0.40 &   d  & 2.2 $\pm$  0.6 & 1111&  		     \\  
     09437 N  &  6.10 $\pm$ 1.10	& 3.46 $\pm$ 0.71 & 4   &	     ...   &	    ...        &      & 6.1 $\pm$  1.3 & 1000&  		     \\
     09437 S  &  3.09 $\pm$ 0.21	& 2.33 $\pm$ 0.39 & 4   &	     ...   &	    ...        &      & 3.1 $\pm$  0.5 & 1000&  		     \\   
    10015    &  3.56 $\pm$ 0.45	& 2.75 $\pm$ 0.48 &     & 1.18 $\pm$ 0.18  &   1.36 $\pm$ 0.10 &   a  & 3.2 $\pm$  0.6 & 1100&  		     \\
    10038    &  2.12 $\pm$ 0.70	& 1.96 $\pm$ 0.61 &     &	     ...   &	    ...        &      & 2.0 $\pm$  0.6 & 1100&  		     \\   
    10257    &  2.66 $\pm$ 0.86	& 1.76 $\pm$ 0.27 &     & 1.55 $\pm$ 0.04  &   1.33 $\pm$ 0.01 &  a,d  & 1.8 $\pm$  0.6 & 1100&  		     \\   
    10409    &  2.92 $\pm$ 0.22	& 2.25 $\pm$ 0.19 &     &	     ...   &	    ...        &      & 2.5 $\pm$  0.5 & 1100&  		     \\
    10567    &  2.88 $\pm$ 1.50	& 3.20 $\pm$ 0.40 & 2   &	     ...   &	    ...        &      & 3.2 $\pm$  0.9 & 1100&  		     \\
    11254    &  3.14 $\pm$ 0.86	& 2.40 $\pm$ 0.57 & 2   &	     ...   &	    ...        &      & 2.6 $\pm$  0.7 & 1100&  		     \\   
    11506    &  1.22 $\pm$ 0.20	& 0.94 $\pm$ 0.17 &     & 0.67 $\pm$ 0.02  &   0.61 $\pm$ 0.02 &   c  & 1.1 $\pm$  0.2 & 1100&  		     \\  
    12042 N  &  1.22 $\pm$ 0.20	& 1.18 $\pm$ 0.29 & 1   &	     ...   &	     ...       &      & 1.2 $\pm$  0.2 & 1100&  		     \\
    12042 S  &  2.32 $\pm$ 0.32	& 1.47 $\pm$ 0.26 & 4   &	     ...   &	     ...       &      & 2.3 $\pm$  0.5 & 1000&  		     \\
    12115     &  2.97 $\pm$ 0.57	& 2.14 $\pm$ 0.36 &     &	     ...   &	     ...       &      & 2.4 $\pm$  0.6 & 1100&  		     \\
    12116     &  1.17 $\pm$ 0.35	& 1.34 $\pm$ 0.25 & 1   & 1.94 $\pm$ 0.88  &   1.12 $\pm$ 0.11 &   b,f & 1.3 $\pm$  0.5 & 1110&  		     \\
    12596     &  			& 2.17 $\pm$ 0.97 & 1   &	  ...	   &   3.13 $\pm$ 0.45 &   c  & 2.7 $\pm$  1.0 & 0101&  		     \\
    13001     &  3.71 $\pm$ 1.00	& 2.89 $\pm$ 1.10 & 4   & 5.79 $\pm$ 0.15  &   3.33 $\pm$ 0.09 &   d  & 5.7 $\pm$  1.5 & 1010&  		     \\
    13229     &  3.41 $\pm$ 0.85	& 2.39 $\pm$ 0.84 & 4   & 1.08 $\pm$ 0.09  &   0.87 $\pm$ 0.04 &   a  & 3.4 $\pm$  1.2 & 1000&  		     \\
    14544 E  &  4.76 $\pm$ 0.56	& 2.63 $\pm$ 0.48 & 4   & 2.66 $\pm$ 0.19  &   1.24 $\pm$ 0.06 &   c  & 4.8 $\pm$  1.2 & 1000&  		     \\
    14544 W  &  3.04 $\pm$ 1.43	& 2.25 $\pm$ 0.51 & 2   & 1.55 $\pm$ 0.16  &   1.09 $\pm$ 0.12 &   a  & 2.3 $\pm$  1.0 & 1100&  		     \\  
    17138    &  1.78 $\pm$ 0.15	& 1.46 $\pm$ 0.25 &     & 1.44 $\pm$ 0.13  &   1.06 $\pm$ 1.14 &   a  & 1.7 $\pm$  0.2 & 1100&  		     \\   
    18093 N  &  1.75 $\pm$ 0.32	& 1.64 $\pm$ 0.35 & 1   & 1.57 $\pm$ 0.18  &   1.13 $\pm$ 0.08 &   a  & 1.8 $\pm$  0.5 & 1000&  		     \\   
    18093 C  &  			&      (1.04)     & 5   & 0.39 $\pm$ 0.21  &   0.26 $\pm$ 0.14 &   c,g &$>$0.3$\pm$  0.2 & 0011&  		     \\ 
    18093 S  &  2.07 $\pm$ 0.53	& 1.46 $\pm$ 0.30 & 4   & 1.89 $\pm$ 0.27  &   1.32 $\pm$ 0.09 &   c  & 2.1 $\pm$  0.7 & 1000&  		     \\  
    21130     &  4.95 $\pm$ 0.48	& 4.56 $\pm$ 1.21 &     &	  ...	   &	    ...        &      & 4.9 $\pm$  0.8 & 1100&  		     \\   
    21453     &  3.68 $\pm$ 0.51	& 2.97 $\pm$ 0.82 &     & 2.59 $\pm$ 0.29  &   1.34 $\pm$ 0.08 &   a  & 3.5 $\pm$  0.7 & 1100&  		     \\   
    22132     &  5.13 $\pm$ 0.66	& 3.12 $\pm$ 1.15 & 2,4  &	  ...	   &   1.11 $\pm$ 0.04 &  a,e  & 5.1 $\pm$  0.7 & 1000&  		     \\  
    22491    &  3.70 $\pm$ 1.00	& 3.42 $\pm$ 2.80 & 2   &	  ...	   &   3.00 $\pm$ 2.40 &   b  & 3.6 $\pm$  1.5 & 1101& 8.29 (a)		     \\
    23128  &  ...			& 1.91 $\pm$ 0.22 &     &	  ...	   &   1.60 $\pm$ 0.22 &      & 1.8 $\pm$  0.3 & 0101&  		     \\
\end{tabular} 
\vskip0.2cm\hskip0.0cm
\begin{minipage}[h]{17.5cm}
\tablefoot{\tiny 
Columns: (1) Identification.
(2) Half-light radius from the 2MASS K-band image using GALFIT. 
(3) Half-light radius from the 2MASS K-band image using the A/2 method (see text). 
(4) Comments associated with the 2MASS images determinations, according to the following code: 1- large uncertainties when applying the A/2 method due to galaxy decomposition; 2- uncertainty due to background value, field stars, and/or complex structure; 3- GALFIT could not obtain a reliable fit; 4- disc with b/a $<$ 0.4, leading to discrepancies between the two methods; 5- unreliable estimates due to background and/or complex structure; 6- point-like source dominating the (unreliable) radius determination; 7- FWHM of the bulge component close to the value for the PSF, making the radii estimates uncertain. 
(5) Half-light radius from the HST H-band image using GALFIT.
(6) Half-light radius from the HST H-band image using the A/2 method. 
(7) Comments associated with the HST determinations, according to the following code: a- severely limited by FoV; b- very uncertain results due to bad background definition; c- limited by FoV; d- HST image obtained with WFC3 or NIC3 (instead of NIC2); e- bad GALFIT fit; f- rich structure in the HST image; g- uncertainties when applying the A/2 method due to galaxy decomposition; h- point-like source dominating the (unreliable) radius determination.
(8) Adopted values. 
(9) Code indication if the individual values in columns 2, 3, 5, and 6 were used in the adopted value (0 = not used, 1 = used). 
(10) Previous determinations with the following code: 
(a) \cite{SCo00}, model fitting (de Vaucoulers or exponential) to F160W images; 
(b) \cite{H06}.}
\end{minipage}
\end{scriptsize}
\end{table*}

\end{document}